\documentclass{aa}

\usepackage{graphicx}
\usepackage{multirow}
\usepackage{xcolor}
\usepackage{txfonts}
\begin{document}

   \title{Infrared spectra of complex organic molecules in astronomically relevant ice matrices}

   \subtitle{III. Methyl formate and its tentative solid-state detection}

   \author{Jeroen Terwisscha van Scheltinga \inst{1,2}
          \and
          Giulia Marcandalli \inst{3}
          \and
          Melissa K. McClure \inst{2}
          \and \\
          Michiel R. Hogerheijde \inst{2,4}
          \and
          Harold Linnartz \inst{1}
          }

   \institute{Laboratory for Astrophysics, Leiden Observatory, Leiden University, P.O. Box 9513, 2300 RA Leiden, The Netherlands\\
              \email{jeroentvs@strw.leidenuniv.nl}
            \and
            Leiden Observatory, Leiden University, P.O. Box 9513, 2300 RA Leiden, The Netherlands
            \and
            Leiden Institute of Chemistry, Leiden University, P.O. Box 9502, 2300 RA Leiden, The Netherlands
            \and
            Anton Pannekoek Institute for Astronomy, University of Amsterdam, Science Park 904, 1098 XH Amsterdam, The Netherlands
             }

   \date{Received 04 March 2021 / Accepted 03 May 2021}

  \abstract
   {Infrared spectroscopy of star and planet forming regions is at the dawn of a new age with the upcoming \textit{James Webb} Space Telescope (JWST). Its high resolution and unprecedented sensitivity allows us to probe the chemical complexity of planet forming regions, such as dense clouds, embedded protostars, and protoplanetary disks, both in the solid state and gas phase. In support of these observations, laboratory spectra are required to identify complex organic molecules in the ices that cover the dust grains in these regions.}
   {This study aims to provide the necessary reference spectra to firmly detect methyl formate (HCOOCH$_3$) in the different evolutionary stages of star and planet forming regions. Methyl formate is mixed in astronomically relevant matrices, and the peak positions, full width at half maximum, and relative band intensities are characterized for different temperatures to provide an analytical tool for astronomers.}
   {Methyl formate is deposited at 15 Kelvin on a cryogenically cooled infrared transmissive window under high-vacuum conditions. Specifically, methyl formate is deposited pure and mixed with CO, H$_2$CO, CH$_3$OH, H$_2$O, and CO:H$_2$CO:CH$_3$OH combined. The sample is linearly heated until all solid-state constituents have desorbed. Throughout the experiment infrared spectra are acquired with a Fourier Transform Infrared spectrometer in the range from 4000--500 cm$^{-1}$ (2.5--20 $\mu$m) at a spectral resolution of 0.5 cm$^{-1}$.}
   {We present the characterization of five solid-state methyl formate vibrational modes in pure and astronomically relevant ice matrices. The five selected vibrational modes, namely the C=O stretch (5.804 $\mu$m), the C$-$O stretch (8.256 $\mu$m), CH$_3$ rocking (8.582 $\mu$m), O$-$CH$_3$ stretching (10.98 $\mu$m), and OCO deformation (13.02 $\mu$m), are best suited for a JWST identification of methyl formate. For each of these vibrational modes, and each of the mixtures the temperature versus spectra heatmaps, peak position versus full width at half maximum and relative band intensities are given. All spectra are publicly available on the Leiden Ice Database. Additionally, the acquired reference spectra of methyl formate are compared with archival \textit{Spitzer} observations of HH~46. A tentative detection of methyl formate provides an upper limit to the column density of $1.7\times10^{17}$ cm$^{-2}$, corresponding to an upper limit relative to water of $\leq 2.2\%$ and $\leq 40\%$ with respect to methanol.}
   {}


   \maketitle
%

\section{Introduction}
The number of identified molecular species in the different evolutionary stages of planet forming systems is continuously increasing \citep{2018_McGuire_det_ApJS..239...17M}. The growing complexity of newly detected species suggests that chemistry in the inter- and circumstellar medium is active and diverse. The majority of these molecules are detected in the gas phase with state-of-the-art facilities such as the Atacama Large Millimeter/submillimeter Array (ALMA), but a dozen are also found in the solid state, e.g., H$_2$O and CH$_3$OH. These are detected through either ground-based telescopes such as the NASA Infrared Telescope Facility (IRTF), or space-based observatories such as the  Infrared Space Observatory (ISO) and \textit{Spitzer} Space Telescope.

\begin{figure*}[ht!]
    \includegraphics[width=0.9475\hsize]{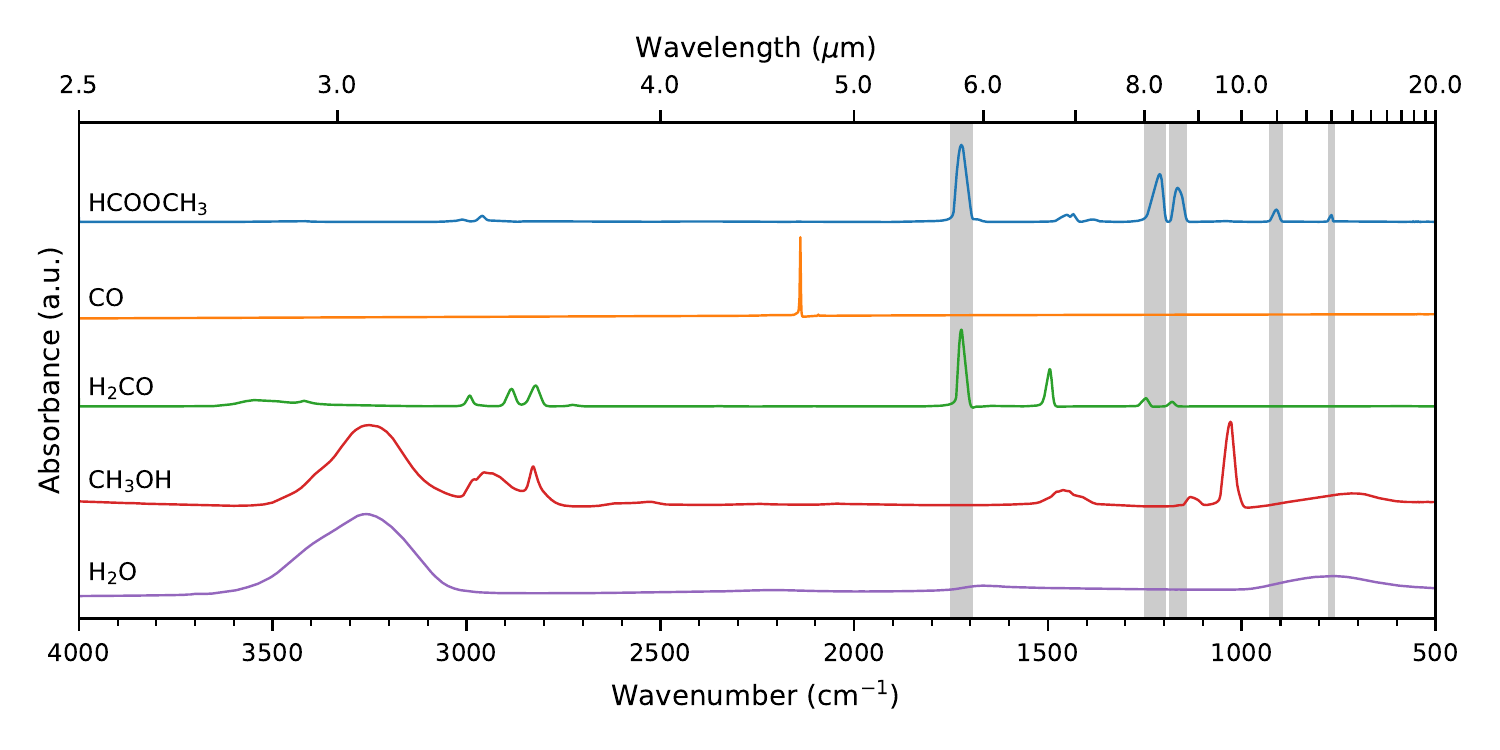} \\
    \caption{Pure normalized spectra of methyl formate (HCOOCH$_3$) and the matrix constituents (CO, H$_2$CO, CH$_3$OH, and H$_2$O) acquired at 15 K. The shaded areas highlight the features of methyl formate investigated in this study.}
    \label{fig:pure}
\end{figure*}

 Infrared facilities probe the vibrational modes of molecules residing in ice-covered dust grains. These grains are present throughout the different evolutionary stages of star and planet forming regions. To date only small species, such as water (H$_2$O), carbon monoxide (CO), carbon dioxide (CO$_2$), methane (CH$_4$), and ammonia (NH$_3$) were identified in the solid state. \citep[e.g.,][]{2015_Boogert_icy_universe_ARA&A..53..541B}. The largest molecule firmly identified to date in the solid-state is methanol (CH$_3$OH), a species that by definition is considered a Complex Organic Molecule (COM). This is interesting as methanol is a stepping stone towards larger COMs, such as acetaldehyde (CH$_3$CHO), ethanol (CH$_3$CH$_2$OH), dimethyl ether (CH$_3$OCH$_3$), methyl formate (HCOOCH$_3$), and glycolaldehyde (HOCH$_2$CHO) as shown by laboratory studies \citep[e.g.,][]{2007_Bennett_GA_MF_ApJ...661..899B, 2009_Oberg_COMs_A&A...504..891O, 2012_Modica_MF_comets_P&SS...73..425M, 2014_MunozCaro_UV_ion_A&A...566A..93M, 2016_KoJu_H-atom_MNRAS.455.1702C, 2017_Fedoseev_Glycerol_ApJ...842...52F}. Of these larger COMs two, acetaldehyde and ethanol ice, have been tentatively detected in the massive embedded protostar W33A \citep{1999_Schutte_W33A_A&A...343..966S, 2011_Oberg_Spitzer_ApJ...740..109O}. COMs are, however, frequently observed in the gas phase with ground-based telescopes such as ALMA and the Institut de Radioastronomie Millimetrique (IRAM) \citep{2015_Oberg_CH3CN_Natur.520..198O, 2016_Jorgensen_PILS_A&A...595A.117J, 2017_Bergner_COM_emb_low_ApJ...841..120B, 2017_Ligterink_CH3NCO_MNRAS.469.2219L, 2017_Ceccarelli_SOLIS_ApJ...850..176C, 2017_McGuire_MOxyMeOH_ApJ...851L..46M, 2018_Ligterink_peptide_MNRAS.480.3628L, 2019_Bogelund_AFGL_A&A...628A...2B, 2019_Bogelund_Nbearing_A&A...624A..82B, 2020_Manigand_16293A_A&A...635A..48M, 2020_Gelder_O-COMs_A&A...639A..87V, 2021_Yang_Peaches_ApJ...910...20Y, 2021_Nazari_COMs_arXiv210403326N}. For example, the recent Perseus ALMA Chemistry Survey (PEACHES) revealed that 58\% of their 50 observed embedded (Class 0/1) protostars have COM emission \citep{2021_Yang_Peaches_ApJ...910...20Y}. Both oxygen- and nitrogen-bearing COMs have been detected, including, but not limited to, methanol (CH$_3$OH), methyl cyanide (CH$_3$CN), formamide (NH$_2$CHO), methyl formate (HCOOCH$_3$), and ethyl cyanide (CH$_3$CH$_2$CN).

The changing physical conditions between the different stages of star and planet formation provide a wide range of different (non)energetic triggers that allow the molecular complexity to increase in both the gas phase and in the solid state. Chemical models attempt to reproduce observed COM abundances, with varying degrees of success. Such models describe the formation and destruction of COMs in laboratory or astronomical settings, and include gas-phase chemistry and/or  solid-state chemistry \citep{2013_Garrod_3P_Hcore_ApJ...765...60G, 2016_Taquet_COM_outbursts_ApJ...821...46T, 2017_Vasyunin_cmod_COM_prestellar_ApJ...842...33V, 2018_Quenard_COMs_peptide_MNRAS.474.2796Q, 2019_Ruaud_cmod_ppd_pJ...885..146R, 2020_Simons_CO+H_mod_A&A...634A..52S, 2020_Jin_COM_chemmod_ApJS..249...26J, 2021_Wang_COM_model_SGB2_arXiv210310376W}. In general, formation of COMs is deemed more likely to occur in the solid state compared to the gas phase. However, gas-phase COM signatures are also detected in cold regions with temperatures well below their thermal desorption value. This requires either unexplored gas-phase reactions or non-thermal desorption of COMs from the solid state. Methyl formate is one such molecules that is expected to form in solid state and yet is observed in cold gas phase environments, e.g., the cold prestellar cores L1689B \citep{2012_Bacmann_L1689B_A&A...541L..12B} and low-mass protostar B1-b \citep{2010_Oberg_B1-b_ApJ...716..825O, 2012_Cernicharo_B1_ApJ...759L..43C}. Besides the observation of methyl formate in cold environments, it is also seen in environments such as hot cores of massive star-forming regions and protostars \citep[see e.g.,][]{2013_Belloche_SB2_A&A...559A..47B, 2016_Jorgensen_PILS_A&A...595A.117J, 2020_Gelder_O-COMs_A&A...639A..87V}.

Several gas-phase formation pathways have been proposed to form methyl formate through ion-molecule and subsequently dissociative recombination reactions. One of the generally used gas-phase formation channels is the formation of protonated methyl formate from protonated methanol and formaldehyde (\ref{eq1}). This protonated methyl formate then recombines with an electron to form methyl formate (\ref{eq2}).

\begin{equation} \label{eq1}
    \mathrm{[CH_3OH_2]^+ + H_2CO \rightarrow [HC(OH)OCH_3]^+ + H_2}
\end{equation}
\begin{equation} \label{eq2}
    \mathrm{[HC(OH)OCH_3]^+ + \textit{e} \rightarrow HCOOCH_3 + H}
\end{equation}

This reaction, with assumed high reaction rates, has been proposed to produce significant amounts of methyl formate within $10^4-10^5$ years in chemical models of hot cores \citep{1991_Millar_O-COMs_Orion_ApJ...369..147M, 1993_Caselli_Chem_diff_ApJ...408..548C, 1995_Charnley_ISM_alcohols_ApJ...448..232C}. \citet{2004_Horn_MF_gas-phase_ApJ...611..605H}, however, investigated several ion-molecule pathways, including Equation~\ref{eq1}, to form protonated methyl formate in a quantum chemical study. They reported that these reactions have significant activation barriers that prohibit these pathways to efficiently contribute to the observed abundances of methyl formate in hot cores. 

The authors tested an alternative association reaction between methyl ions and formic acid (HCOOH) (\ref{eq3})

\begin{equation} \label{eq3}
    \mathrm{CH_3^+ + HCOOH \rightarrow [HC(OH)OCH_3]^+}
\end{equation}
which was found to be more efficient. However, despite its higher rate, this pathway is still not able to reproduce the observed methyl formate abundances in hot cores. 

\begin{table*}[ht!]
    \centering
    \caption[]{Selected transitions of methyl formate for solid-state identification.}
    \begin{tabular}{l l l l l l l}
    \hline
    \hline
    \noalign{\smallskip}
    \multicolumn{2}{c}{\multirow{2}{*}{Species}} & \multirow{2}{*}{Formula} & \multirow{2}{*}{Mode} & \multicolumn{2}{c}{Peak position*} & $A'$ \\
    &       &       & & cm$^{-1}$ & $\mu$m & cm molec.$^{-1}$ \\
    \hline
    \\
    \multirow{5}{*}{\includegraphics[height=2.0cm]{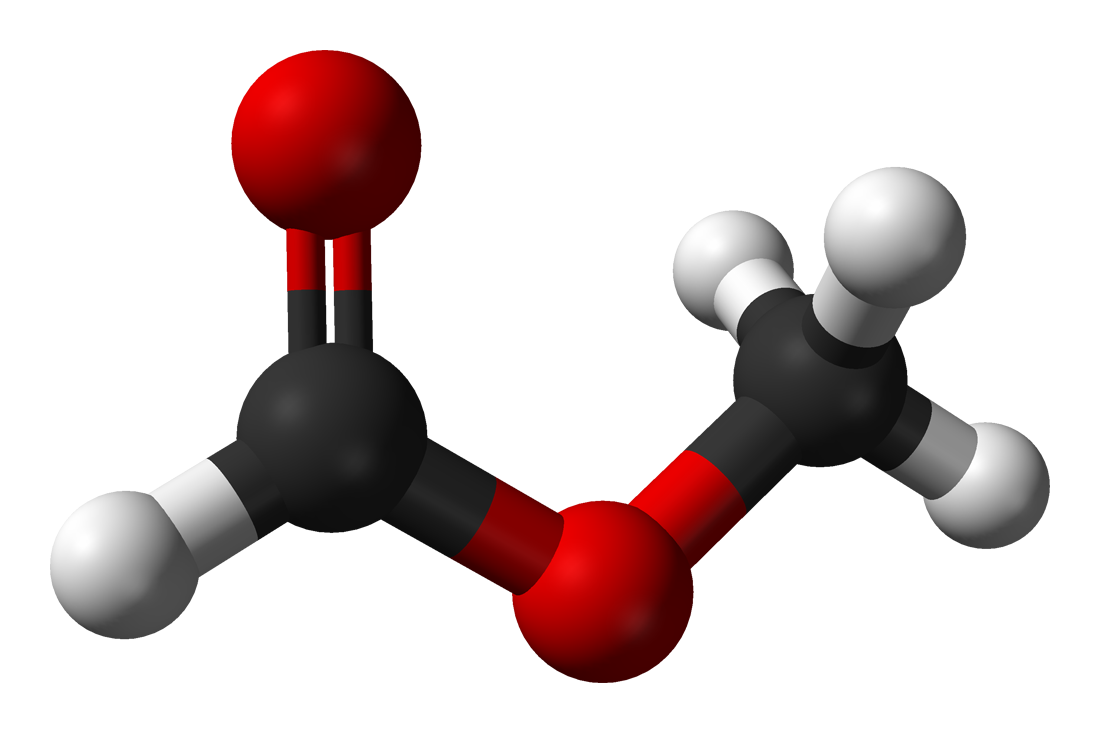}} & \multirow{5}{*}{Methyl formate} & \multirow{5}{*}{HCOOCH$_{3}$} & C$=$O stretch. & 1723.1 & 5.804 & 4.96$\times$10$^{-17,\ a}$ \\
    & & & C$-$O stretch. & 1211.3 & 8.256 & 2.93$\times$10$^{-17,\ a}$ \\
    & & & CH$_3$ rock. & 1165.3 & 8.582 & 1.96$\times$10$^{-17,\ a}$ \\
    & & & O$-$CH$_3$ stretch. & 910.7 & 10.98 & 4.82$\times$10$^{-18,\ a}$ \\
    & & & OCO deform. & 768.3 & 13.02 & 1.25$\times$10$^{-18,\ a}$ \\
    \\
    \hline
    \noalign{\smallskip}
    \end{tabular}
    \label{tab.selectbands}
    \\
\tablefoot{*Peak position of the pure molecule at 15~K. $^{a}$ Taken from \citet{2010_Modica_MF_cosmic_ion_A&A...519A..22M}.}
\end{table*}

Recent modeling attempts by \citet{2015_balucani_COMs_gp_MNRAS.449L..16B} suggest that non-thermally desorped methanol ice could potentially feed formation of COMs in the gas phase. The gas-phase methanol reacts with a hydroxyl radical to form methoxy (\ref{eq:OCH3}). These methoxy radicals can then subsequently react with methyl radicals to form dimethyl ether (CH$_3$OCH$_3$) through reaction~\ref{eq:DME}. Methyl formate is formed through the oxidation of the CH$_2$OCH$_3$ radicals (\ref{eq:MF}), which are created through hydrogen abstraction of dimethyl ether (\ref{eq:DME_rad1}~\&~\ref{eq:DME_rad2}).

\begin{equation} \label{eq:OCH3}
    \mathrm{CH_3OH + {^{\bullet}OH} \rightarrow {^{\bullet}OCH_3} + H_2O}
\end{equation}
\begin{equation} \label{eq:DME}
    \mathrm{^{\bullet}OCH_3 + {^{\bullet}CH_3} \rightarrow CH_3OCH_3\ +}\ h\nu 
\end{equation}
\begin{equation} \label{eq:DME_rad1}
    \mathrm{CH_3OCH_3 + F \rightarrow {^{\bullet}CH_2OCH_3} + HF}
\end{equation}
\begin{equation} \label{eq:DME_rad2}
    \mathrm{CH_3OCH_3 + Cl \rightarrow {^{\bullet}CH_2OCH_3} + HCl}
\end{equation}
\begin{equation} \label{eq:MF}
    \mathrm{{^{\bullet}CH_2OCH_3} + O \rightarrow HCOOCH_3 + H}
\end{equation}

With these reactions incorporated in their model, it is able to retrieve abundances that are in agreement with upper limits derived from observations.

Besides gas-phase reactions it is also possible to form methyl formate in the solid state through both energetic and non-energetic processes. The associated radicals to form methyl formate in the solid state are formyl and methoxy (\ref{eq5}), and both radicals can be formed through energetic and non-energetic processing.

\begin{equation} \label{eq5}
    \mathrm{^\bullet CHO + {^{\bullet}OCH_3} \rightarrow HCOOCH_3}
\end{equation}
\citet{2009_Oberg_COMs_A&A...504..891O} investigated the formation of different COMs through the UV irradiation of pure CH$_3$OH ice and CH$_3$OH mixed with either CO or CH$_4$. In their work the formation of methyl formate was confirmed and reaction~\ref{eq5} was proposed as the underlying reaction pathway. Apart from UV-photon triggered formation it is also possible to form methyl formate through cosmic ray irradiation \citep{2010_Modica_MF_cosmic_ion_A&A...519A..22M}. In their work pure methanol and mixtures of methanol and CO were irradiated with 200 keV protons, and methyl formate was identified through its fundamental CH$_3$ rocking vibrational mode in the infrared. Methyl formate can also be formed in the solid state through a non-energetic process involving hydrogen addition and abstraction reactions. The hydrogenation of CO is known to form H$_2$CO and CH$_3$OH \citep{1994_Hiraoka_hydro_CPL...229..408H, 2002_Hiraoka_hydro_ApJ...577..265H, 2002_Watanabe_hydro_ApJ...571L.173W, 2004_Hidaka_H2CO_ApJ...614.1124H, 2004_Watanabe_hydro_ApJ...616..638W, 2009_Fuchs_hydro_A&A...505..629F}. The work by \citet{2016_KoJu_H-atom_MNRAS.455.1702C} showed that the intermediate radicals - HCO, CH$_3$O, and CH$_2$OH - can also react with each other in radical-radical recombinations and form COMs, such as methyl formate, glycolaldehyde, and ethylene glycol. As is evident from these laboratory experiments, there are sufficient pathways to form COMs in the ices that cover dust grains.

Observational results suggest that the ices on dust grains are comprised of two layers, a polar and an apolar layer \citep[e.g.,][]{2015_Boogert_icy_universe_ARA&A..53..541B}. The radicals that form methyl formate in the solid state are part of the CO hydrogenation family. It is thus likely that methyl formate is mixed with CO, H$_2$CO, and CH$_3$OH and therefore resides in the apolar layer on interstellar dust grains. 

The observational identification of COMs will receive a new impetus with the upcoming \textit{James Webb} Space Telescope (JWST). It will be possible to probe these icy grains with higher spatial and spectral resolution than previously available. This creates an opportunity to look for COMs in the different evolutionary stages of star and planet forming regions. In preparation to these observations, this manuscript is the third in a series where COMs are spectroscopically characterized in the infrared for different astronomically relevant ice matrices \citep{2018_TvS_COMs_A&A...611A..35T,2020_Rachid_acetone_A&A...639A...4R}. Each molecule has it own spectroscopic signature that consists of a series of vibrational modes and allows for unambiguous identifications. In the solid state vibrational transition properties are influenced by the physical structure, temperature, and surrounding molecules \citep{2005_Palumbo_h2o_morph_JPhCS...6..211P, 2007_Dawes_morph_JChPh.126x4711D, 2007_Oberg_H2O_CO2_A&A...462.1187O, 2009_Oberg_segre_A&A...505..183O, 2012_Bossa_porous_A&A...545A..82B, 2014_Isokoski_porosity_PCCP...16.3456I, 2017_Jiao_CO2_ASW_ApJ...837...65H}. These changes in transition properties are reflected in the spectroscopic shape of the transitions, i.e., peak position, full width at half maximum (FWHM), and apparent band strengths. The characterization of these changes offers not only a tool to identify species, but also to characterize temperatures and mixing rates. For this systematic laboratory data are needed. Here the spectroscopic data for methyl formate are presented.

The paper is structured as follows. Section 2 describes the experimental setup and how it was used to measure the spectra. In Section 3 the results are discussed and Section 4 explores how these features can be used to interpret astronomical observations. In Section 5 we conclude with a summary.

\section{Experimental}
\subsection{Setup}

\begin{figure*}[ht!]
    \begin{tabular}{cc}
        \vspace{0.5cm}
        \includegraphics[width=0.475\hsize]{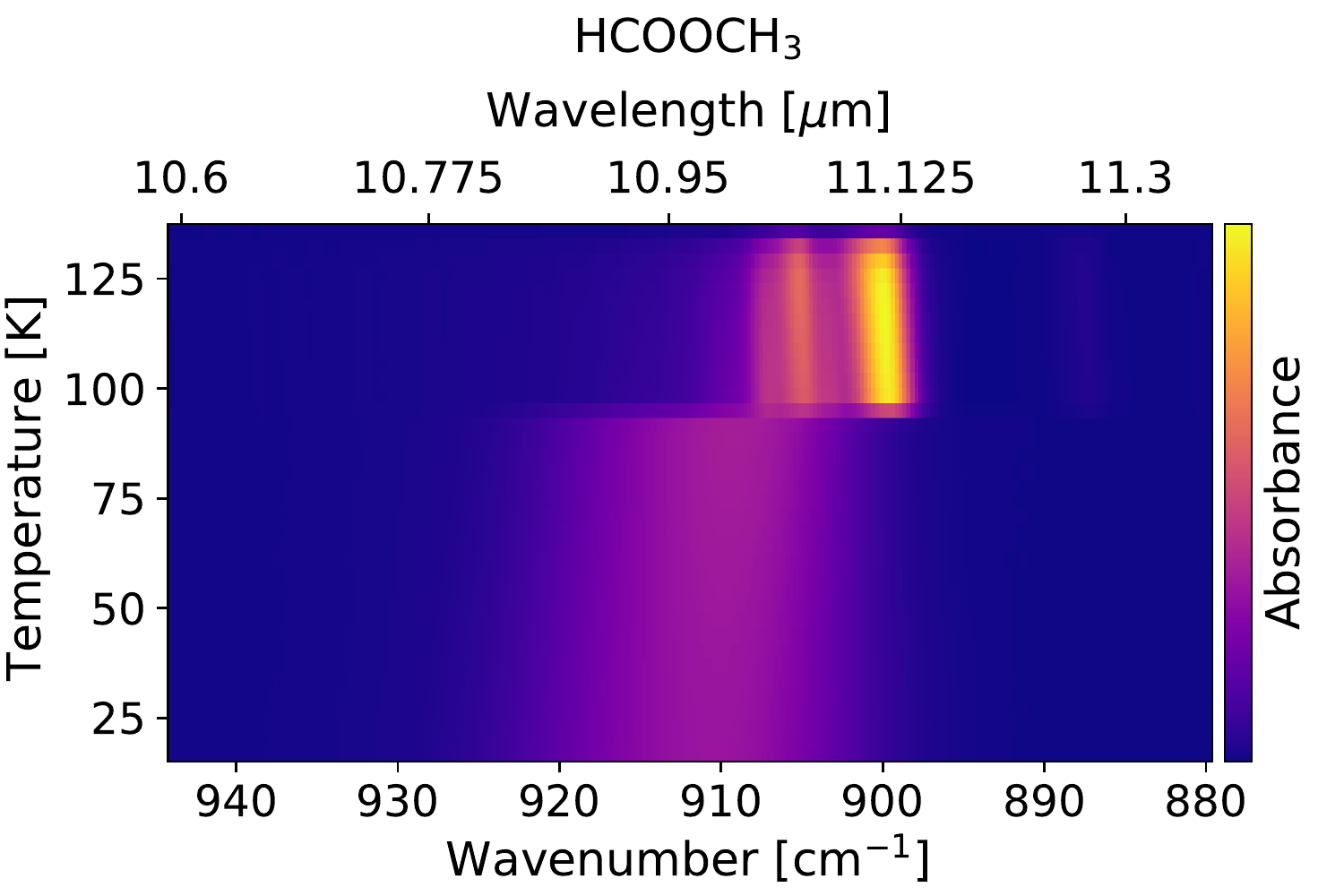} & \includegraphics[width=0.475\hsize]{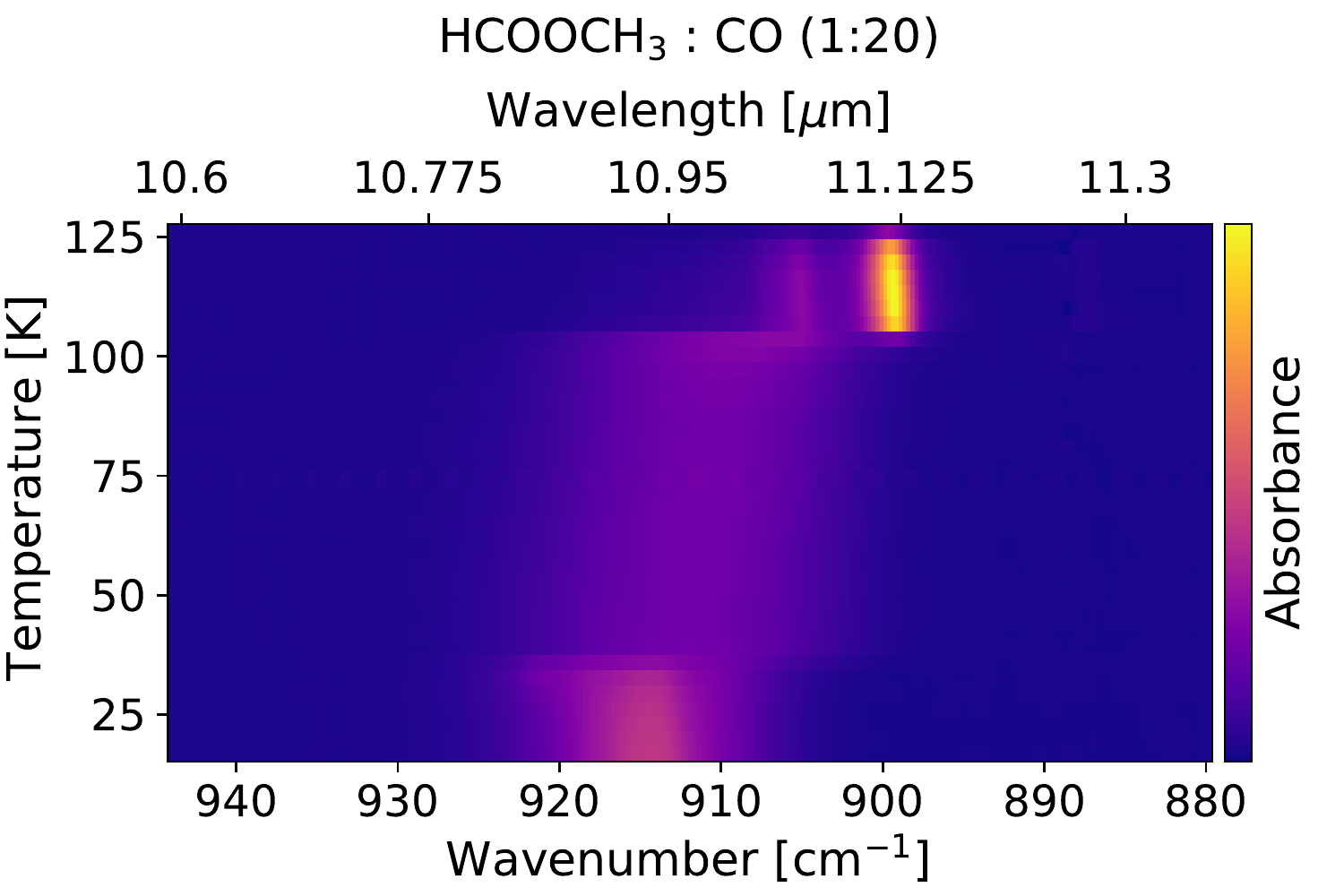} \\
        \vspace{0.5cm}
        \includegraphics[width=0.475\hsize]{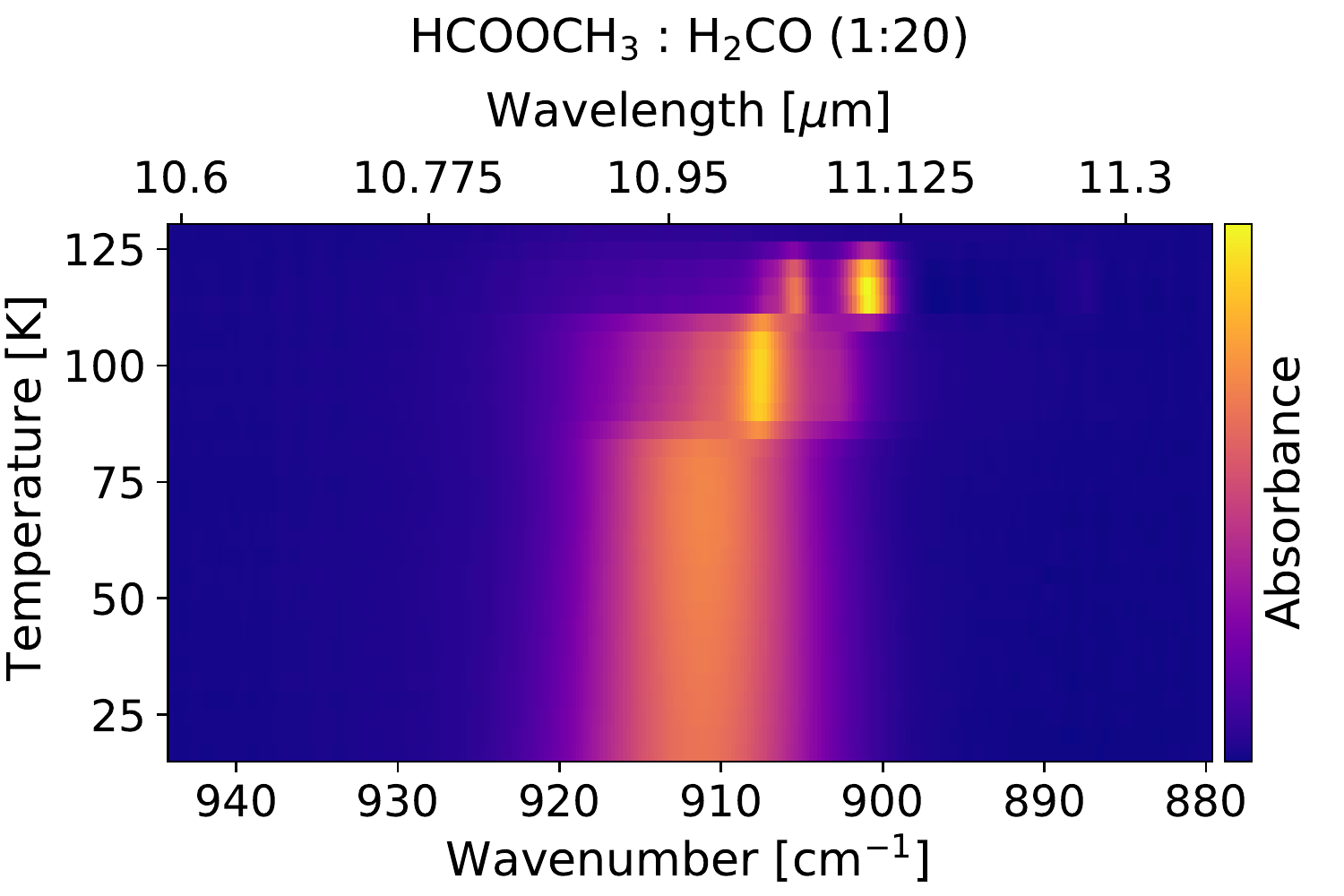} & \includegraphics[width=0.475\hsize]{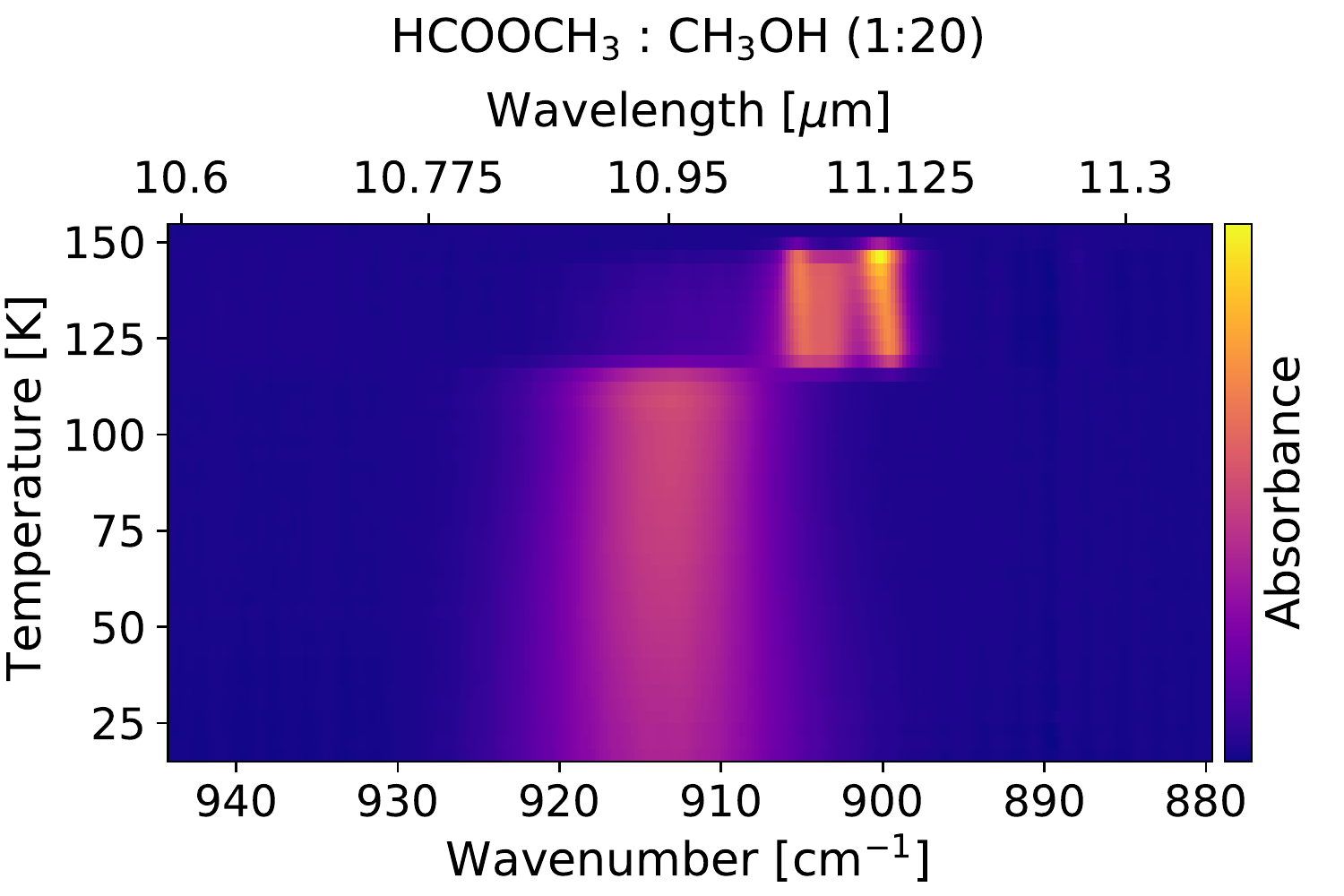} \\
        \includegraphics[width=0.475\hsize]{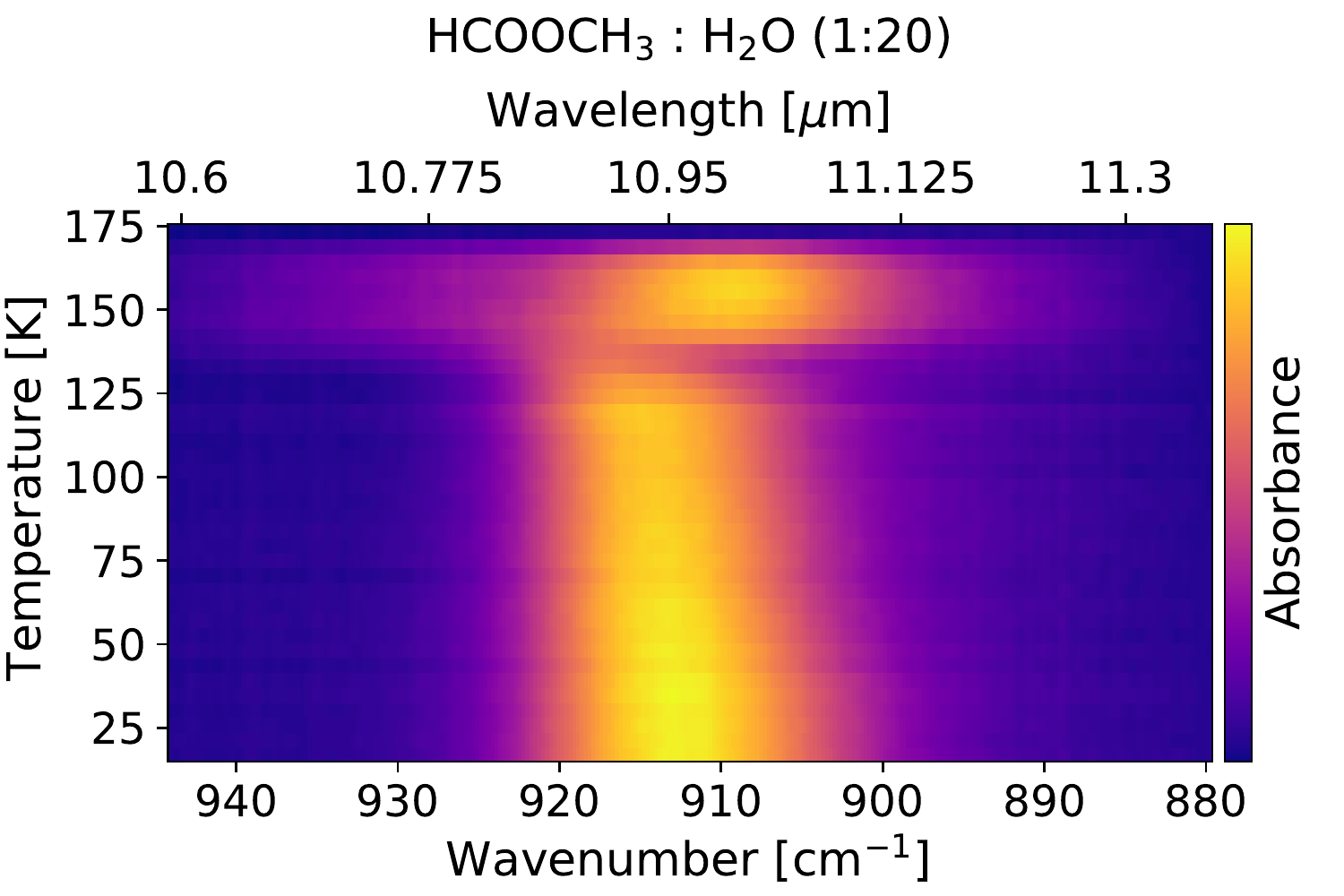} & \includegraphics[width=0.475\hsize]{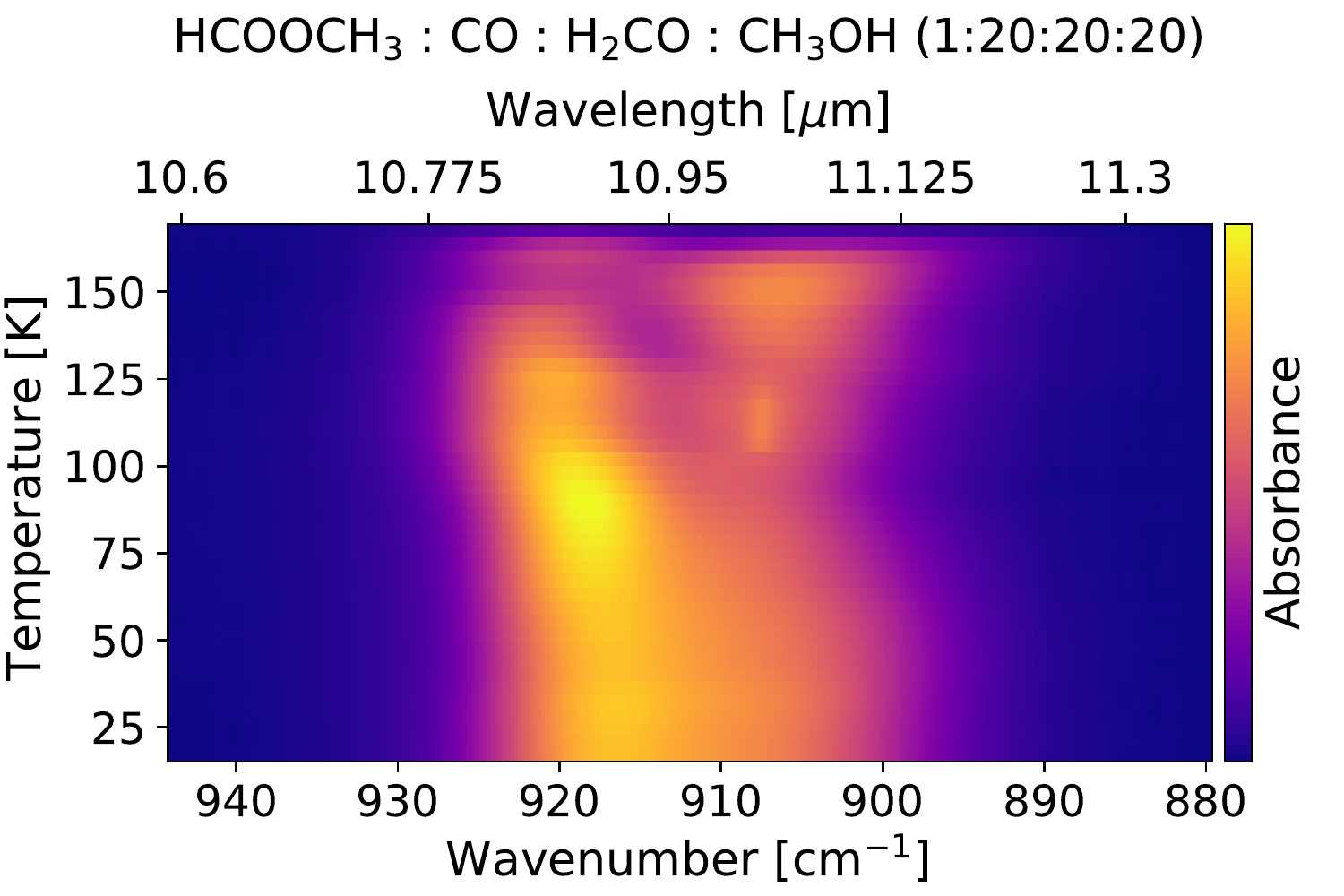} \\
    \end{tabular}
\caption{Temperature versus Spectra (TvS) heatmaps of the O$-$CH$_3$ stretching mode (910.5 cm$^{-1}$ / 10.98 $\mu$m) of HCOOCH$_3$ in the astronomically relevant ice matrices as a function of temperature.}
\label{fig:MF_910_heatmaps}
\end{figure*}

The experiments in this manuscript are measured on the High-Vacuum (HV) setup in the Laboratory for Astrophysics at Leiden Observatory. This setup consists of a stainless steel chamber which is evacuated to a pressure as low as 10$^{-7}$ mbar at room temperature. The sample on which the ice is deposited is a ZnSe window which can be cooled down to 12 K by a closed cycle helium refrigerator. The temperature of the sample is monitored and controlled with a LakeShore 330 temperature controller and silicon diode temperature sensor. The spectra are acquired with a Fourier Transform InfraRed (FTIR) Varian 670-IR spectrometer. This system covers the infrared range from $4000-500$ cm$^{-1}$ which corresponds to $2.5-20\ \mu$m. The best achievable spectral resolution is up to 0.1 cm$^{-1}$, but in this work a resolution of 0.5 cm$^{-1}$ is used which is more than sufficient. Gas mixtures for sample deposition are prepared in a separate multiline gas mixing system, which is able to produce gas mixtures with a maximum error of $< 10\%$. The specific gas and liquids used in this study are methyl formate (Sigma Aldrich, 99.0\%), carbon monoxide (Linde gas, 99.997\%), paraformaldehyde (Sigma Aldrich, 95.0\%), methanol (Sigma Aldrich, $\geq$ 99.8\%), and Milli-Q H$_2$O (Type I). The paraformaldehyde is exteriorly heated with water at 80$\ ^{\circ}$C to enhance depolymerization and increase the vapor pressure of pure formaldehyde during sample preparation. During deposition, samples containing formaldehyde are similarly heated to prevent polymerization back into paraformaldehyde. For more details on the experimental setup see \citep{2018_TvS_COMs_A&A...611A..35T} and references therein.

\begin{figure*}[ht!]
    \includegraphics[width=0.975\hsize]{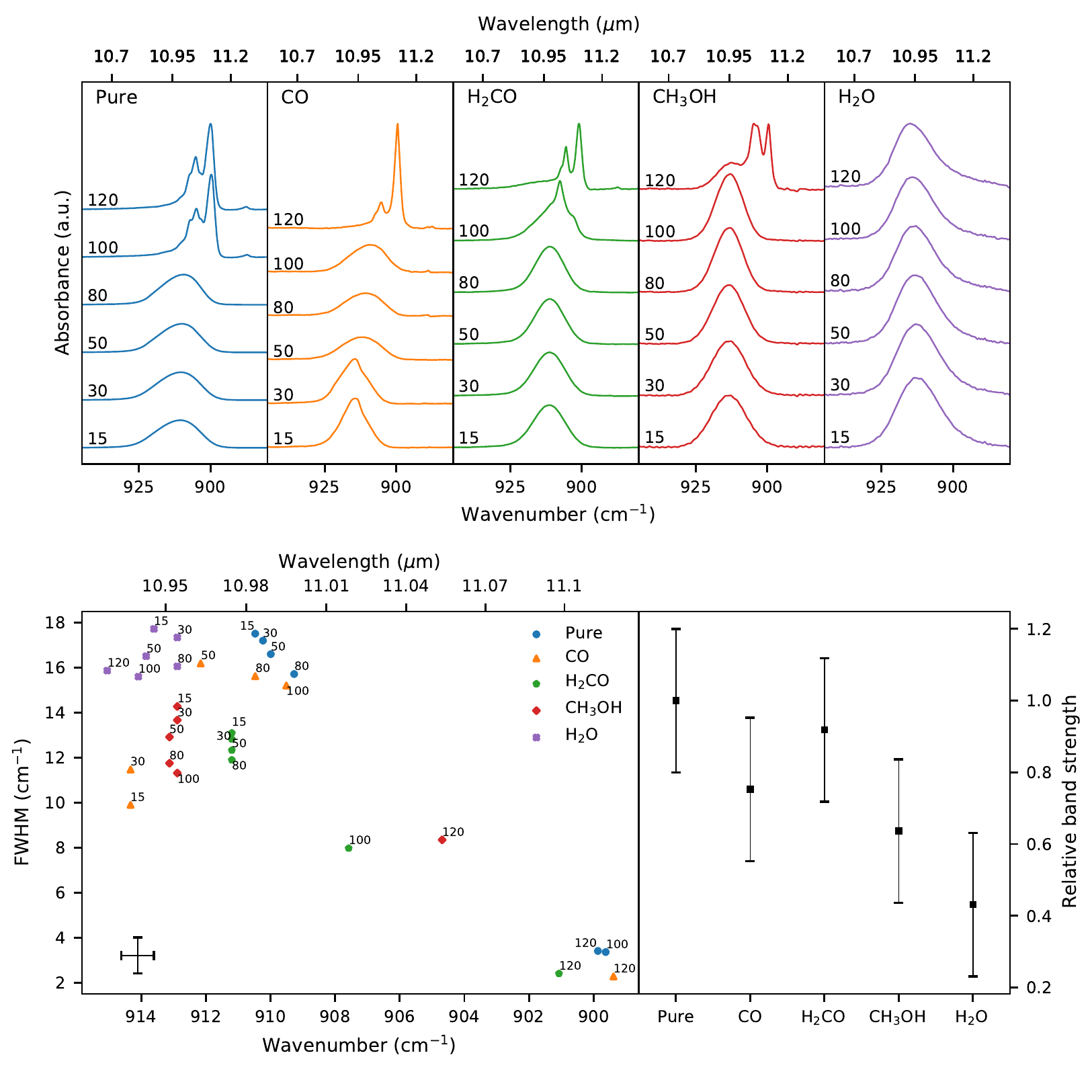}
    \caption{Top panel: Absorption feature of the O$-$CH$_3$ stretching vibrational mode (910.5 cm$^{-1}$ / 10.98 $\mu$m) for each of the mixtures at selected temperatures. Bottom left panel: FWHM as a function of peak position at the selected temperatures. The error bar of each individual point is given in the bottom left.} Bottom right panel: The relative band strengths for methyl formate in the different ice matrices at 15 K with respect to the pure band strength.
    \label{fig:MF_910}
\end{figure*}

\subsection{Measurement protocol}
The different gas samples, pure or mixed, are prepared in the gas mixing system and stored in a 2L glass bulb at a pressure of 20 mbar. The ratio at which the matrix constituents are mixed with methyl formate is 20:1. This ratio ensures a balance between sufficient dilution to investigate spectroscopic changes and adequate signal-to-noise for quantitative analysis of the spectroscopic features. Any further dilution is not expected to result in strong deviation of the observed results for methyl formate. This bulb is then connected to the HV setup and the gas is subsequently dosed onto the substrate through a needle valve. The 20 mbar pressure in a 2L glass bulb ensures a large enough reservoir to have a constant deposition rate over time. During deposition the growth of the ice is traced with the FTIR spectrometer. The measured absorption features can be translated to a column density, $N_{species}$, according to the following relation:

\begin{table*}[ht!]
    \centering
    \caption{Peak position and FWHM of the methyl formate O$-$CH$_{3}$ stretching mode at 15~K in various matrices.}
    \begin{tabular}{l c l l l l}
        \hline
        \hline
        \multirow{2}{*}{Mixture} & Temperature &  \multicolumn{2}{c}{$\lambda_{\rm{peak}}$} &  \multicolumn{2}{c}{FWHM} \\ 
        & (K) & (cm$^{-1}$) & ($\mu$m) & (cm$^{-1}$)  &  ($\mu$m) \\ 
        \hline
        HCOOCH$_3$              & \multirow{5}{*}{15}   & 910.47 & 10.983 & 17.51 & 0.2108\\
        HCOOCH$_3$ : CO         &                       & 914.33 & 10.937 & \phantom{1}9.909 & 0.1184\\
        HCOOCH$_3$ : H$_2$CO    &                       & 911.20 & 10.975 & 13.10 & 0.1577\\
        HCOOCH$_3$ : CH$_3$OH   &                       & 912.88 & 10.954 & 14.28 & 0.1711\\
        HCOOCH$_3$ : H$_2$O     &                       & 913.61 & 10.946 & 17.72 & 0.2131\\
        \hline
        \noalign{\smallskip}
    \end{tabular}
    \tablefoot{Excerpt from Table~\ref{tab.MF_O-CH3_str}.}
    \label{tab.example}
\end{table*}

\begin{equation} \label{eq:col_den}
N_{species} = \ln(10)\frac{\int_{band}\log_{10}\left(\frac{I_0(\tilde{\nu})}{I(\tilde{\nu})}\right) d\tilde{\nu}}{A'}
\end{equation}
where $\int_{band}\log_{10}\left(\frac{I_0(\tilde{\nu})}{I(\tilde{\nu})}\right) d\tilde{\nu}$ is the area of an absorption feature with $I_0(\tilde{\nu})$ and $I(\tilde{\nu})$ as the incoming flux and transmitted flux through the sample, respectively, and the apparent band strength is given by $A'$. The deposition is stopped when the column density reaches a value of approximately $2.0 \times 10^{18}$ molecules cm$^{-2}$. This equals to 2000 monolayers (ML), where one ML equals 10$^{15}$ molecules cm$^{-2}$. The 2000 ML coverage ensures that over time the measurement is not significantly contaminated by H$_2$O background gas deposition in the setup which is 30 ML h$^{-1}$. A typical deposition takes one hour which results in a contamination by background H$_2$O deposition of 1.5$\%$. In order to derive the column densities, the apparent band strengths for each of the molecules used in this study are taken from the literature and those of methyl formate are given in Table~\ref{tab.selectbands} \citep{2010_Modica_MF_cosmic_ion_A&A...519A..22M, 2015_Bouilloud_bnd_str_MNRAS.451.2145B}. It should be noted that high sample coverage combined with strong absorption features may lead to saturation and inaccurate determination of column densities. To avoid this the column density is traced with multiple features during deposition.

After deposition spectra are acquired with a resolution of 0.5 cm$^{-1}$ and scan average of 256. Once a spectrum has been acquired at 15 K the sample is linearly heated and spectra are continuously acquired until the ice has fully desorbed from the sample. The heating ramp, 0.5 K min$^{-1}$, and the time it takes to measure one spectrum, $\sim$ 7 minutes, results in an acquisition every 3.5 K. This combination allows for an optimal balance between signal-to-noise and temperature resolution per spectrum. These spectra allow characterization of their peak position and FWHM as a function of temperature.

\subsection{Analysis}
The amount of spectra acquired per experiment, i.e., for a specific gas mixture, is roughly 35. A selected few spectra at characteristic temperatures, 15, 30, 50, 80, 100, and 120 K, are shown in this manuscript and all other spectra are publicly available from the Leiden Ice Database \footnote{https://icedb.strw.leidenuniv.nl}. Each spectrum is baseline subtracted with a Cubic Spline using fit points that are chosen by eye. In the pure spectra the points are chosen in featureless regions, and in the diluted sample spectra the main constituent features are incorporated in the baseline. This allows for more accurate determination of the peak position and FWHM of the selected methyl formate transitions. As the sample is heated the molecules start orientating themselves in more energetically favorable positions, which changes the vibrational transition properties. This translates itself into changing absorption feature shape and absorption wavelength. Upon ice crystallization, for example, spectral features may narrow. An in depth investigation of how vibrational modes change in the solid state with changing environmental ice conditions is beyond the scope of this work. In this work the transitions are characterized by the peak positions and the FWHMs. Certain transitions, e.g., O--CH$_3$ stretching mode, are a combination of different transitions, which overlap when the sample is in amorphous form but become independently visible when it crystallizes. In some specific cases it is evident that there are two or more features, but at half maximum the features are still blended. In these cases the peak position of the strongest transition is given with the combined FWHM and is indicated by an asterisk. The assignments of the vibrational modes investigated in this work are given in Table~\ref{tab.selectbands} and are taken from literature \citep{2010_Modica_MF_cosmic_ion_A&A...519A..22M}.

\section{Results and discussion}
In this section we discuss the analysis of the spectra of methyl formate in pure form and mixed with astronomically relevant matrices. The spectra of methyl formate and separate matrix constituents are shown in Fig.~\ref{fig:pure}. The shaded areas highlight the absorption features of pure methyl formate that are potential candidates for a JWST identification. The O$-$CH$_3$ stretching mode (910.5 cm$^{-1}$ / 10.98 $\mu$m) analysis is shown here, and the results for the other vibrational modes mentioned in Table~\ref{tab.selectbands} are given in the appendices.
In order to visualize the large amount of data acquired, several types of figures and tables are presented in this work. Temperature versus Spectra (TvS) heatmaps are made for a quick and general overview of the effects of temperature and different neighbouring molecules on a single transition, see e.g., Fig.~\ref{fig:MF_910_heatmaps}. Each panel visualizes a different ice sample, i.e., pure methyl formate or methyl formate mixed with abundant or chemically related precursor species. A horizontal cut at a given temperature would give a regular absorption spectrum at that given experimental temperature. Methyl formate  crystallizes at approximately 90 K and this can be seen in the top left panel by the sudden shift and intensity change. In the top right panel the mixture of methyl formate with CO shows that when CO is still present in the solid state, T < 35 K, the O$-$CH$_3$ stretching mode feature is blue shifted and narrower. After CO desorption this band returns to its `original' position. However, crystallization now occurs after 100 K, and the secondary peak is less prominent compared to the pure sample. Even though CO is no longer in the solid state, it has left an imprint on the absorption spectrum of methyl formate. In aid of future astronomical identifications, a more in-depth characterization of the absorption features is given for the five selected transitions, see Fig.~\ref{fig:pure}. For these transitions the FWHMs as function of peak position and the relative band intensities are given for the selected temperatures, see e.g., Fig.~\ref{fig:MF_910}. Lastly, exact peak position, FWHM, and relative band intensity are given in the tables found in the appendices. The error on the peak position and FWHM are 0.5 and 0.8 cm$^{-1}$, respectively.

\subsection{The O$-$CH$_\text{3}$ stretching mode} \label{ssec:O-CH3_stretch}
The O$-$CH$_3$ stretching vibrational mode is positioned at 910.7 cm$^{-1}$ (10.98 $\mu$m) in a pure methyl formate ice sample at 15 K. The TvS heatmaps of this vibrational mode are shown in Fig.~\ref{fig:MF_910_heatmaps}. The pure TvS heatmap shows that around 100 K solid-state methyl formate changes from an amorphous structure to a crystalline structure. This changes the vibrational properties and expresses itself in peak splitting and a red shift to 899.6 cm$^{-1}$ (11.12 $\mu$m) for the main peak and 904.9 cm$^{-1}$ (11.05 $\mu$m) for the secondary peak. In general, when methyl formate is diluted in one of the matrix constituents (Table~\ref{tab.example}) the O$-$CH$_3$ stretching mode is blue shifted and the FWHM decreases with the exception of the H$_2$O matrix where the FWHM stays comparable. In Fig.~\ref{fig:MF_910} the O$-$CH$_3$ stretching vibrational band is shown and the FWHM versus peak position for each of the characteristic temperatures, i.e., 15, 30, 50, 80, 100, and 120 K. The bottom right panel of Fig.~\ref{fig:MF_910} shows the relative band strength compared to the pure band strength in different matrices at 15 K.

The combined apolar (CO:H$_2$CO:CH$_3$OH) matrix analysis is omitted for the O$-$CH$_3$ stretching vibrational mode as H$_2$CO and CH$_3$OH have a vibrational matrix interaction at approximately 915 cm$^{-1}$ (10.9 $\mu$m). This matrix interaction is interesting as it takes place between two main interstellar ice constituents. It is likely that this interaction scales with mixing ratio and depends on whether the ice is fully mixed or layered. We are not aware of an earlier mention of this interaction. A more detailed investigation is beyond the scope of this work.

\subsection{The OCO deformation mode} \label{ssec:OCO_deform}
The OCO deformation mode is positioned at 768.3 cm$^{-1}$ (13.02 $\mu$m) in a pure methyl formate ice sample at 15 K. As crystallization occurs around 90 K the mode splits into two modes. The main peak is blue shifted to 769.0 cm$^{-1}$ (13.00 $\mu$m) and the secondary peak to 775.2 cm$^{-1}$ (12.90 $\mu$m). This is the weakest of the transitions selected for analysis but it is in a favorable position as it only overlaps with the H$_2$O libration mode and CH$_3$OH torsion mode. These two modes are easily fitted into the baseline and subtracted. Acetaldehyde (CH$_3$CHO) has a vibrational mode at similar wavenumber, but, the presence or absence of other vibrational modes of methyl formate will allow distinction from acetaldehyde. The TvS heatmaps of the OCO deformation mode in the pure and astronomically relevant matrices are given in Fig.~\ref{fig:MF_769_heatmaps}. The general trend is that the FWHM decreases compared to the pure spectrum of methyl formate when mixed with other astronomically relevant constituents. An exception to this is the H$_2$O matrix: here the FWHM is almost doubled by the interaction with the H$_2$O matrix, and the peak position blue shifts nearly 10 cm$^{-1}$. The matrices containing H$_2$CO blue shift the peak position by a couple wavenumbers. This is visually summarized in Fig.~\ref{fig:MF_769} and the specifics are given in Table~\ref{tab.MF_OCO_def}.

\subsection{The CH$_\text{3}$ rocking mode} \label{ssec:CH3_rock}
The CH$_3$ rocking vibrational mode is positioned at 1165.3 cm$^{-1}$ (8.582 $\mu$m) in a pure methyl formate ice sample at 15 K. The mode splits into multiple features after crystallization with the main peak at 1165.8 cm$^{-1}$ (8.578 $\mu$m) and secondary peaks at 1177.1 cm$^{-1}$ (8.496 $\mu$m) and 1158.8 cm$^{-1}$ (8.630 $\mu$m). This evolution of the vibrational mode as function of temperature in pure form and mixed astronomically relevant matrices is shown in Fig.~\ref{fig:MF_1165_heatmaps}. The CH$_3$ rocking mode is blended in the mixtures containing H$_2$CO with the CH$_2$ wagging vibrational mode of H$_2$CO. This mode of H$_2$CO is positioned at 1178 cm$^{-1}$ (8.489 $\mu$m) which is in the blue wing of the methyl formate CH$_3$ rocking mode. This blending results in a larger FWHM, but it is still smaller than the FWHM of the pure methyl formate spectra, as shown in Fig.~\ref{fig:MF_1165}. One should be cautious using the derived peak positions, FWHMs, and relative band strength intensities for the mixtures containing H$_2$CO. These values are influenced by and depend on the ratio to which H$_2$CO and HCOOCH$_3$ are deposited, in this case 20:1. The mixtures containing CH$_3$OH show a feature around 1130 cm$^{-1}$ (8.85 $\mu$m) which is the CH$_3$ vibrational rocking mode of CH$_3$OH. Although this feature of CH$_3$OH is close to the CH$_3$ rocking mode of methyl formate, there is no overlap in this study. Any signal associated with the CH$_3$OH mode is omitted from the analysis.

\begin{figure*}[ht!]
    \includegraphics[width=1\hsize]{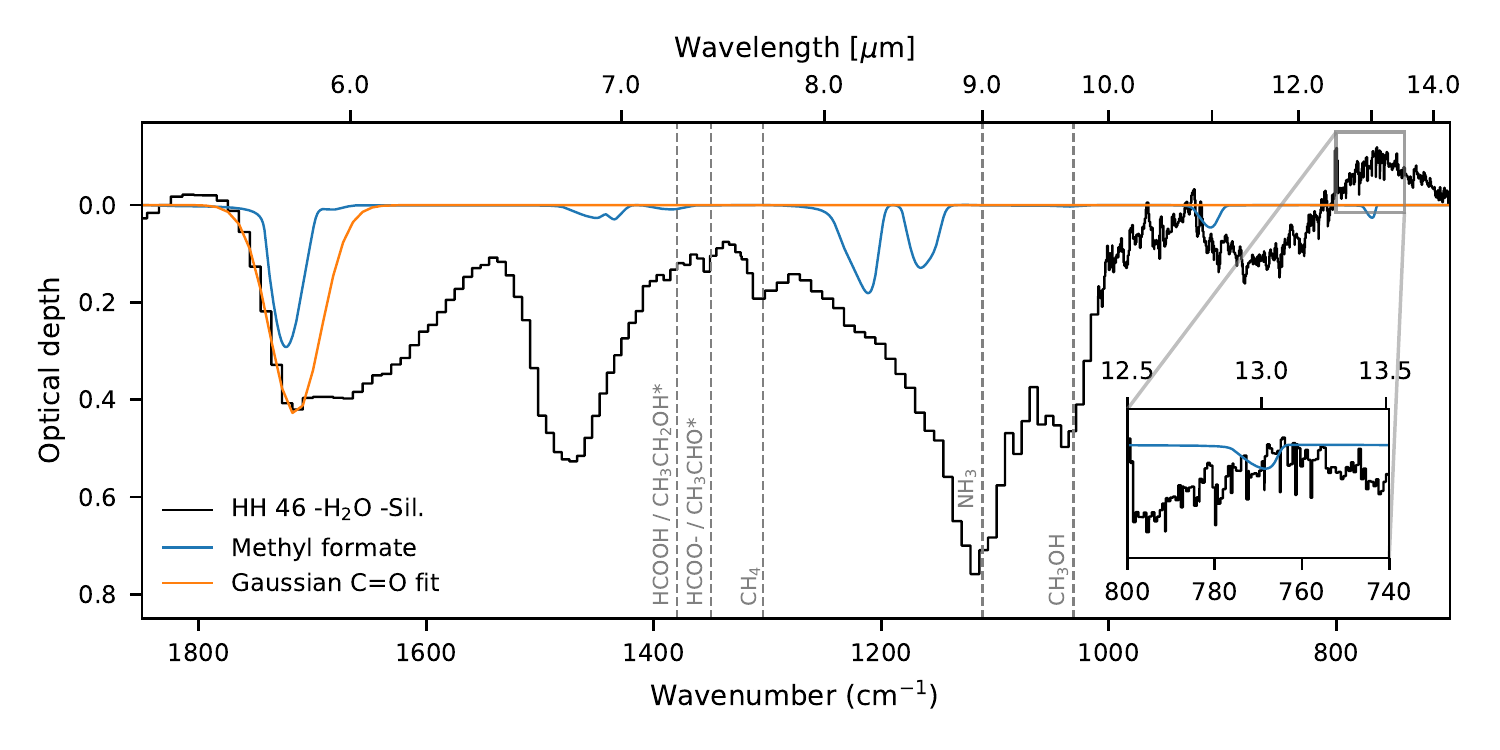}
    \caption{\textit{Spitzer} spectrum of HH~46 with subtraction of the continuum, H$_2$O, and silicates. Additionally, the spectrum of amorphous methyl formate and a Gaussian fit to estimate all molecules containing a C=O bond around 5.8 $\mu$m are shown in blue and orange, respectively. The inset in the bottom right zooms in on the 13 $\mu$m feature where the laboratory spectrum is vertically offset to match the optical depth of the HH~46 spectrum. Vertical dashed lines indicate locations of previously detected species, where the asterisk indicates tentative detections.}
    \label{fig:HH_46_full}
\end{figure*}

\subsection{The C$-$O stretching mode} \label{ssec:C-O_stretch}
The C$-$O vibrational stretching mode is positioned at 1211.3 cm$^{-1}$ (8.256 $\mu$m) in a pure methyl formate ice sample at 15 K. After crystallization the peak is slightly red shifted to 1210.4 cm$^{-1}$ (8.262 $\mu$m), no band splitting is found, and the FWHM decreases. However, there is a significant broad blue wing below the half maximum. The TvS heatmaps of the pure and mixed methyl formate spectra as function of temperature are shown in the Fig.~\ref{fig:MF_1212_heatmaps}. The CH$_2$ rocking mode of H$_2$CO is positioned at approximately 1247 cm$^{-1}$ (8.019 $\mu$m), which introduces a slight overlap with the blue wing of the C$-$O stretching mode. This CH$_2$ rocking mode of H$_2$CO is omitted from analysis by introducing a cut-off at 1234.5 cm$^{-1}$ (8.100 $\mu$m). The spectra, peak positions, and corresponding FWHMs at the characteristic temperatures of the C$-$O stretching mode are given in Fig.~\ref{fig:MF_1212}. Most remarkable are the spectra of methyl formate mixed in H$_2$O where at 15 K the peak position is blue shifted to 1233.7 cm$^{-1}$ (8.106 $\mu$m) and the FWHM is 40 cm$^{-1}$ which is 10 cm$^{-1}$ larger compared to the pure spectra. This feature therefore allows us to tell whether methyl formate resides in a polar or an apolar matrix environment. Notably, at low temperatures the apolar mixture has a FWHM of approximately 50 cm$^{-1}$ due to blending with the CH$_2$ rocking mode of H$_2$CO.

\subsection{The C=O stretching mode} \label{ssec:C=O_stretch}
The C=O stretching mode is the strongest vibrational mode of methyl formate at 1723.1 cm$^{-1}$ (5.804 $\mu$m) in a pure methyl formate ice sample at 15 K, as shown in Fig.~\ref{fig:pure}. After crystallization the main peak shifts to the red to a new position of 1698.0 cm$^{-1}$ (5.889 $\mu$m). A secondary peak is seen at 1709.8 cm$^{-1}$ (5.849 $\mu$m), and a blue-wing plateau is seen. The C=O stretching mode is, however, a common vibrational mode that is also seen in other molecules, such as formaldehyde (H$_2$CO), formic acid (HCOOH), formamide (NH$_2$CHO), acetaldehyde (CH$_3$CHO), and acetone (CH$_3$COCH$_3$). The TvS heatmaps and detailed analysis, i.e., FWHM versus peak position and relative band strength intensities, are shown in Fig.~\ref{fig:MF_1723_heatmaps} and Fig.~\ref{fig:MF_1723}, respectively. The C=O stretching mode of methyl formate mixed in H$_2$O shows a red-wing access and has an approximately 20\% larger relative band strength. This is due to a less precise baseline subtraction of the bending mode of H$_2$O which overlaps with this feature, see Fig.~\ref{fig:pure}.

Since there is an overlap with the C=O stretch of H$_2$CO the matrices containing H$_2$CO are omitted in our analysis. In general a feature observed around 1720 cm$^{-1}$ (5.81 $\mu$m) will likely consist of multiple different organic molecules with a C=O bond. This makes it difficult to identify a specific molecule on this vibrational mode alone. This region is, however, ideal to get an estimate on the total amount of molecules in the solid state containing a C=O functional group in their chemical structure. As such this feature also can act as a diagnostic tool for a family of molecules. This concept is explored in Section~\ref{ssec:CO_func}.

\section{Infrared observations}
The experiments and data analysis performed in this study will be a valuable analytical tool for astronomers using JWST for ice surveys. The spectroscopic characterization of methyl formate, and other COMs in the solid state \citep[see e.g.,][]{2018_TvS_COMs_A&A...611A..35T, 2020_Rachid_acetone_A&A...639A...4R, 2020_Hudson_acetaldehyde_MNRAS.492..283H, 2020_Hudson_DME_AcSpA.23318217H, 2020_Gerakines_OptCon_ApJ...901...52G}, is a prerequisite for their identification in astronomical environments such as dense clouds, embedded protostars, and inclined protoplanetary disks. Previous observations with ISO and \textit{Spitzer} have observed such sources and their spectral resolution and sensitivity allowed for the detection of some smaller solid-state species, including methanol \citep[e.g.,][]{2015_Boogert_icy_universe_ARA&A..53..541B}. A few COMs have been tentatively identified in the massive embedded protostar W33A, namely acetaldehyde, ethanol, and formic acid \citep{1999_Schutte_W33A_A&A...343..966S, 2011_Oberg_Spitzer_ApJ...740..109O}. The reanalysis of the ISO W33A spectrum with laboratory spectra in \citet{2018_TvS_COMs_A&A...611A..35T} shows that it is likely possible to determine with JWST in which astronomically relevant ice matrix these COMs reside.

\begin{figure*}[ht!]
    \includegraphics[width=1\hsize]{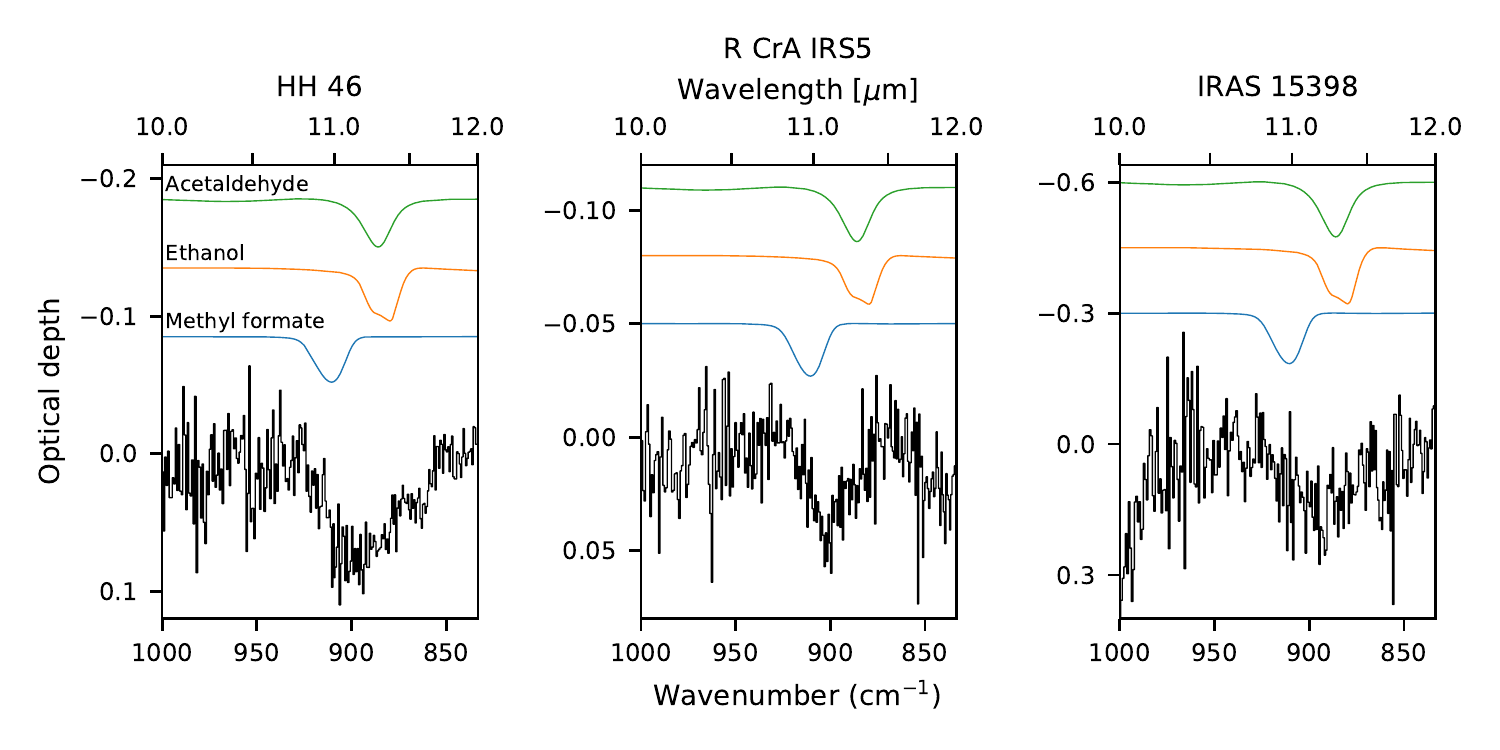}
    \caption{\textit{Spitzer} CASSIS pipeline spectra of R~Cra~IRS5, HH~46, and IRAS 15398 (from left to right) in the 10-14 $\mu$m region. The spectra on top are derived from labortory data for pure acetaldehyde, ethanol, and methyl formate recorded at 15 K.}
    \label{fig:11_um_spectra}
\end{figure*}

The infrared fingerprint region covered here spans the 2000--400 cm$^{-1}$ (5--25 $\mu$m) region. An unambiguous identification of a species in this region from a single vibrational band is challenging, but the presence of multiple vibrational bands considerably helps in the assignment. Methyl formate has five potential candidates for identification in the fingerprint region. These are C=O stretching mode (5.804 $\mu$m), the C$-$O stretching (8.256 $\mu$m), CH$_3$ rocking (8.582 $\mu$m), O$-$CH$_3$ stretching (10.98 $\mu$m), and OCO deformation (13.02 $\mu$m) mode. The C=O stretching mode (5.804 $\mu$m) is the strongest absorption feature, but it overlaps with absorption bands of other molecules containing a C=O in their chemical structure. This makes it less favorable to identify methyl formate ice, but it should be visible when other features of methyl formate are observed. The combination of the O$-$CH$_3$ stretching (10.98 $\mu$m) and OCO deformation (13.02 $\mu$m) modes is most suited for an unambiguous solid-state methyl formate identification. These vibrational modes have the lowest apparent band strengths of the selected transitions but do not have interfering overlap with other species. They will be superimposed on the H$_2$O libration and CH$_3$OH torsion modes, though. These are broad features on which the superimposed methyl formate modes are clearly visible. The C$-$O stretching (8.256 $\mu$m) and CH$_3$ rocking (8.582 $\mu$m) modes have more favorable apparent band strengths, but they overlap with two of the vibrational modes of H$_2$CO. Since formaldehyde is chemically related to the radicals that form methyl formate, it is reasonable to assume these two species are mixed in the ice and thus identification on solely these two modes is uncertain. From this it becomes clear that an identification of methyl formate ice is not straight forward {from one transition. Unambiguous identification of methyl formate requires multiple transitions, even if some are partially overlapping with features of other species. For overlapping species it is important to consider the relative absorbance ratios of the individual transitions, see Appendix~\ref{app:ratios}.}

\subsection{Methyl formate in HH~46} \label{ssec:MF_HH46}
Current available infrared spectra lack the spectral resolution and sensitivity to firmly identify methyl formate ice. However, observations towards some protostars are found to have an additional absorption on the red wing of the 10 $\mu$m silicate feature, between 11--12 $\mu$m. The O$-$CH$_3$ stretching at 10.98 $\mu$m is a potential candidate for this observed absorption feature. HH~46, a low-mass protostar, is one of such protostars and we use its spectrum to derive an upper limit of frozen methyl formate in this source. This upper limit is approximated through a best fit by eye of a laboratory spectrum of methyl formate matched to the HH~46 spectrum observed by \textit{Spitzer} \citep{2008_Boogert_c2d_ApJ...678..985B}. These authors corrected the HH~46 spectrum by subtraction of a standard interstellar silicate spectrum \citep[GCS 3;][]{2004_Kemper_silicates_ApJ...609..826K} and a laboratory spectrum of pure H$_2$O ice at a temperature of 40 K \citep{1993_Hudgins_ices_ApJS...86..713H}. As the spectral quality does not allow us to make a distinction in which matrix the methyl formate resides a pure amorphous spectrum is used. Fig.~\ref{fig:HH_46_full} shows the \textit{Spitzer} spectrum of HH~46 from roughly 5.5 to 14 $\mu$m with the laboratory methyl formate spectrum that is fitted by eye to the observed spectrum of HH~46. Additionally, the spectral locations of some previously detected ice species, e.g., NH$_3$, CH$_4$, and CH$_3$OH, are shown in Fig.~\ref{fig:HH_46_full} by vertically dashed lines.

The C$=$O stretch (5.804 $\mu$m), C$-$O stretch (8.256 $\mu$m), and CH$_3$ rocking (8.582 $\mu$m) modes of methyl formate match with the observation but do not reproduce the full pattern. It should be noted that in the blue wing of the astronomical 9 $\mu$m feature there appears to be two weak bumps at approximately 8.2 and 8.6 $\mu$m that line up with the C$-$O stretch and CH$_3$ rocking modes of methyl formate. From 11--12 $\mu$m there is a broad absorption that seems to be composed of multiple features, of which at least two are positioned at 11.1 and 11.6 $\mu$m. The O$-$CH$_3$ stretching mode at approximately 11 $\mu$m fits the blue wing of the 11.1 $\mu$m feature. The weakest transition, OCO deformation (13.02 $\mu$m), is in this case the important transition that gives a strict upper limit on the methyl formate column density. As shown in the inset of Fig.~\ref{fig:HH_46_full} there is a potential absorption feature around 12.98 $\mu$m. The inset also shows the methyl formate spectrum at that wavelength. The laboratory spectrum is slightly red shifted and wider compared to the absorption feature in HH~46. However, the FWHM versus peak position analysis, see Fig.~\ref{fig:MF_769}, of this mode shows that this feature can be attributed to methyl formate. An upper limit of $1.7\cdot10^{17}$ cm$^{-2}$ is found for methyl formate in HH~46. With respect to water, the upper limit abundances of methyl formate is $\leq 2.2\%$, given a H$_2$O column density of $7.8\cdot10^{18}$ cm$^{-2}$, and $\leq 40\%$ with respect to methanol, given a column density of $4.3\cdot10^{17}$ cm$^{-2}$ \citep{2008_Boogert_c2d_ApJ...678..985B}.

\subsection{The 11 $\mu$m feature} \label{ssec:11um_reg}
Infrared absorption features around 7 $\mu$m towards protostars show potential for COM identification with JWST, e.g., in W33A the absorption features at 7.25 and 7.41 $\mu$m are tentatively identified as ethanol and acetaldehyde, respectively \citep{1999_Schutte_W33A_A&A...343..966S, 2011_Oberg_Spitzer_ApJ...740..109O, 2018_TvS_COMs_A&A...611A..35T}. Another such potential regions is the 11 $\mu$m region, since some protostars have an additional feature on the 10 $\mu$m silicate feature. Here we briefly look at \textit{Spitzer} IRS Short-High spectra taken from the CASSIS database of three protostars, i.e., HH~46, R~CrA~IRS5, and IRAS~15398 \citep{2011_Lebouteiller_CASSIS_ApJS..196....8L, 2015_Lebouteiller_CASSIS2_ApJS..218...21L}. Fig.~\ref{fig:11_um_spectra} shows these normalized absorption features and how they compare to one another. In HH~46 the feature seems to be composed out of at least two components with their respective peaks at approximately 11.1 and 11.6 $\mu$m. For R~CrA~IRS5 only the feature at 11.1 $\mu$m is seen, but with comparable optical depth. The third source is IRAS~15398, which has a broader feature compared to the other two. The optical depth is substantially higher, but the signal-to-noise is worse.
Throughout the literature different possible carriers for features in this region have been proposed. \citet{2000_Bregman_PAH_11um_ApJ...544L..75B} attribute the observed absorption feature at 11.2 $\mu$m in Monoceros R2 IRS 3 to a C--H out-of-plane vibrational mode of Polycyclic Aromatic Hydrocarbon (PAH) molecules. Surveys with both ground- and space-based telescopes showed that across the different environments, i.e., interstellar medium, young stars, and evolved stars, the 11 $\mu$m feature can be attributed to crystalline silicates, specifically forsterite \citep{2005_Kessler_11.3_ApJ...622..404K, 2016_Wright_YSO_11um_MNRAS.457.1593W, 2020_DoDuy_11.1um_MNRAS.493.4463D}. Based on the data obtained in this and previous studies, we propose a third carrier as a possible candidate for the absorption feature between 11--12 $\mu$m, namely COMs. 
On top off the infrared spectra of the three sources, Fig.~\ref{fig:11_um_spectra} shows the laboratory spectra of three COMs: acetaldehyde (CH$_3$CHO), ethanol (CH$_3$CH$_2$OH), and methyl formate (HCOOCH$_3$). The laboratory spectra are offset and scaled with arbitrary factors to show the overlap with the observed absorption features between 11--12 $\mu$m. A combination of these and other COMs with vibrational transitions in this region could, partially or fully, be responsible for the absorption features observed.

\subsection{The C=O functional group} \label{ssec:CO_func}
Different molecules with a common functional group (i.e., a CH$_3$, NH$_2$, or CO bond) will exhibit some similar spectral features in the infrared as the involved vibrational bands are the same. The C=O stretching mode is such a common functional group in COMs containing carbon and oxygen atoms. Although each molecule will have this feature at its distinctive peak position in general they are relatively close to each other and will overlap when an ice is composed of different C=O containing molecules. The apparent band strengths of the different C=O stretching modes can be averaged to make an estimate on the total amount of molecules with a C=O in their chemical structure. In this way a spectral feature is not assigned to a specific molecule but to a specific class of molecules that generally are chemically related as well. This average apparent band strength is approximated by averaging the apparent band strengths of the molecules mentioned in Section~\ref{ssec:C=O_stretch} in a one-to-one ratio \citep{1999_Schutte_W33A_A&A...343..966S, 2010_Modica_MF_cosmic_ion_A&A...519A..22M, 2015_Bouilloud_bnd_str_MNRAS.451.2145B, 2018_Hudson_Acetone_AcSpA.193...33H}. For the HH~46 spectrum we have fitted a Gaussian at 5.831 $\mu$m with a FWHM of 0.0544 $\mu$m, see Fig.~\ref{fig:HH_46_full}. The average apparent band strength of the C=O band is derived to be approximately $3.2\cdot10^{-17}$ cm$^{-1}$ resulting in a column density of $7.6\cdot10^{17}$ cm$^{-2}$ using a Gaussian fit of the integrated band area. The abundance of molecules with C=O in their chemical structure for HH~46 with respect to solid-state water is 9.5\%. The upper limit to the column density of methyl formate in HH~46 from Section~\ref{ssec:MF_HH46}, together with the estimate of molecules containing a C=O functional group, provide an upper limit of $\leq 23\%$ to the contribution of methyl formate to the family of molecules containing the C=O bond.

\section{Conclusions}
In this study we present the characterization of solid-state methyl formate in the infrared. Spectra for pure methyl formate and mixed in astronomical matrix environments are analysed and the TvS heatmaps, peak positions versus FWHM, and relative band intensities are given for five selected methyl formate transitions. We conclude the following:
\begin{enumerate}
    \item Of the five selected transitions, the combination of the O$-$CH$_3$ stretching (10.98 $\mu$m) and OCO deformation (13.02 $\mu$m) modes is most suited for unambiguous detection of methyl formate ice. The other three transitions, C$=$O stretch (5.804 $\mu$m), C$-$O stretch (8.256 $\mu$m), and CH$_3$ rocking (8.582 $\mu$m) potentially overlap with chemically related species. However, these overlapping features are still diagnostic when considering their relative intensities compared to the O$-$CH$_3$ stretching and OCO deformation modes.
    \item The peak position versus FWHM analysis shows that, compared to the pure spectra, the apolar mixtures generally blue shift and decrease the FWHM of the selected transitions. The features in a water matrix are clearly distinguishable from the apolar constituents, e.g., the C$-$O stretching mode in a water matrix is blue shifted by approximately 20 cm$^{-1}$.
    \item Comparing the reference spectra of methyl formate with \textit{Spitzer} observations of HH~46 provides a column density upper limit of $1.7\cdot10^{17}$ cm$^{-2}$. With respect to water the upper limit abundances of methyl formate is $\leq 2.2\%$ and $\leq 40\%$ with respect to methanol.
    \item The observed 11 $\mu$m feature has previously been attributed to polycyclic aromatic hydrocarbons or crystalline silicates. Several COMs show features in this region and may be contributing carriers of this absorption feature.
    \item The total amount of molecules containing a functional C=O group in their chemical structure is estimated to be $7.6\cdot10^{17}$ cm$^{-2}$ which in respect to water equals to 9.5\% for the embedded protostar HH~46.
\end{enumerate}
\begin{acknowledgements}
The authors thank the anonymous referee for the constructive feedback on this manuscript. This research was funded through the Dutch Astrochemistry II program of the Netherlands Organization for Scientific Research (648.000.025) and NOVA, the Netherlands Research School for Astronomy.
\end{acknowledgements}

%
\bibliographystyle{aa} 
\bibliography{JTvS_MF.bib} 

\begin{thebibliography}{70}
\expandafter\ifx\csname natexlab\endcsname\relax\def\natexlab#1{#1}\fi

\bibitem[{{Bacmann} {et~al.}(2012){Bacmann}, {Taquet}, {Faure}, {Kahane}, \&
  {Ceccarelli}}]{2012_Bacmann_L1689B_A&A...541L..12B}
{Bacmann}, A., {Taquet}, V., {Faure}, A., {Kahane}, C., \& {Ceccarelli}, C.
  2012, \aap, 541, L12

\bibitem[{{Balucani} {et~al.}(2015){Balucani}, {Ceccarelli}, \&
  {Taquet}}]{2015_balucani_COMs_gp_MNRAS.449L..16B}
{Balucani}, N., {Ceccarelli}, C., \& {Taquet}, V. 2015, \mnras, 449, L16

\bibitem[{{Belloche} {et~al.}(2013){Belloche}, {M{\"u}ller}, {Menten},
  {Schilke}, \& {Comito}}]{2013_Belloche_SB2_A&A...559A..47B}
{Belloche}, A., {M{\"u}ller}, H.~S.~P., {Menten}, K.~M., {Schilke}, P., \&
  {Comito}, C. 2013, \aap, 559, A47

\bibitem[{{Bennett} \& {Kaiser}(2007)}]{2007_Bennett_GA_MF_ApJ...661..899B}
{Bennett}, C.~J. \& {Kaiser}, R.~I. 2007, \apj, 661, 899

\bibitem[{{Bergner} {et~al.}(2017){Bergner}, {{\"O}berg}, {Garrod}, \&
  {Graninger}}]{2017_Bergner_COM_emb_low_ApJ...841..120B}
{Bergner}, J.~B., {{\"O}berg}, K.~I., {Garrod}, R.~T., \& {Graninger}, D.~M.
  2017, \apj, 841, 120

\bibitem[{{B{\o}gelund} {et~al.}(2019{\natexlab{a}}){B{\o}gelund}, {Barr},
  {Taquet}, {Ligterink}, {Persson}, {Hogerheijde}, \& {van
  Dishoeck}}]{2019_Bogelund_AFGL_A&A...628A...2B}
{B{\o}gelund}, E.~G., {Barr}, A.~G., {Taquet}, V., {et~al.} 2019{\natexlab{a}},
  \aap, 628, A2

\bibitem[{{B{\o}gelund} {et~al.}(2019{\natexlab{b}}){B{\o}gelund}, {McGuire},
  {Hogerheijde}, {van Dishoeck}, \&
  {Ligterink}}]{2019_Bogelund_Nbearing_A&A...624A..82B}
{B{\o}gelund}, E.~G., {McGuire}, B.~A., {Hogerheijde}, M.~R., {van Dishoeck},
  E.~F., \& {Ligterink}, N. F.~W. 2019{\natexlab{b}}, \aap, 624, A82

\bibitem[{{Boogert} {et~al.}(2015){Boogert}, {Gerakines}, \&
  {Whittet}}]{2015_Boogert_icy_universe_ARA&A..53..541B}
{Boogert}, A.~C.~A., {Gerakines}, P.~A., \& {Whittet}, D. C.~B. 2015, \araa,
  53, 541

\bibitem[{{Boogert} {et~al.}(2008){Boogert}, {Pontoppidan}, {Knez}, {Lahuis},
  {Kessler-Silacci}, {van Dishoeck}, {Blake}, {Augereau}, {Bisschop},
  {Bottinelli}, {Brooke}, {Brown}, {Crapsi}, {Evans}, {Fraser}, {Geers},
  {Huard}, {J{\o}rgensen}, {{\"O}berg}, {Allen}, {Harvey}, {Koerner}, {Mundy},
  {Padgett}, {Sargent}, \& {Stapelfeldt}}]{2008_Boogert_c2d_ApJ...678..985B}
{Boogert}, A.~C.~A., {Pontoppidan}, K.~M., {Knez}, C., {et~al.} 2008, \apj,
  678, 985

\bibitem[{{Bossa} {et~al.}(2012){Bossa}, {Isokoski}, {de Valois}, \&
  {Linnartz}}]{2012_Bossa_porous_A&A...545A..82B}
{Bossa}, J.~B., {Isokoski}, K., {de Valois}, M.~S., \& {Linnartz}, H. 2012,
  \aap, 545, A82

\bibitem[{{Bouilloud} {et~al.}(2015){Bouilloud}, {Fray}, {B{\'e}nilan},
  {Cottin}, {Gazeau}, \& {Jolly}}]{2015_Bouilloud_bnd_str_MNRAS.451.2145B}
{Bouilloud}, M., {Fray}, N., {B{\'e}nilan}, Y., {et~al.} 2015, \mnras, 451,
  2145

\bibitem[{{Bregman} {et~al.}(2000){Bregman}, {Hayward}, \&
  {Sloan}}]{2000_Bregman_PAH_11um_ApJ...544L..75B}
{Bregman}, J.~D., {Hayward}, T.~L., \& {Sloan}, G.~C. 2000, \apjl, 544, L75

\bibitem[{{Caselli} {et~al.}(1993){Caselli}, {Hasegawa}, \&
  {Herbst}}]{1993_Caselli_Chem_diff_ApJ...408..548C}
{Caselli}, P., {Hasegawa}, T.~I., \& {Herbst}, E. 1993, \apj, 408, 548

\bibitem[{{Ceccarelli} {et~al.}(2017){Ceccarelli}, {Caselli}, {Fontani},
  {Neri}, {L{\'o}pez-Sepulcre}, {Codella}, {Feng}, {Jim{\'e}nez-Serra},
  {Lefloch}, {Pineda}, {Vastel}, {Alves}, {Bachiller}, {Balucani}, {Bianchi},
  {Bizzocchi}, {Bottinelli}, {Caux}, {Chac{\'o}n-Tanarro}, {Choudhury},
  {Coutens}, {Dulieu}, {Favre}, {Hily-Blant}, {Holdship}, {Kahane}, {Jaber
  Al-Edhari}, {Laas}, {Ospina}, {Oya}, {Podio}, {Pon}, {Punanova}, {Quenard},
  {Rimola}, {Sakai}, {Sims}, {Spezzano}, {Taquet}, {Testi}, {Theul{\'e}},
  {Ugliengo}, {Vasyunin}, {Viti}, {Wiesenfeld}, \&
  {Yamamoto}}]{2017_Ceccarelli_SOLIS_ApJ...850..176C}
{Ceccarelli}, C., {Caselli}, P., {Fontani}, F., {et~al.} 2017, \apj, 850, 176

\bibitem[{{Cernicharo} {et~al.}(2012){Cernicharo}, {Marcelino}, {Roueff},
  {Gerin}, {Jim{\'e}nez-Escobar}, \& {Mu{\~n}oz
  Caro}}]{2012_Cernicharo_B1_ApJ...759L..43C}
{Cernicharo}, J., {Marcelino}, N., {Roueff}, E., {et~al.} 2012, \apjl, 759, L43

\bibitem[{{Charnley} {et~al.}(1995){Charnley}, {Kress}, {Tielens}, \&
  {Millar}}]{1995_Charnley_ISM_alcohols_ApJ...448..232C}
{Charnley}, S.~B., {Kress}, M.~E., {Tielens}, A.~G.~G.~M., \& {Millar}, T.~J.
  1995, \apj, 448, 232

\bibitem[{{Chuang} {et~al.}(2016){Chuang}, {Fedoseev}, {Ioppolo}, {van
  Dishoeck}, \& {Linnartz}}]{2016_KoJu_H-atom_MNRAS.455.1702C}
{Chuang}, K.~J., {Fedoseev}, G., {Ioppolo}, S., {van Dishoeck}, E.~F., \&
  {Linnartz}, H. 2016, \mnras, 455, 1702

\bibitem[{{Dawes} {et~al.}(2007){Dawes}, {Mukerji}, {Davis}, {Holtom}, {Webb},
  {Sivaraman}, {Hoffmann}, {Shaw}, \&
  {Mason}}]{2007_Dawes_morph_JChPh.126x4711D}
{Dawes}, A., {Mukerji}, R.~J., {Davis}, M.~P., {et~al.} 2007, \jcp, 126, 244711

\bibitem[{{Do-Duy} {et~al.}(2020){Do-Duy}, {Wright}, {Fujiyoshi}, {Glasse},
  {Siebenmorgen}, {Smith}, {Stecklum}, \&
  {Sterzik}}]{2020_DoDuy_11.1um_MNRAS.493.4463D}
{Do-Duy}, T., {Wright}, C.~M., {Fujiyoshi}, T., {et~al.} 2020, \mnras, 493,
  4463

\bibitem[{{Fedoseev} {et~al.}(2017){Fedoseev}, {Chuang}, {Ioppolo}, {Qasim},
  {van Dishoeck}, \& {Linnartz}}]{2017_Fedoseev_Glycerol_ApJ...842...52F}
{Fedoseev}, G., {Chuang}, K.~J., {Ioppolo}, S., {et~al.} 2017, \apj, 842, 52

\bibitem[{{Fuchs} {et~al.}(2009){Fuchs}, {Cuppen}, {Ioppolo}, {Romanzin},
  {Bisschop}, {Andersson}, {van Dishoeck}, \&
  {Linnartz}}]{2009_Fuchs_hydro_A&A...505..629F}
{Fuchs}, G.~W., {Cuppen}, H.~M., {Ioppolo}, S., {et~al.} 2009, \aap, 505, 629

\bibitem[{{Garrod}(2013)}]{2013_Garrod_3P_Hcore_ApJ...765...60G}
{Garrod}, R.~T. 2013, \apj, 765, 60

\bibitem[{{Gerakines} \&
  {Hudson}(2020)}]{2020_Gerakines_OptCon_ApJ...901...52G}
{Gerakines}, P.~A. \& {Hudson}, R.~L. 2020, \apj, 901, 52

\bibitem[{{He} {et~al.}(2017){He}, {Emtiaz}, \&
  {Vidali}}]{2017_Jiao_CO2_ASW_ApJ...837...65H}
{He}, J., {Emtiaz}, S.~M., \& {Vidali}, G. 2017, \apj, 837, 65

\bibitem[{{Hidaka} {et~al.}(2004){Hidaka}, {Watanabe}, {Shiraki}, {Nagaoka}, \&
  {Kouchi}}]{2004_Hidaka_H2CO_ApJ...614.1124H}
{Hidaka}, H., {Watanabe}, N., {Shiraki}, T., {Nagaoka}, A., \& {Kouchi}, A.
  2004, \apj, 614, 1124

\bibitem[{{Hiraoka} {et~al.}(1994){Hiraoka}, {Ohashi}, {Kihara}, {Yamamoto},
  {Sato}, \& {Yamashita}}]{1994_Hiraoka_hydro_CPL...229..408H}
{Hiraoka}, K., {Ohashi}, N., {Kihara}, Y., {et~al.} 1994, Chemical Physics
  Letters, 229, 408

\bibitem[{{Hiraoka} {et~al.}(2002){Hiraoka}, {Sato}, {Sato}, {Sogoshi},
  {Yokoyama}, {Takashima}, \& {Kitagawa}}]{2002_Hiraoka_hydro_ApJ...577..265H}
{Hiraoka}, K., {Sato}, T., {Sato}, S., {et~al.} 2002, \apj, 577, 265

\bibitem[{{Horn} {et~al.}(2004){Horn}, {M{\o}llendal}, {Sekiguchi}, {Uggerud},
  {Roberts}, {Herbst}, {Viggiano}, \&
  {Fridgen}}]{2004_Horn_MF_gas-phase_ApJ...611..605H}
{Horn}, A., {M{\o}llendal}, H., {Sekiguchi}, O., {et~al.} 2004, \apj, 611, 605

\bibitem[{{Hudgins} {et~al.}(1993){Hudgins}, {Sandford}, {Allamandola}, \&
  {Tielens}}]{1993_Hudgins_ices_ApJS...86..713H}
{Hudgins}, D.~M., {Sandford}, S.~A., {Allamandola}, L.~J., \& {Tielens},
  A.~G.~G.~M. 1993, \apjs, 86, 713

\bibitem[{{Hudson} \&
  {Ferrante}(2020)}]{2020_Hudson_acetaldehyde_MNRAS.492..283H}
{Hudson}, R.~L. \& {Ferrante}, R.~F. 2020, \mnras, 492, 283

\bibitem[{{Hudson} {et~al.}(2018){Hudson}, {Gerakines}, \&
  {Ferrante}}]{2018_Hudson_Acetone_AcSpA.193...33H}
{Hudson}, R.~L., {Gerakines}, P.~A., \& {Ferrante}, R.~F. 2018, Spectrochimica
  Acta Part A: Molecular Spectroscopy, 193, 33

\bibitem[{{Hudson} {et~al.}(2020){Hudson}, {Yarnall}, \&
  {Coleman}}]{2020_Hudson_DME_AcSpA.23318217H}
{Hudson}, R.~L., {Yarnall}, Y.~Y., \& {Coleman}, F.~M. 2020, Spectrochimica
  Acta Part A: Molecular Spectroscopy, 233, 118217

\bibitem[{{Isokoski} {et~al.}(2014){Isokoski}, {Bossa}, {Triemstra}, \&
  {Linnartz}}]{2014_Isokoski_porosity_PCCP...16.3456I}
{Isokoski}, K., {Bossa}, J.~B., {Triemstra}, T., \& {Linnartz}, H. 2014,
  Physical Chemistry Chemical Physics (Incorporating Faraday Transactions), 16,
  3456

\bibitem[{{Jin} \& {Garrod}(2020)}]{2020_Jin_COM_chemmod_ApJS..249...26J}
{Jin}, M. \& {Garrod}, R.~T. 2020, \apjs, 249, 26

\bibitem[{{J{\o}rgensen} {et~al.}(2016){J{\o}rgensen}, {van der Wiel},
  {Coutens}, {Lykke}, {M{\"u}ller}, {van Dishoeck}, {Calcutt}, {Bjerkeli},
  {Bourke}, {Drozdovskaya}, {Favre}, {Fayolle}, {Garrod}, {Jacobsen},
  {{\"O}berg}, {Persson}, \& {Wampfler}}]{2016_Jorgensen_PILS_A&A...595A.117J}
{J{\o}rgensen}, J.~K., {van der Wiel}, M.~H.~D., {Coutens}, A., {et~al.} 2016,
  \aap, 595, A117

\bibitem[{{Kemper} {et~al.}(2004){Kemper}, {Vriend}, \&
  {Tielens}}]{2004_Kemper_silicates_ApJ...609..826K}
{Kemper}, F., {Vriend}, W.~J., \& {Tielens}, A.~G.~G.~M. 2004, \apj, 609, 826

\bibitem[{{Kessler-Silacci} {et~al.}(2005){Kessler-Silacci}, {Hillenbrand},
  {Blake}, \& {Meyer}}]{2005_Kessler_11.3_ApJ...622..404K}
{Kessler-Silacci}, J.~E., {Hillenbrand}, L.~A., {Blake}, G.~A., \& {Meyer},
  M.~R. 2005, \apj, 622, 404

\bibitem[{{Lebouteiller} {et~al.}(2015){Lebouteiller}, {Barry}, {Goes},
  {Sloan}, {Spoon}, {Weedman}, {Bernard-Salas}, \&
  {Houck}}]{2015_Lebouteiller_CASSIS2_ApJS..218...21L}
{Lebouteiller}, V., {Barry}, D.~J., {Goes}, C., {et~al.} 2015, \apjs, 218, 21

\bibitem[{{Lebouteiller} {et~al.}(2011){Lebouteiller}, {Barry}, {Spoon},
  {Bernard-Salas}, {Sloan}, {Houck}, \&
  {Weedman}}]{2011_Lebouteiller_CASSIS_ApJS..196....8L}
{Lebouteiller}, V., {Barry}, D.~J., {Spoon}, H.~W.~W., {et~al.} 2011, \apjs,
  196, 8

\bibitem[{{Ligterink} {et~al.}(2017){Ligterink}, {Coutens}, {Kofman},
  {M{\"u}ller}, {Garrod}, {Calcutt}, {Wampfler}, {J{\o}rgensen}, {Linnartz}, \&
  {van Dishoeck}}]{2017_Ligterink_CH3NCO_MNRAS.469.2219L}
{Ligterink}, N.~F.~W., {Coutens}, A., {Kofman}, V., {et~al.} 2017, \mnras, 469,
  2219

\bibitem[{{Ligterink} {et~al.}(2018){Ligterink}, {Terwisscha van Scheltinga},
  {Taquet}, {J{\o}rgensen}, {Cazaux}, {van Dishoeck}, \&
  {Linnartz}}]{2018_Ligterink_peptide_MNRAS.480.3628L}
{Ligterink}, N.~F.~W., {Terwisscha van Scheltinga}, J., {Taquet}, V., {et~al.}
  2018, \mnras, 480, 3628

\bibitem[{{Manigand} {et~al.}(2020){Manigand}, {J{\o}rgensen}, {Calcutt},
  {M{\"u}ller}, {Ligterink}, {Coutens}, {Drozdovskaya}, {van Dishoeck}, \&
  {Wampfler}}]{2020_Manigand_16293A_A&A...635A..48M}
{Manigand}, S., {J{\o}rgensen}, J.~K., {Calcutt}, H., {et~al.} 2020, \aap, 635,
  A48

\bibitem[{{McGuire}(2018)}]{2018_McGuire_det_ApJS..239...17M}
{McGuire}, B.~A. 2018, \apjs, 239, 17

\bibitem[{{McGuire} {et~al.}(2017){McGuire}, {Shingledecker}, {Willis},
  {Burkhardt}, {El-Abd}, {Motiyenko}, {Brogan}, {Hunter}, {Margul{\`e}s},
  {Guillemin}, {Garrod}, {Herbst}, \&
  {Remijan}}]{2017_McGuire_MOxyMeOH_ApJ...851L..46M}
{McGuire}, B.~A., {Shingledecker}, C.~N., {Willis}, E.~R., {et~al.} 2017,
  \apjl, 851, L46

\bibitem[{{Millar} {et~al.}(1991){Millar}, {Herbst}, \&
  {Charnley}}]{1991_Millar_O-COMs_Orion_ApJ...369..147M}
{Millar}, T.~J., {Herbst}, E., \& {Charnley}, S.~B. 1991, \apj, 369, 147

\bibitem[{{Modica} \&
  {Palumbo}(2010)}]{2010_Modica_MF_cosmic_ion_A&A...519A..22M}
{Modica}, P. \& {Palumbo}, M.~E. 2010, \aap, 519, A22

\bibitem[{{Modica} {et~al.}(2012){Modica}, {Palumbo}, \&
  {Strazzulla}}]{2012_Modica_MF_comets_P&SS...73..425M}
{Modica}, P., {Palumbo}, M.~E., \& {Strazzulla}, G. 2012, \planss, 73, 425

\bibitem[{{Mu{\~n}oz Caro} {et~al.}(2014){Mu{\~n}oz Caro}, {Dartois}, {Boduch},
  {Rothard}, {Domaracka}, \&
  {Jim{\'e}nez-Escobar}}]{2014_MunozCaro_UV_ion_A&A...566A..93M}
{Mu{\~n}oz Caro}, G.~M., {Dartois}, E., {Boduch}, P., {et~al.} 2014, \aap, 566,
  A93

\bibitem[{{Nazari} {et~al.}(2021){Nazari}, {van Gelder}, {van Dishoeck},
  {Tabone}, {van 't Hoff}, {Ligterink}, {Beuther}, {Boogert}, {Garatti},
  {Klaassen}, {Linnartz}, {Taquet}, \&
  {Tychoniec}}]{2021_Nazari_COMs_arXiv210403326N}
{Nazari}, P., {van Gelder}, M.~L., {van Dishoeck}, E.~F., {et~al.} 2021, arXiv
  e-prints, arXiv:2104.03326

\bibitem[{{{\"O}berg} {et~al.}(2011){{\"O}berg}, {Boogert}, {Pontoppidan}, {van
  den Broek}, {van Dishoeck}, {Bottinelli}, {Blake}, \&
  {Evans}}]{2011_Oberg_Spitzer_ApJ...740..109O}
{{\"O}berg}, K.~I., {Boogert}, A.~C.~A., {Pontoppidan}, K.~M., {et~al.} 2011,
  \apj, 740, 109

\bibitem[{{{\"O}berg} {et~al.}(2010){{\"O}berg}, {Bottinelli}, {J{\o}rgensen},
  \& {van Dishoeck}}]{2010_Oberg_B1-b_ApJ...716..825O}
{{\"O}berg}, K.~I., {Bottinelli}, S., {J{\o}rgensen}, J.~K., \& {van Dishoeck},
  E.~F. 2010, \apj, 716, 825

\bibitem[{{{\"O}berg} {et~al.}(2009{\natexlab{a}}){{\"O}berg}, {Fayolle},
  {Cuppen}, {van Dishoeck}, \& {Linnartz}}]{2009_Oberg_segre_A&A...505..183O}
{{\"O}berg}, K.~I., {Fayolle}, E.~C., {Cuppen}, H.~M., {van Dishoeck}, E.~F.,
  \& {Linnartz}, H. 2009{\natexlab{a}}, \aap, 505, 183

\bibitem[{{{\"O}berg} {et~al.}(2007){{\"O}berg}, {Fraser}, {Boogert},
  {Bisschop}, {Fuchs}, {van Dishoeck}, \&
  {Linnartz}}]{2007_Oberg_H2O_CO2_A&A...462.1187O}
{{\"O}berg}, K.~I., {Fraser}, H.~J., {Boogert}, A.~C.~A., {et~al.} 2007, \aap,
  462, 1187

\bibitem[{{{\"O}berg} {et~al.}(2009{\natexlab{b}}){{\"O}berg}, {Garrod}, {van
  Dishoeck}, \& {Linnartz}}]{2009_Oberg_COMs_A&A...504..891O}
{{\"O}berg}, K.~I., {Garrod}, R.~T., {van Dishoeck}, E.~F., \& {Linnartz}, H.
  2009{\natexlab{b}}, \aap, 504, 891

\bibitem[{{{\"O}berg} {et~al.}(2015){{\"O}berg}, {Guzm{\'a}n}, {Furuya}, {Qi},
  {Aikawa}, {Andrews}, {Loomis}, \&
  {Wilner}}]{2015_Oberg_CH3CN_Natur.520..198O}
{{\"O}berg}, K.~I., {Guzm{\'a}n}, V.~V., {Furuya}, K., {et~al.} 2015, \nat,
  520, 198

\bibitem[{{Palumbo}(2005)}]{2005_Palumbo_h2o_morph_JPhCS...6..211P}
{Palumbo}, M.~E. 2005, in Journal of Physics Conference Series, Vol.~6, Journal
  of Physics Conference Series, 211--216

\bibitem[{{Qu{\'e}nard} {et~al.}(2018){Qu{\'e}nard}, {Jim{\'e}nez-Serra},
  {Viti}, {Holdship}, \& {Coutens}}]{2018_Quenard_COMs_peptide_MNRAS.474.2796Q}
{Qu{\'e}nard}, D., {Jim{\'e}nez-Serra}, I., {Viti}, S., {Holdship}, J., \&
  {Coutens}, A. 2018, \mnras, 474, 2796

\bibitem[{{Rachid} {et~al.}(2020){Rachid}, {Terwisscha van Scheltinga},
  {Koletzki}, \& {Linnartz}}]{2020_Rachid_acetone_A&A...639A...4R}
{Rachid}, M.~G., {Terwisscha van Scheltinga}, J., {Koletzki}, D., \&
  {Linnartz}, H. 2020, \aap, 639, A4

\bibitem[{{Ruaud} \& {Gorti}(2019)}]{2019_Ruaud_cmod_ppd_pJ...885..146R}
{Ruaud}, M. \& {Gorti}, U. 2019, \apj, 885, 146

\bibitem[{{Schutte} {et~al.}(1999){Schutte}, {Boogert}, {Tielens}, {Whittet},
  {Gerakines}, {Chiar}, {Ehrenfreund}, {Greenberg}, {van Dishoeck}, \& {de
  Graauw}}]{1999_Schutte_W33A_A&A...343..966S}
{Schutte}, W.~A., {Boogert}, A.~C.~A., {Tielens}, A.~G.~G.~M., {et~al.} 1999,
  \aap, 343, 966

\bibitem[{{Simons} {et~al.}(2020){Simons}, {Lamberts}, \&
  {Cuppen}}]{2020_Simons_CO+H_mod_A&A...634A..52S}
{Simons}, M.~A.~J., {Lamberts}, T., \& {Cuppen}, H.~M. 2020, \aap, 634, A52

\bibitem[{{Taquet} {et~al.}(2016){Taquet}, {Wirstr{\"o}m}, \&
  {Charnley}}]{2016_Taquet_COM_outbursts_ApJ...821...46T}
{Taquet}, V., {Wirstr{\"o}m}, E.~S., \& {Charnley}, S.~B. 2016, \apj, 821, 46

\bibitem[{{Terwisscha van Scheltinga} {et~al.}(2018){Terwisscha van
  Scheltinga}, {Ligterink}, {Boogert}, {van Dishoeck}, \&
  {Linnartz}}]{2018_TvS_COMs_A&A...611A..35T}
{Terwisscha van Scheltinga}, J., {Ligterink}, N.~F.~W., {Boogert}, A.~C.~A.,
  {van Dishoeck}, E.~F., \& {Linnartz}, H. 2018, \aap, 611, A35

\bibitem[{{van Gelder} {et~al.}(2020){van Gelder}, {Tabone}, {Tychoniec}, {van
  Dishoeck}, {Beuther}, {Boogert}, {Caratti o Garatti}, {Klaassen}, {Linnartz},
  {M{\"u}ller}, \& {Taquet}}]{2020_Gelder_O-COMs_A&A...639A..87V}
{van Gelder}, M.~L., {Tabone}, B., {Tychoniec}, {\L}., {et~al.} 2020, \aap,
  639, A87

\bibitem[{{Vasyunin} {et~al.}(2017){Vasyunin}, {Caselli}, {Dulieu}, \&
  {Jim{\'e}nez-Serra}}]{2017_Vasyunin_cmod_COM_prestellar_ApJ...842...33V}
{Vasyunin}, A.~I., {Caselli}, P., {Dulieu}, F., \& {Jim{\'e}nez-Serra}, I.
  2017, \apj, 842, 33

\bibitem[{{Wang} {et~al.}(2021){Wang}, {Du}, {Semenov}, {Wang}, \&
  {Li}}]{2021_Wang_COM_model_SGB2_arXiv210310376W}
{Wang}, Y., {Du}, F., {Semenov}, D., {Wang}, H., \& {Li}, J. 2021, arXiv
  e-prints, arXiv:2103.10376

\bibitem[{{Watanabe} \& {Kouchi}(2002)}]{2002_Watanabe_hydro_ApJ...571L.173W}
{Watanabe}, N. \& {Kouchi}, A. 2002, \apjl, 571, L173

\bibitem[{{Watanabe} {et~al.}(2004){Watanabe}, {Nagaoka}, {Shiraki}, \&
  {Kouchi}}]{2004_Watanabe_hydro_ApJ...616..638W}
{Watanabe}, N., {Nagaoka}, A., {Shiraki}, T., \& {Kouchi}, A. 2004, \apj, 616,
  638

\bibitem[{{Wright} {et~al.}(2016){Wright}, {Do Duy}, \&
  {Lawson}}]{2016_Wright_YSO_11um_MNRAS.457.1593W}
{Wright}, C.~M., {Do Duy}, T., \& {Lawson}, W. 2016, \mnras, 457, 1593

\bibitem[{{Yang} {et~al.}(2021){Yang}, {Sakai}, {Zhang}, {Murillo}, {Zhang},
  {Higuchi}, {Zeng}, {L{\'o}pez-Sepulcre}, {Yamamoto}, {Lefloch}, {Bouvier},
  {Ceccarelli}, {Hirota}, {Imai}, {Oya}, {Sakai}, \&
  {Watanabe}}]{2021_Yang_Peaches_ApJ...910...20Y}
{Yang}, Y.-L., {Sakai}, N., {Zhang}, Y., {et~al.} 2021, \apj, 910, 20

\end{thebibliography}
%

\begin{appendix}

\section{Integrated absorbance ratios of the selected methyl formate transitions} \label{app:ratios}
\begin{table*}[ht]
    \centering
    \caption{The integrated absorbance ratios of methyl formate for each of the selected transitions and temperatures, for both pure and mixed with astronomically relevant matrices, with respect to the C$-$O stretching mode (1211.3 cm$^{-1}$ / 8.256 $\mu$m).}
    \begin{tabular}{l c c c c c c}
        \hline
        \hline
        \multirow{2}{*}{Mixture} & Temperature &  OCO deform. & O$-$CH$_{3}$ stretch. & CH$_3$ rock. & C$-$O stretch. & C=O stretch.\\ 
        & (K) & (13.02 $\mu$m) & (10.98 $\mu$m) & (8.582 $\mu$m) & (8.256 $\mu$m) & (5.804 $\mu$m) \\ 
        \hline
        HCOOCH$_3$              & \multirow{6}{*}{15}   & 0.038 & 	0.150 & 	0.602 & 	1.000 & 	1.659\\
        HCOOCH$_3$ : CO         &                       & 0.032 & 	0.141 & 	0.567 & 	1.000 & 	1.458\\
        HCOOCH$_3$ : H$_2$CO    &                       & 0.038 & 	0.178 & 	0.983 & 	1.000 & 	$-$**\\
        HCOOCH$_3$ : CH$_3$OH   &                       & 0.030 & 	0.123 & 	0.693 & 	1.000 & 	1.934\\
        HCOOCH$_3$ : H$_2$O     &                       & 0.023 & 	0.075 & 	0.226 & 	1.000 & 	2.291\\
        HCOOCH$_3$ : Apolar     &                       & 0.046 & 	$-$* & 	0.723 & 	1.000 & 	$-$**\\
        \hline
        HCOOCH$_3$              & \multirow{6}{*}{30}   & 0.038 & 	0.150 & 	0.598 & 	1.000 & 	1.669\\
        HCOOCH$_3$ : CO         &                       & 0.035 & 	0.140 & 	0.566 & 	1.000 & 	1.451\\
        HCOOCH$_3$ : H$_2$CO    &                       & 0.038 & 	0.180 & 	0.983 & 	1.000 & 	$-$**\\
        HCOOCH$_3$ : CH$_3$OH   &                       & 0.029 & 	0.131 & 	0.714 & 	1.000 & 	1.972\\
        HCOOCH$_3$ : H$_2$O     &                       & 0.022 & 	0.076 & 	0.224 & 	1.000 & 	2.253\\
        HCOOCH$_3$ : Apolar     &                       & 0.043 & 	$-$* & 	0.727 & 	1.000 & 	$-$**\\
        \hline
        HCOOCH$_3$              & \multirow{6}{*}{50}   & 0.038 & 	0.151 & 	0.592 & 	1.000 & 	1.683\\
        HCOOCH$_3$ : CO         &                       & 0.029 & 	0.142 & 	0.467 & 	1.000 & 	1.396\\
        HCOOCH$_3$ : H$_2$CO    &                       & 0.037 & 	0.183 & 	0.979 & 	1.000 & 	$-$**\\
        HCOOCH$_3$ : CH$_3$OH   &                       & 0.027 & 	0.143 & 	0.759 & 	1.000 & 	2.055\\
        HCOOCH$_3$ : H$_2$O     &                       & 0.023 & 	0.069 & 	0.221 & 	1.000 & 	2.181\\
        HCOOCH$_3$ : Apolar     &                       & 0.039 & 	$-$* & 	0.759 & 	1.000 & 	$-$**\\
        \hline
        HCOOCH$_3$              & \multirow{6}{*}{80}   & 0.039 & 	0.154 & 	0.595 & 	1.000 & 	1.718\\
        HCOOCH$_3$ : CO         &                       & 0.028 & 	0.143 & 	0.461 & 	1.000 & 	1.416\\
        HCOOCH$_3$ : H$_2$CO    &                       & 0.037 & 	0.189 & 	0.965 & 	1.000 & 	$-$**\\
        HCOOCH$_3$ : CH$_3$OH   &                       & 0.029 & 	0.153 & 	0.796 & 	1.000 & 	2.145\\
        HCOOCH$_3$ : H$_2$O     &                       & 0.022 & 	0.066 & 	0.239 & 	1.000 & 	2.075\\
        HCOOCH$_3$ : Apolar     &                       & 0.035 & 	$-$* & 	0.769 & 	1.000 & 	$-$**\\
        \hline
        HCOOCH$_3$              & \multirow{6}{*}{100}  & 0.088 & 	0.362 & 	0.436 & 	1.000 & 	1.967\\
        HCOOCH$_3$ : CO         &                       & 0.032 & 	0.156 & 	0.569 & 	1.000 & 	1.591\\
        HCOOCH$_3$ : H$_2$CO    &                       & 0.036 & 	0.237 & 	0.718 & 	1.000 & 	$-$**\\
        HCOOCH$_3$ : CH$_3$OH   &                       & 0.026 & 	0.156 & 	0.797 & 	1.000 & 	2.188\\
        HCOOCH$_3$ : H$_2$O     &                       & 0.021 & 	0.062 & 	0.268 & 	1.000 & 	2.035\\
        HCOOCH$_3$ : Apolar     &                       & 0.034 & 	$-$* & 	0.696 & 	1.000 & 	$-$**\\
        \hline
        HCOOCH$_3$              & \multirow{6}{*}{120}  & 0.089 & 	0.366 & 	0.439 & 	1.000 & 	1.992\\
        HCOOCH$_3$ : CO         &                       & 0.026 & 	0.122 & 	0.389 & 	1.000 & 	1.625\\
        HCOOCH$_3$ : H$_2$CO    &                       & 0.015 & 	0.218 & 	0.378 & 	1.000 & 	$-$**\\
        HCOOCH$_3$ : CH$_3$OH   &                       & 0.111 & 	0.496 & 	1.597 & 	1.000 & 	4.295\\
        HCOOCH$_3$ : H$_2$O     &                       & 0.020 & 	0.071 & 	0.320 & 	1.000 & 	2.007\\
        HCOOCH$_3$ : Apolar     &                       & 0.026 & 	$-$* & 	0.578 & 	1.000 & 	$-$**\\
        \hline
        \noalign{\smallskip}
    \end{tabular}
    \tablefoot{* Can not be determined due to H$_2$CO and CH$_3$OH matrix interaction, see Section~\ref{ssec:O-CH3_stretch}. ** Due to the overlapping C=O stretching modes of H$_2$CO and HCOOCH$_3$ the relative intensity can not be determined, see Section~\ref{ssec:C=O_stretch}.}
    \label{tab.int_area}
\end{table*}

\section{The OCO deformation mode}
\begin{figure*}[ht!]
    \begin{tabular}{cc}
        \vspace{0.5cm}
        \includegraphics[width=0.475\hsize]{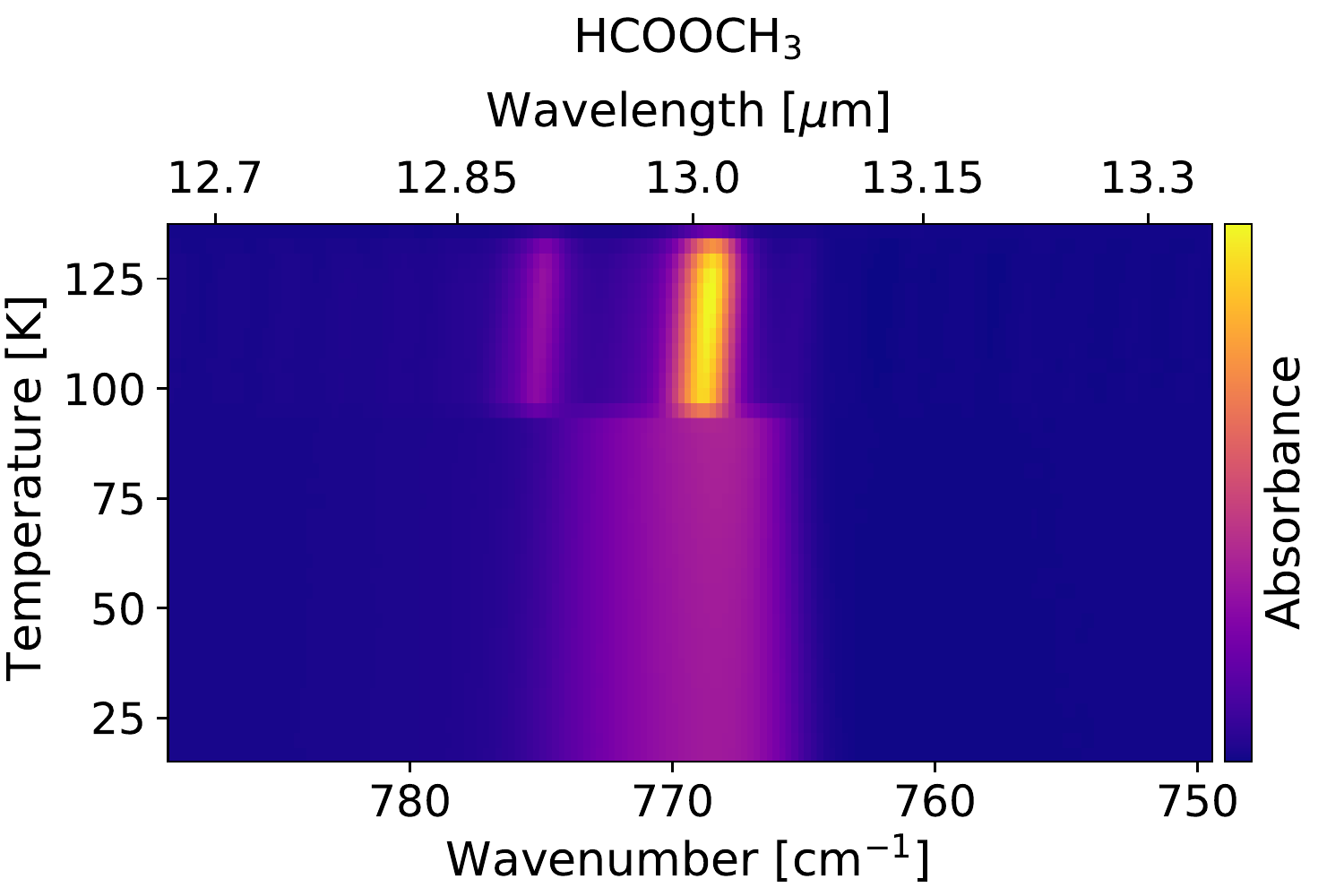} & \includegraphics[width=0.475\hsize]{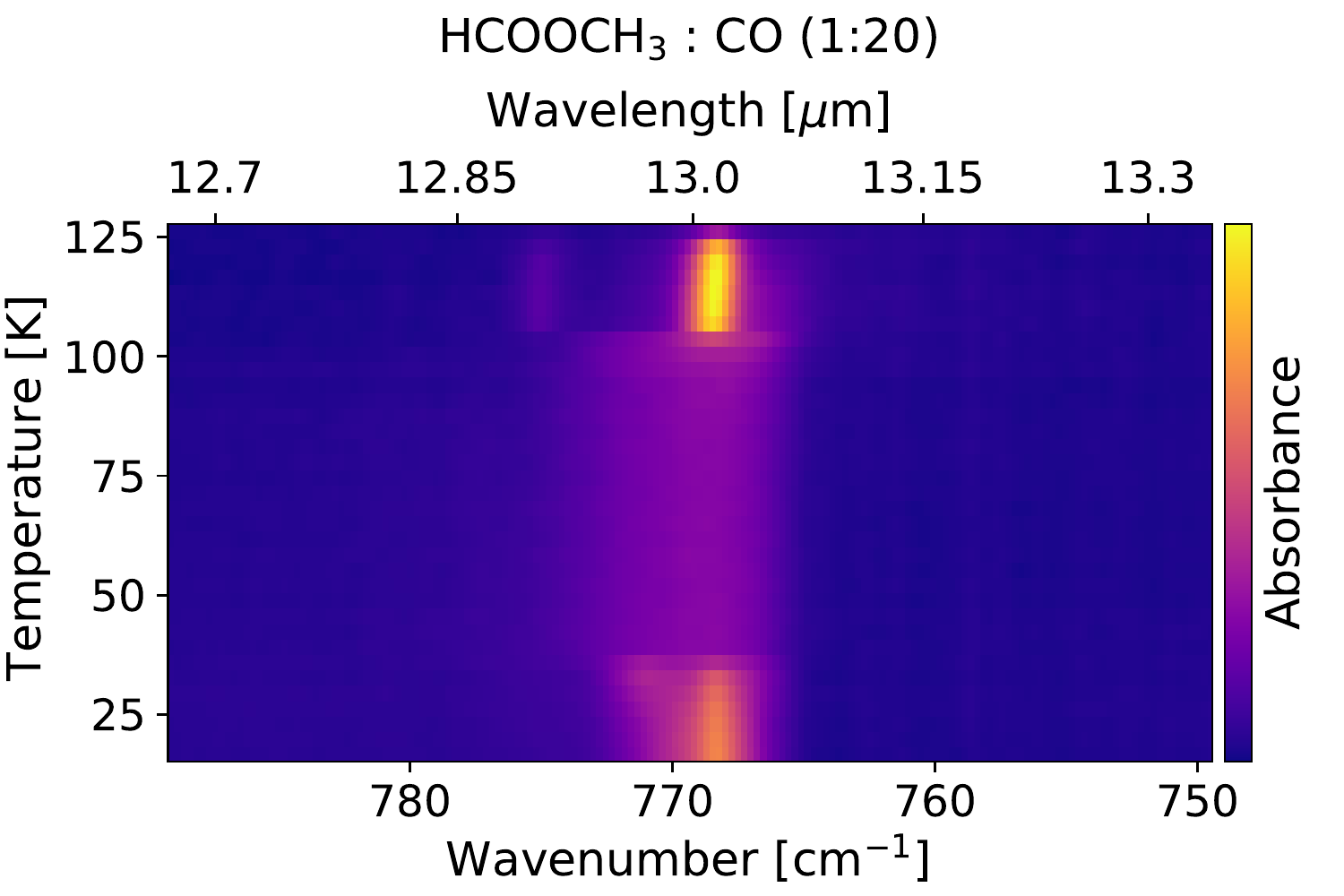} \\
        \vspace{0.5cm}
        \includegraphics[width=0.475\hsize]{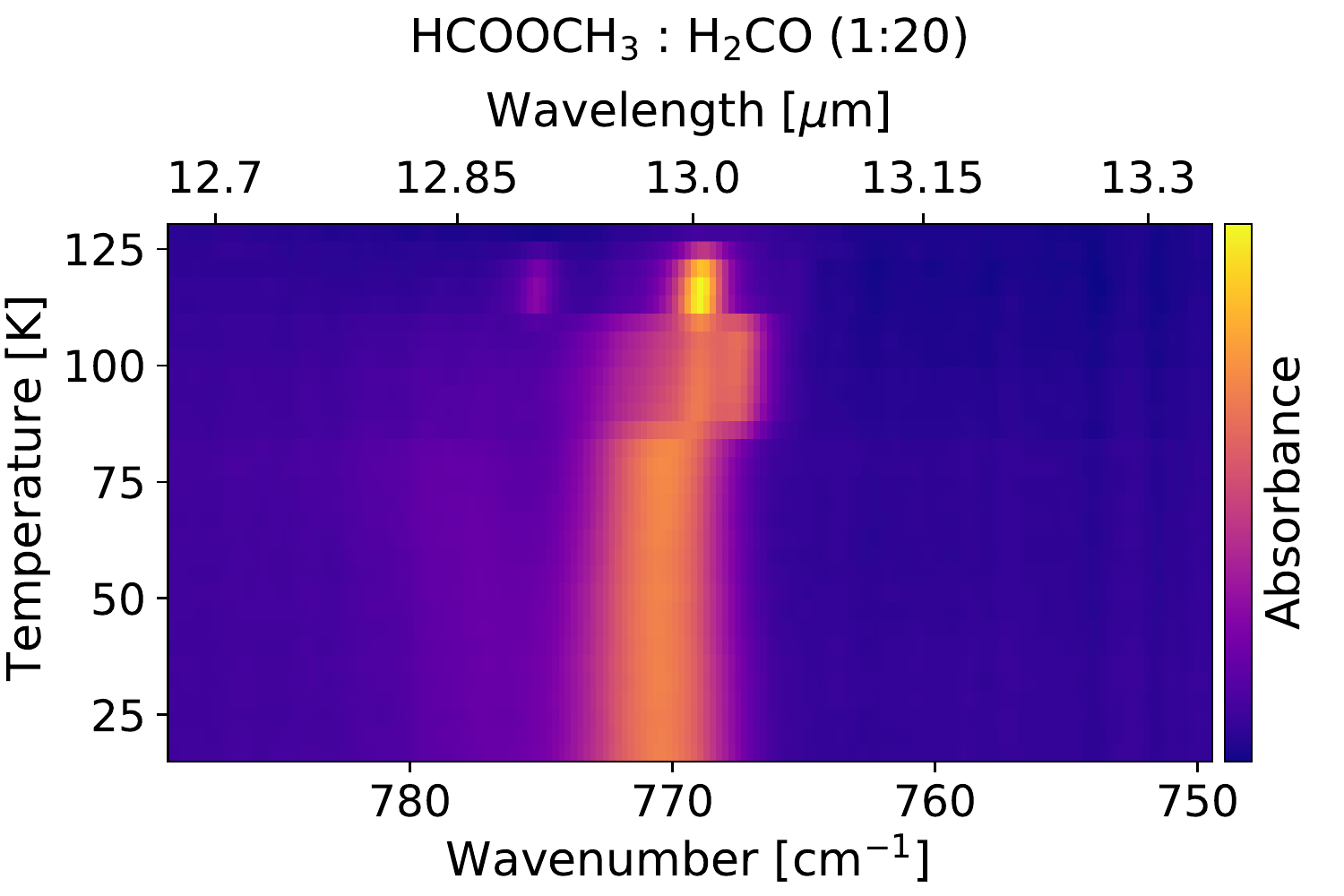} & \includegraphics[width=0.475\hsize]{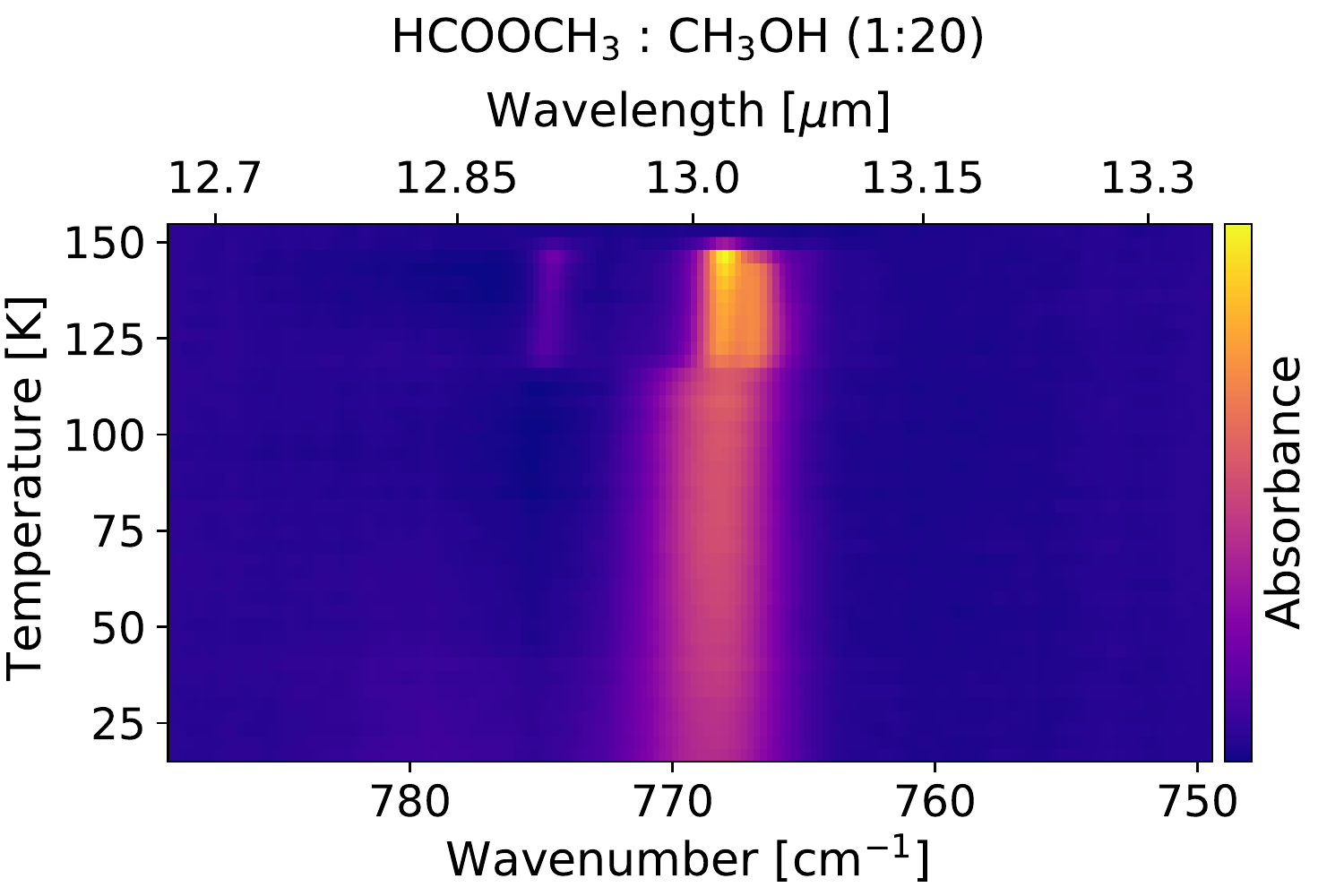} \\
        \includegraphics[width=0.475\hsize]{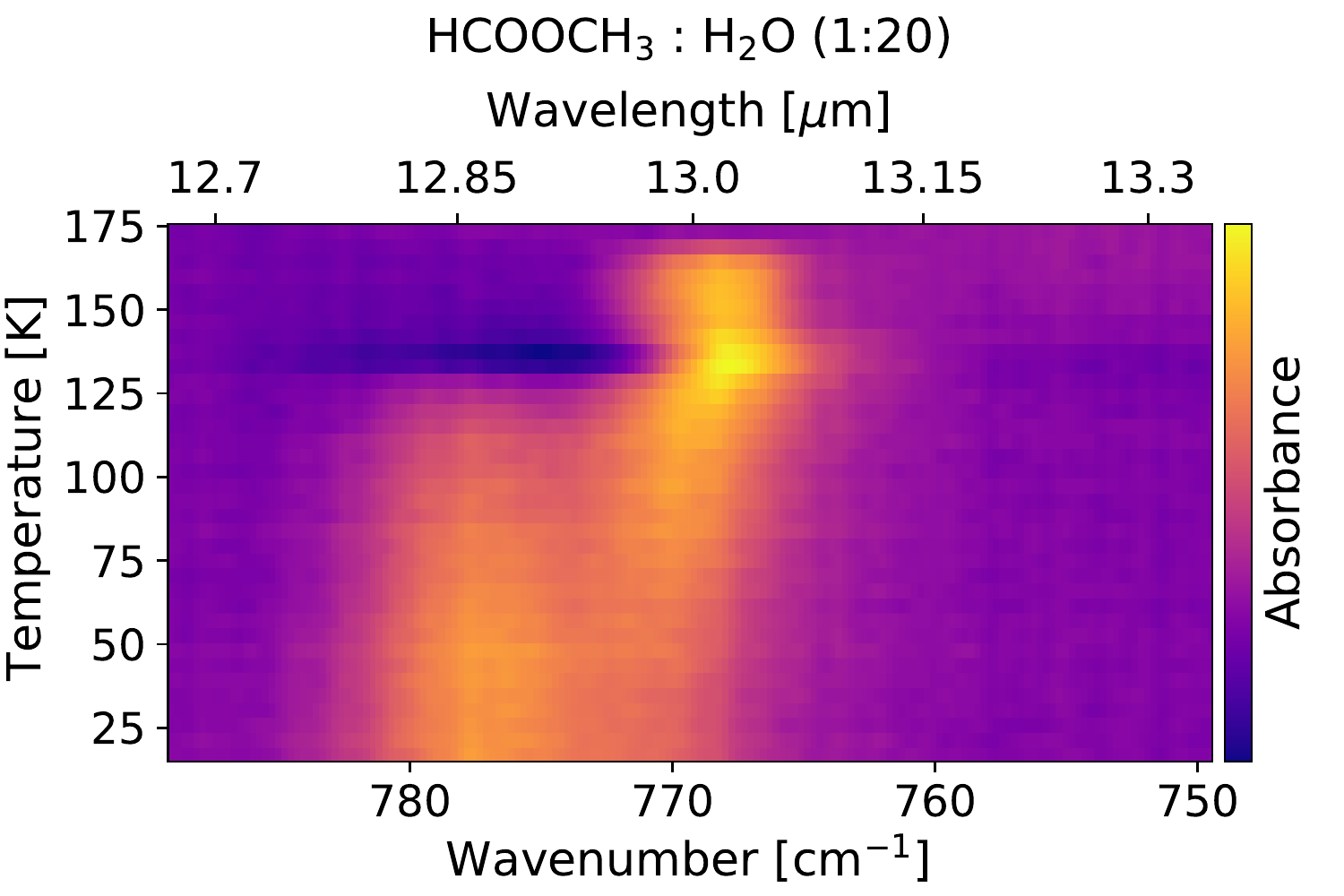} & \includegraphics[width=0.475\hsize]{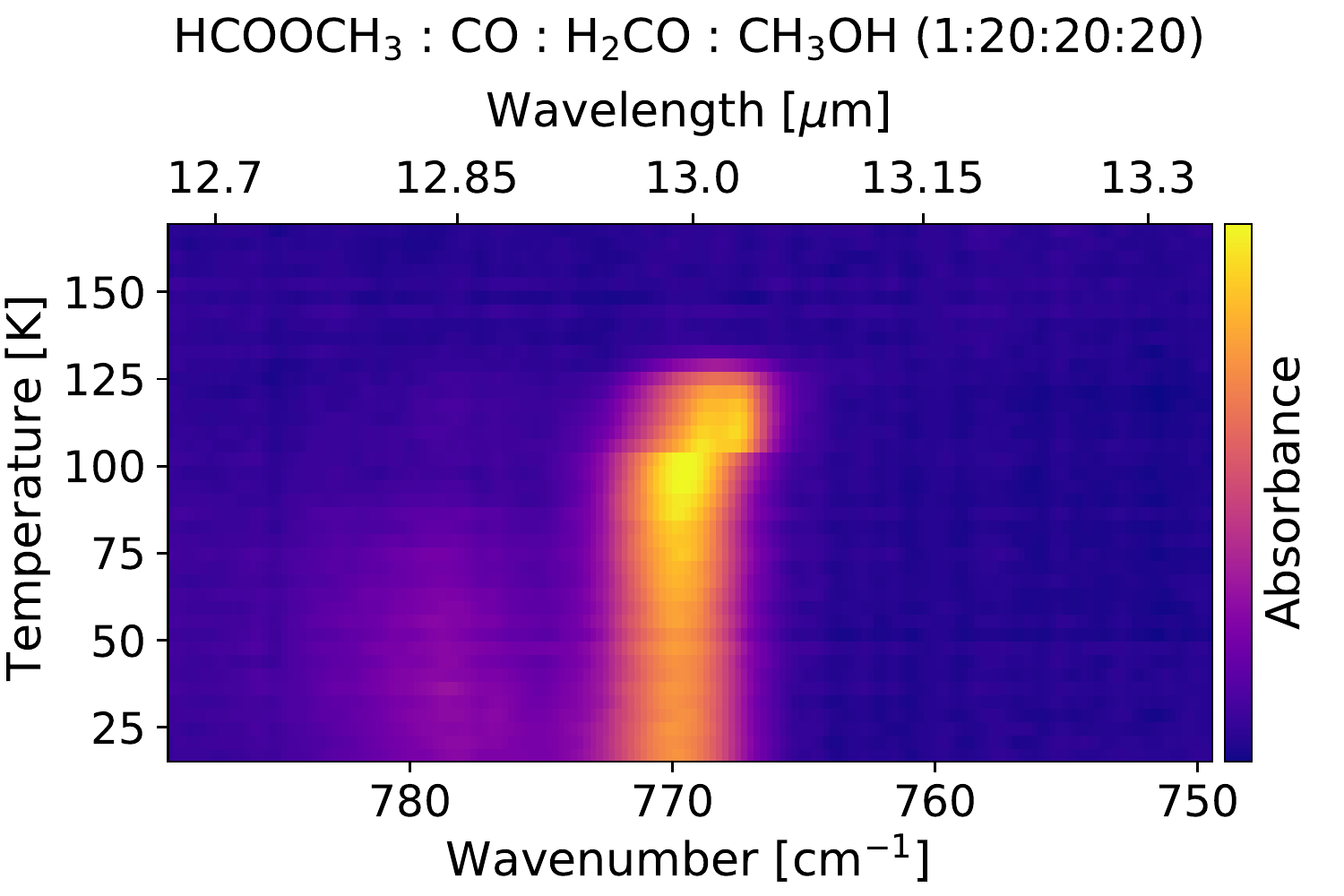} \\
    \end{tabular}
\caption{Temperature versus Spectra (TvS) heatmaps of the OCO deformation mode (768.3 cm$^{-1}$ / 13.02 $\mu$m) of HCOOCH$_3$ in the astronomically relevant ice matrices as a function of temperature.}
\label{fig:MF_769_heatmaps}
\end{figure*}

\begin{figure*}[ht!]
    \includegraphics[width=0.975\hsize]{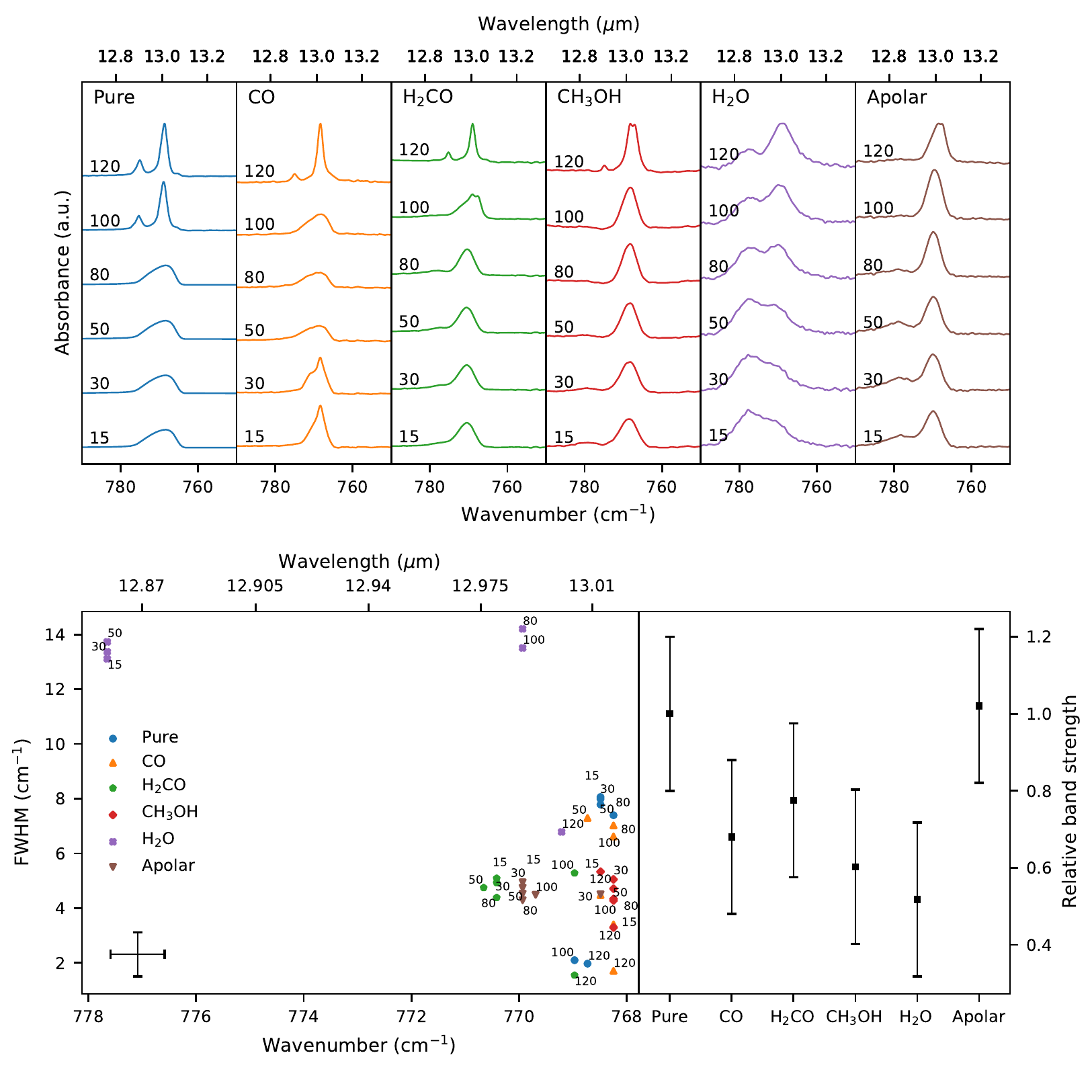}
    \caption{Top panel: Absorption feature of the OCO deformation vibrational mode (768.3 cm$^{-1}$ / 13.02 $\mu$m) for each of the mixtures at selected temperatures. Bottom left panel: FWHM as a function of peak position at the selected temperatures. The error bar of each individual point is given in the bottom left. Bottom right panel: The relative band strengths for methyl formate in the different ice matrices at 15 K with respect to the pure band strength.}
    \label{fig:MF_769}
\end{figure*}

\begin{table*}
    \centering
    \caption{Peak position and FWHM of the methyl formate OCO deformation stretching mode (768.3 cm$^{-1}$ / 13.02 $\mu$m) in various matrices.}
    \begin{tabular}{l c l l l l}
        \hline
        \hline
        \multirow{2}{*}{Mixture} & Temperature &  \multicolumn{2}{c}{$\lambda_{\rm{peak}}$} &  \multicolumn{2}{c}{FWHM} \\ 
        & (K) & (cm$^{-1}$) & ($\mu$m) & (cm$^{-1}$)  &  ($\mu$m) \\ 
        \hline
        HCOOCH$_3$              & \multirow{6}{*}{15}   & 768.48 & 13.012 & \phantom{1}8.061 & 0.1361\\
        HCOOCH$_3$ : CO         &                       & 768.25 & 13.017 & \phantom{1}3.404 & 0.0576\\
        HCOOCH$_3$ : H$_2$CO    &                       & 770.42 & 12.980 & \phantom{1}5.089 & 0.0857\\
        HCOOCH$_3$ : CH$_3$OH   &                       & 768.49 & 13.013 & \phantom{1}5.329 & 0.0902\\
        HCOOCH$_3$ : H$_2$O     &                       & 777.65 & 12.859 & 13.11* & 0.2184*\\
        HCOOCH$_3$ : Apolar     &                       & 769.94 & 12.988 & \phantom{1}4.941 & 0.0833\\
        \hline
        HCOOCH$_3$              & \multirow{6}{*}{30}   & 768.49 & 13.012 & \phantom{1}7.987 & 0.1349\\
        HCOOCH$_3$ : CO         &                       & 768.49 & 13.012 & \phantom{1}4.479 & 0.0757\\
        HCOOCH$_3$ : H$_2$CO    &                       & 770.42 & 12.980 & \phantom{1}4.927 & 0.0829\\
        HCOOCH$_3$ : CH$_3$OH   &                       & 768.25 & 13.017 & \phantom{1}5.050 & 0.0855\\
        HCOOCH$_3$ : H$_2$O     &                       & 777.65 & 12.859 & 13.37* & 0.2228*\\
        HCOOCH$_3$ : Apolar     &                       & 769.94 & 12.988 & \phantom{1}4.743 & 0.0800\\
        \hline
        HCOOCH$_3$              & \multirow{6}{*}{50}   & 768.49 & 13.012 & \phantom{1}7.785 & 0.1314\\
        HCOOCH$_3$ : CO         &                       & 768.73 & 13.008 & \phantom{1}7.296 & 0.1231\\
        HCOOCH$_3$ : H$_2$CO    &                       & 770.66 & 12.976 & \phantom{1}4.747 & 0.0799\\
        HCOOCH$_3$ : CH$_3$OH   &                       & 768.25 & 13.017 & \phantom{1}4.709 & 0.0797\\
        HCOOCH$_3$ : H$_2$O     &                       & 777.65 & 12.859 & 13.73* & 0.2289*\\
        HCOOCH$_3$ : Apolar     &                       & 769.94 & 12.988 & \phantom{1}4.523 & 0.0763\\
        \hline
        HCOOCH$_3$              & \multirow{6}{*}{80}   & 768.25 & 13.017 & \phantom{1}7.396 & 0.1249\\
        HCOOCH$_3$ : CO         &                       & 768.25 & 13.017 & \phantom{1}7.025 & 0.1186\\
        HCOOCH$_3$ : H$_2$CO    &                       & 770.42 & 12.980 & \phantom{1}4.385 & 0.0739\\
        HCOOCH$_3$ : CH$_3$OH   &                       & 768.25 & 13.017 & \phantom{1}4.336 & 0.0734\\
        HCOOCH$_3$ : H$_2$O     &                       & 769.94 & 12.988 & 14.21* & 0.2374*\\
        HCOOCH$_3$ : Apolar     &                       & 769.94 & 12.988 & \phantom{1}4.290 & 0.0724\\
        \hline
        HCOOCH$_3$              & \multirow{7}{*}{100}  & 768.97 & 13.004 & \phantom{1}2.095 & 0.0354\\
                                &                       & 775.24 & 12.899 & $-$ & $-$ \\
        HCOOCH$_3$ : CO         &                       & 768.25 & 13.017 & \phantom{1}6.623 & 0.1119\\
        HCOOCH$_3$ : H$_2$CO    &                       & 768.97 & 13.004 & \phantom{1}5.281 & 0.0892\\
        HCOOCH$_3$ : CH$_3$OH   &                       & 768.25 & 13.017 & \phantom{1}4.279 & 0.0725\\
        HCOOCH$_3$ : H$_2$O     &                       & 769.94 & 12.988 & 13.51* & 0.2261*\\
        HCOOCH$_3$ : Apolar     &                       & 769.69 & 12.992 & \phantom{1}4.484 & 0.0757\\
        \hline
        HCOOCH$_3$              & \multirow{10}{*}{120}  & 768.73 & 13.008 & 1.971 & 0.0333\\
                                &                       & 775.00 & 13.008 & $-$ & $-$ \\
        HCOOCH$_3$ : CO         &                       & 768.45 & 13.017 & 1.710 & 0.0287\\
                                &                       & 775.24 & 12.899 & $-$ & $-$ \\
        HCOOCH$_3$ : H$_2$CO    &                       & 768.97 & 13.004 & 1.546 & 0.0262\\
                                &                       & 775.24 & 12.899 & $-$ & $-$ \\
        HCOOCH$_3$ : CH$_3$OH   &                       & 768.25 & 13.017 & 3.289 & 0.0558\\
                                &                       & 775.24 & 12.899 & $-$ & $-$ \\
        HCOOCH$_3$ : H$_2$O     &                       & 769.21 & 13.000 & 6.784 & 0.1147\\
        HCOOCH$_3$ : Apolar     &                       & 768.49 & 13.013 & 4.505 & 0.0762\\
        \hline
        \noalign{\smallskip}
    \end{tabular}
    \tablefoot{* FWHM result of two or more blended peaks. ** Due to sever blending with the matrix constituent no FWHM determination.}
    \label{tab.MF_OCO_def}
\end{table*}

\section{The O$-$CH3 stretching mode}
\begin{table*}
    \centering
    \caption{Peak position and FWHM of the methyl formate O$-$CH$_{3}$ stretching mode (910.5 cm$^{-1}$ / 10.98 $\mu$m) in various matrices.}
    \begin{tabular}{l c l l l l}
        \hline
        \hline
        \multirow{2}{*}{Mixture} & Temperature &  \multicolumn{2}{c}{$\lambda_{\rm{peak}}$} &  \multicolumn{2}{c}{FWHM} \\ 
        & (K) & (cm$^{-1}$) & ($\mu$m) & (cm$^{-1}$)  &  ($\mu$m) \\ 
        \hline
        HCOOCH$_3$              & \multirow{5}{*}{15}   & 910.47 & 10.983 & 17.51 & 0.2108\\
        HCOOCH$_3$ : CO         &                       & 914.33 & 10.937 & \phantom{1}9.909 & 0.1184\\
        HCOOCH$_3$ : H$_2$CO    &                       & 911.20 & 10.975 & 13.10 & 0.1577\\
        HCOOCH$_3$ : CH$_3$OH   &                       & 912.88 & 10.954 & 14.28 & 0.1711\\
        HCOOCH$_3$ : H$_2$O     &                       & 913.61 & 10.946 & 17.72 & 0.2131\\
        \hline
        HCOOCH$_3$              & \multirow{5}{*}{30}   & 910.23 & 10.986 & 17.20 & 0.2071\\
        HCOOCH$_3$ : CO         &                       & 914.33 & 10.937 & 11.48 & 0.1371\\
        HCOOCH$_3$ : H$_2$CO    &                       & 911.20 & 10.975 & 12.83 & 0.1544\\
        HCOOCH$_3$ : CH$_3$OH   &                       & 912.88 & 10.954 & 13.67 & 0.1638\\
        HCOOCH$_3$ : H$_2$O     &                       & 912.88 & 10.954 & 17.34 & 0.2085\\
        \hline
        HCOOCH$_3$              & \multirow{5}{*}{50}   & 909.99 & 10.989 & 16.60 & 0.2000\\
        HCOOCH$_3$ : CO         &                       & 912.16 & 10.963 & 16.19 & 0.1945\\
        HCOOCH$_3$ : H$_2$CO    &                       & 911.20 & 10.975 & 12.36 & 0.1485\\
        HCOOCH$_3$ : CH$_3$OH   &                       & 913.12 & 10.951 & 12.92 & 0.1548\\
        HCOOCH$_3$ : H$_2$O     &                       & 913.85 & 10.943 & 16.51 & 0.1982\\
        \hline
        HCOOCH$_3$              & \multirow{5}{*}{80}   & 909.27 & 10.998 & 15.72 & 0.1896\\
        HCOOCH$_3$ : CO         &                       & 910.47 & 10.983 & 15.63 & 0.1882\\
        HCOOCH$_3$ : H$_2$CO    &                       & 911.20 & 10.975 & 11.90 & 0.1434\\
        HCOOCH$_3$ : CH$_3$OH   &                       & 913.12 & 10.951 & 11.13 & 0.1335\\
        HCOOCH$_3$ : H$_2$O     &                       & 912.88 & 10.954 & 16.06 & 0.1927\\
        \hline
        HCOOCH$_3$              & \multirow{6}{*}{100}  & 899.63 & 11.116 & \phantom{1}3.374 & 0.0417\\
                                &                       & 904.93 & 11.051 & $-$ & $-$ \\
        HCOOCH$_3$ : CO         &                       & 909.51 & 10.995 & 15.21 & 0.1836\\
        HCOOCH$_3$ : H$_2$CO    &                       & 907.58 & 11.018 & \phantom{1}7.982 & 0.0966\\
        HCOOCH$_3$ : CH$_3$OH   &                       & 912.88 & 10.954 & 11.32 & 0.1357\\
        HCOOCH$_3$ : H$_2$O     &                       & 914.09 & 10.940 & 15.60 & 0.1871\\
        \hline
        HCOOCH$_3$              & \multirow{9}{*}{120}  & 899.87 & 11.113 & \phantom{1}3.422 & 0.0422\\
                                &                       & 905.17 & 11.048 & $-$ & $-$ \\
        HCOOCH$_3$ : CO         &                       & 899.38 & 11.119 & \phantom{1}2.295 & 0.0284\\
                                &                       & 905.17 & 11.048 & $-$ & $-$ \\
        HCOOCH$_3$ : H$_2$CO    &                       & 901.07 & 11.098 & \phantom{1}2.418 & 0.0298\\
                                &                       & 905.41 & 11.045 & $-$ & $-$ \\
        HCOOCH$_3$ : CH$_3$OH   &                       & 904.69 & 11.054 & \phantom{1}8.354* & 0.1025*\\
                                &                       & 899.63 & 11.116 & $-$ & $-$ \\
        HCOOCH$_3$ : H$_2$O     &                       & 915.05 & 10.928 & 15.87 & 0.1901\\
        \hline
        \noalign{\smallskip}
    \end{tabular}
    \tablefoot{* FWHM result of two or more blended peaks. ** Due to sever blending with the matrix constituent no FWHM determination.}
    \label{tab.MF_O-CH3_str}
\end{table*}

\section{The CH$_3$ rocking mode}
\begin{figure*}[ht!]
    \begin{tabular}{cc}
        \vspace{0.5cm}
        \includegraphics[width=0.475\hsize]{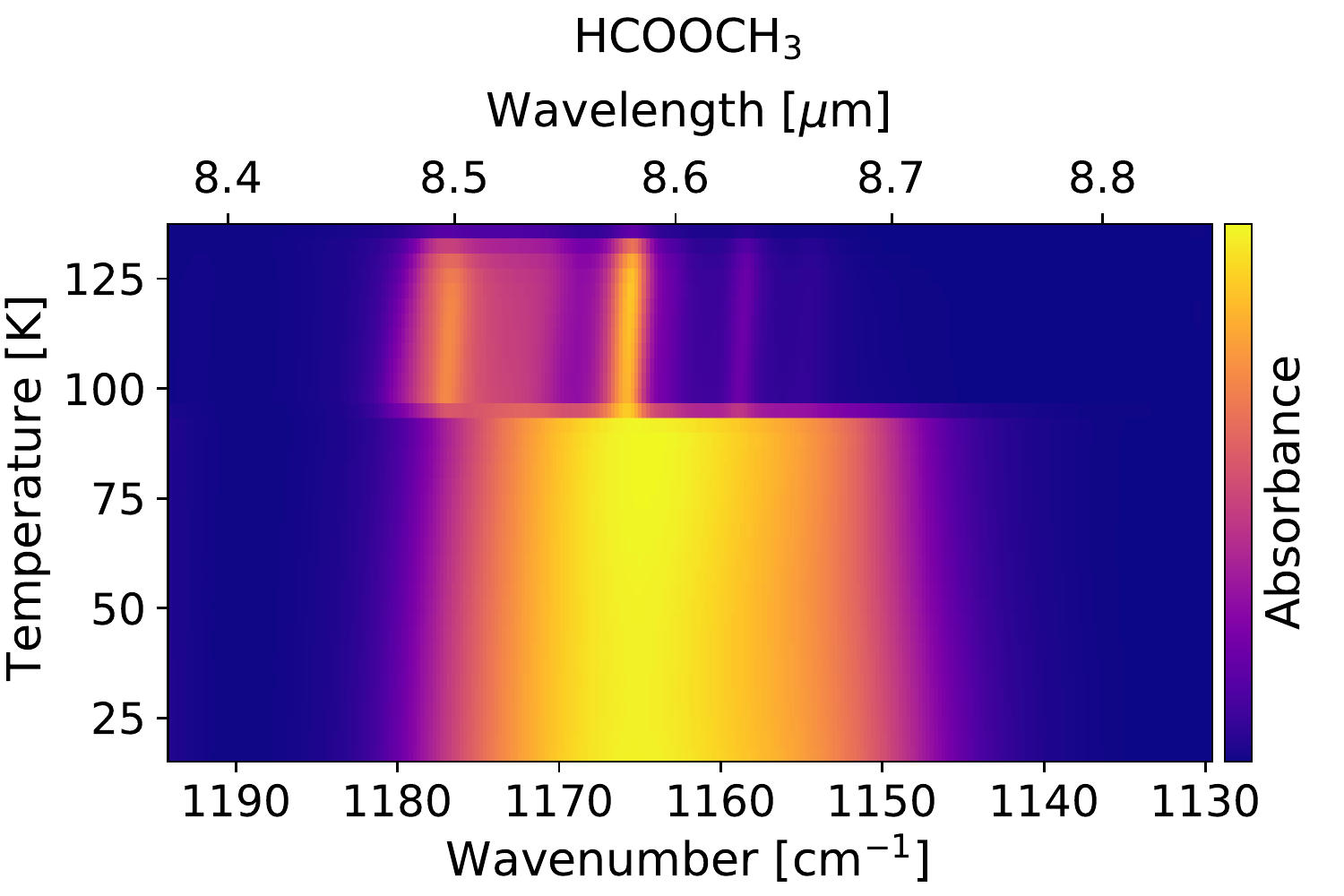} & \includegraphics[width=0.475\hsize]{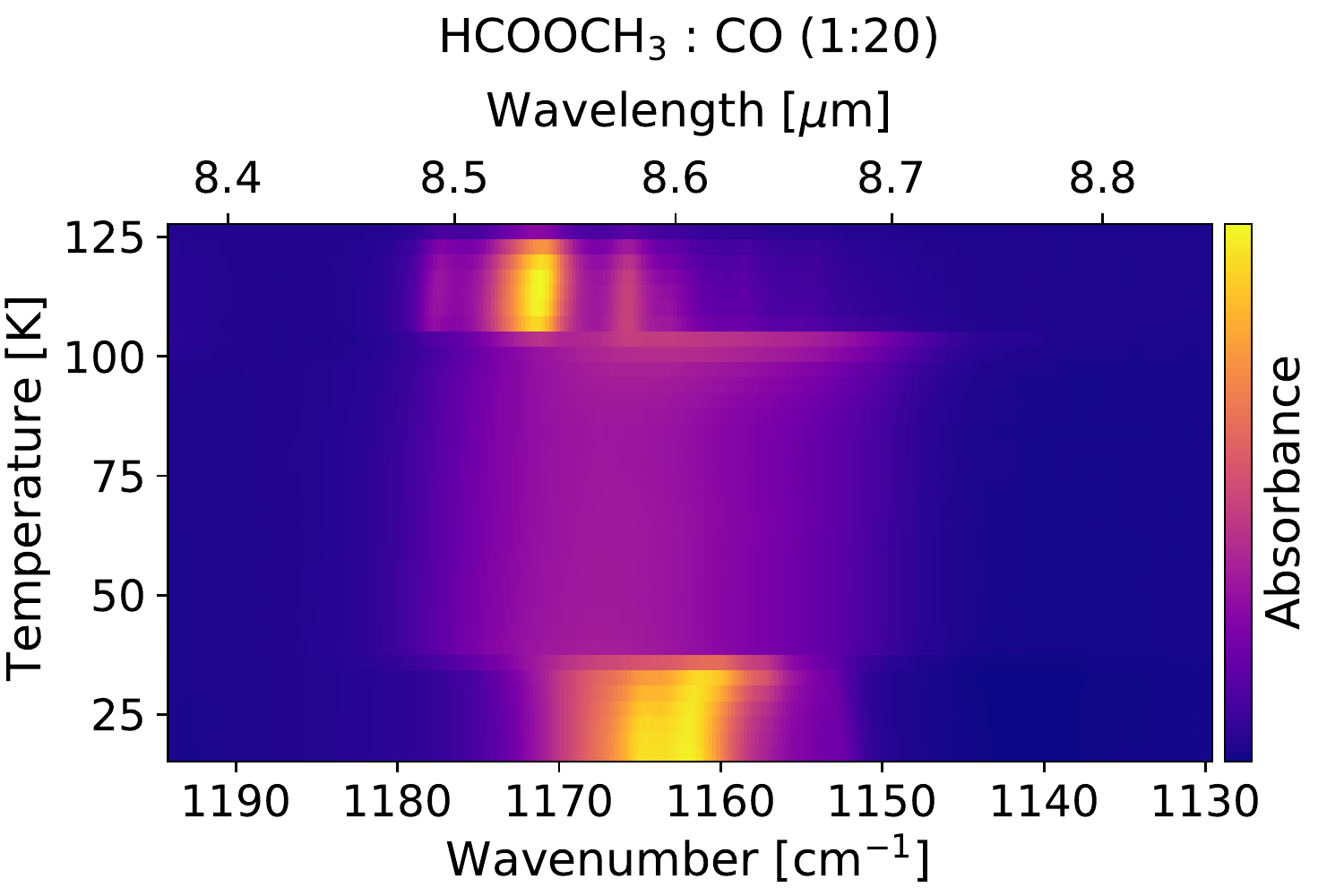} \\
        \vspace{0.5cm}
        \includegraphics[width=0.475\hsize]{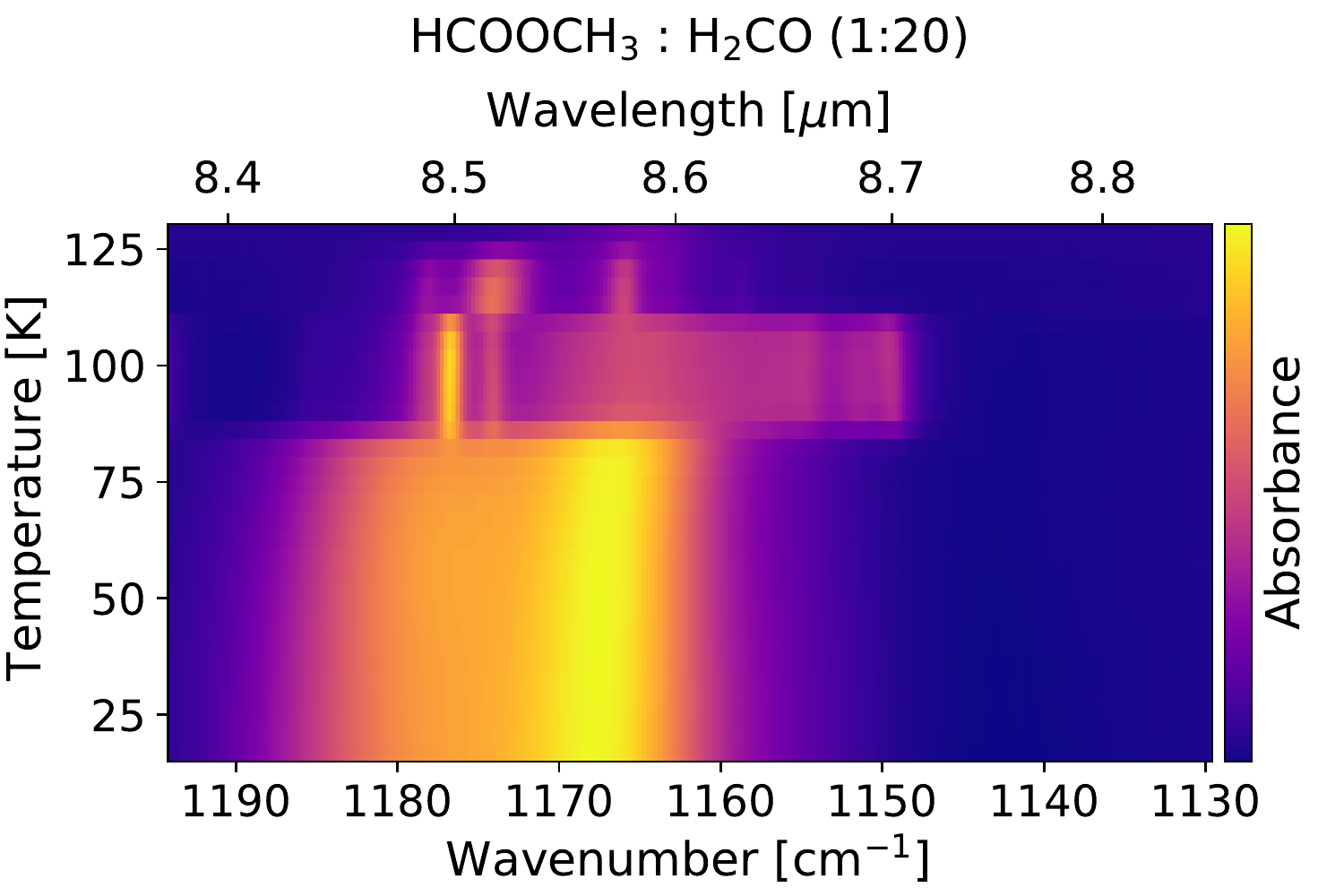} & \includegraphics[width=0.475\hsize]{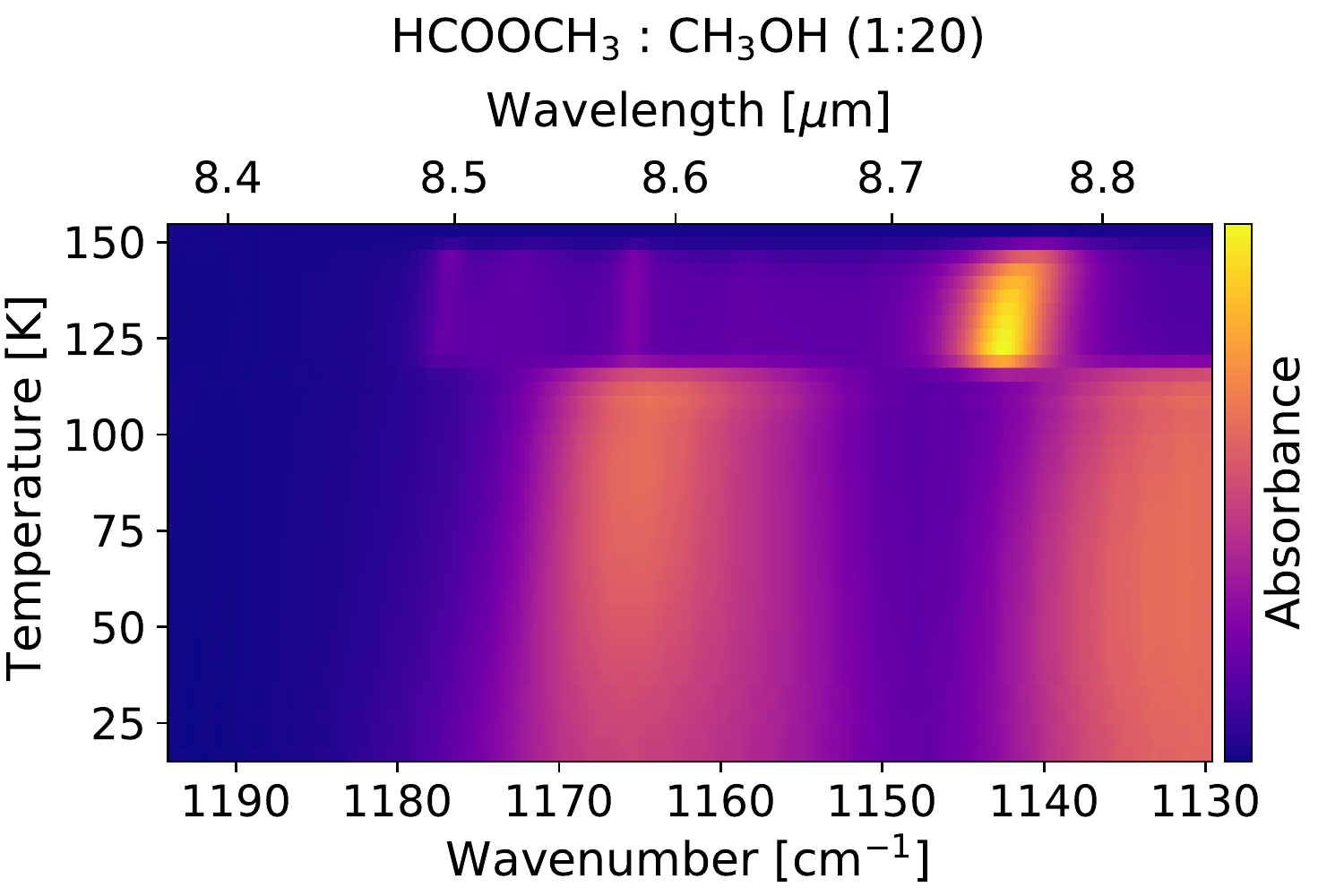} \\
        \includegraphics[width=0.475\hsize]{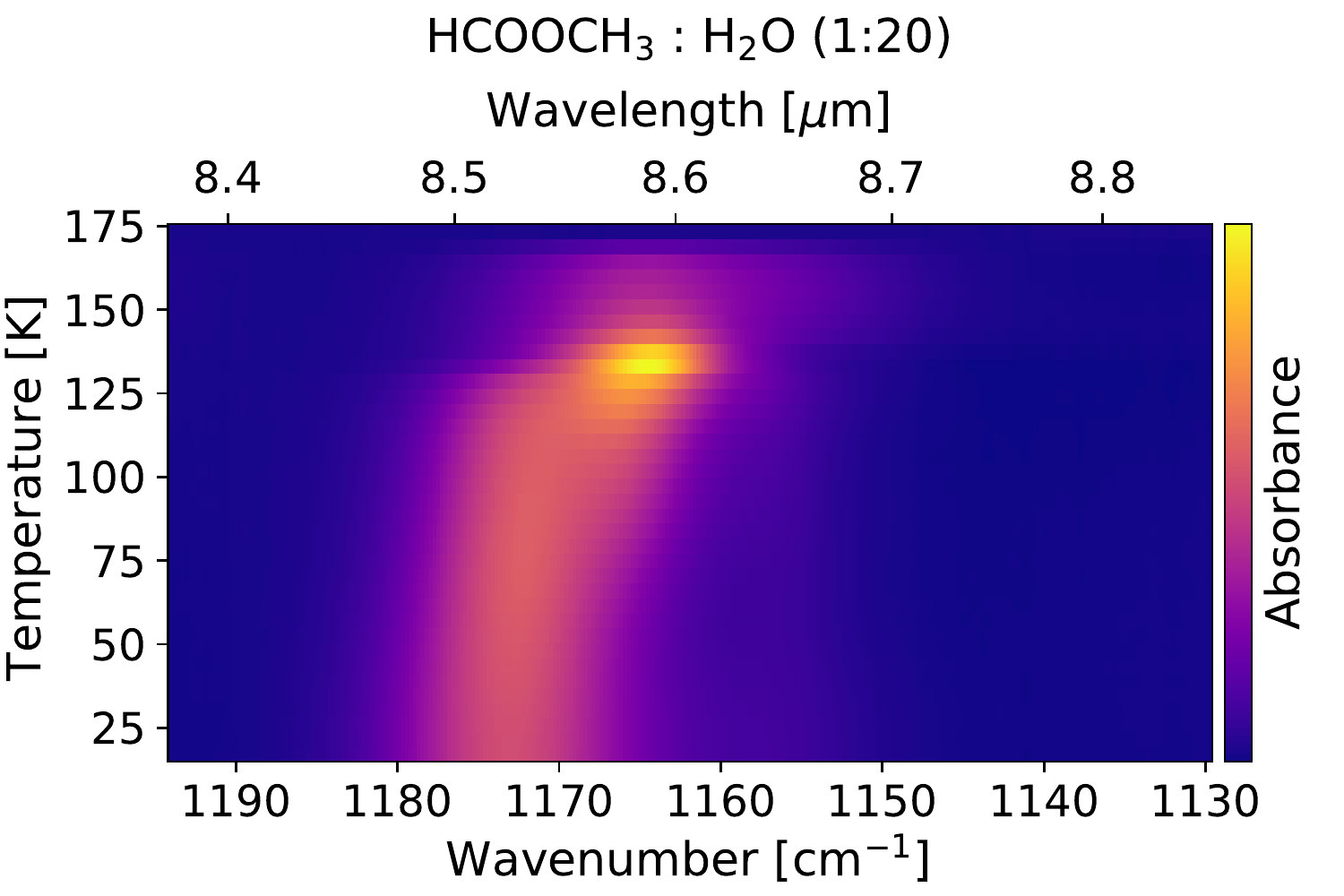} & \includegraphics[width=0.475\hsize]{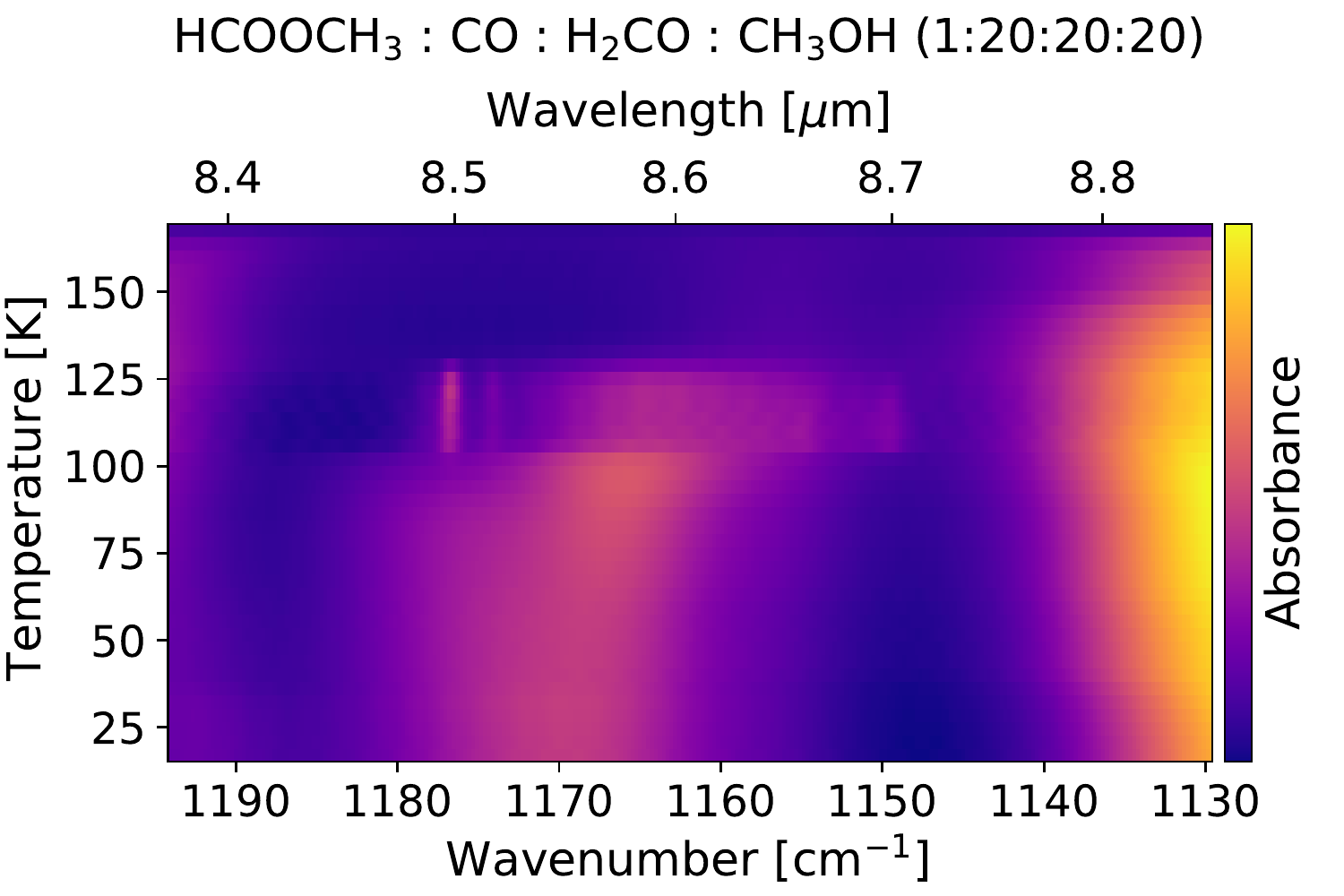} \\
    \end{tabular}
\caption{Temperature versus Spectra (TvS) heatmaps of the CH$_3$ rocking mode (1165.3 cm$^{-1}$ / 8.582 $\mu$m) of HCOOCH$_3$ in the astronomically relevant ice matrices as a function of temperature.}
\label{fig:MF_1165_heatmaps}
\end{figure*}

\begin{figure*}[ht!]
    \includegraphics[width=0.975\hsize]{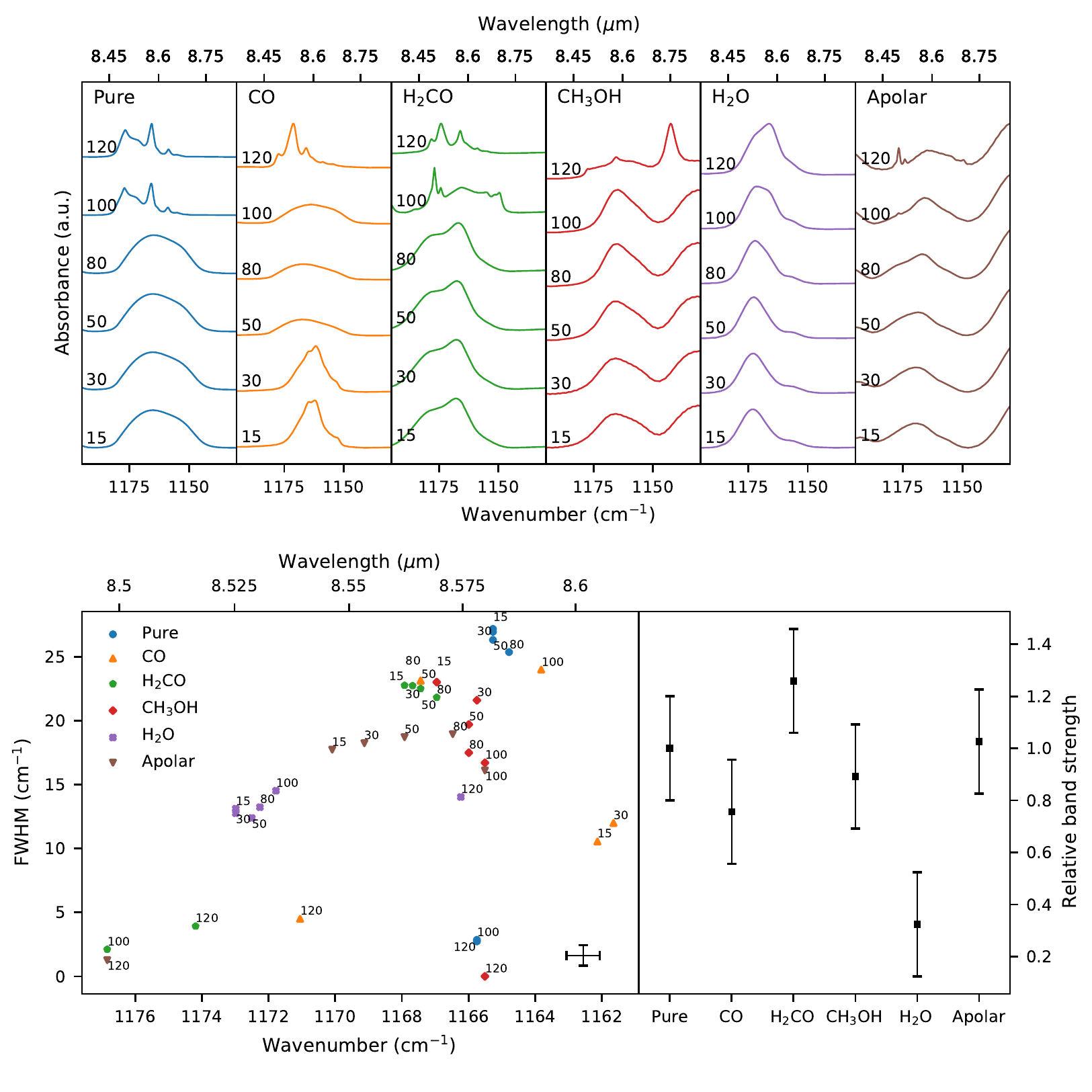}
    \caption{Top panel: Absorption feature of the CH$_3$ rocking vibrational mode (1165.3 cm$^{-1}$ / 8.582 $\mu$m) for each of the mixtures at selected temperatures. Bottom left panel: FWHM as a function of peak position at the selected temperatures. The error bar of each individual point is given in the bottom right. Bottom right panel: The relative band strengths for methyl formate in the different ice matrices at 15 K with respect to the pure band strength.}
    \label{fig:MF_1165}
\end{figure*}

\begin{table*}
    \centering
    \caption{Peak position and FWHM of the methyl formate CH$_3$ rocking mode (1165.3 cm$^{-1}$ / 8.582 $\mu$m) in various matrices.}
    \begin{tabular}{l c l l l l}
        \hline
        \hline
        \multirow{2}{*}{Mixture} & Temperature &  \multicolumn{2}{c}{$\lambda_{\rm{peak}}$} &  \multicolumn{2}{c}{FWHM} \\ 
        & (K) & (cm$^{-1}$) & ($\mu$m) & (cm$^{-1}$)  &  ($\mu$m) \\ 
        \hline
        HCOOCH$_3$              & \multirow{6}{*}{15}   & 1165.3 & 8.5817 & 27.17 & 0.2009\\
        HCOOCH$_3$ : CO         &                       & 1162.1 & 8.6048 & 10.53 & 0.0777\\
        HCOOCH$_3$ : H$_2$CO    &                       & 1167.9 & 8.5622 & 22.75 & 0.1654*\\
        HCOOCH$_3$ : CH$_3$OH   &                       & 1167.0 & 8.5693 & 23.01 & 0.1701\\
        HCOOCH$_3$ : H$_2$O     &                       & 1173.0 & 8.5253 & 13.11 & 0.0953\\
        HCOOCH$_3$ : Apolar     &                       & 1170.1 & 8.5463 & 17.72 & 0.1293\\
        \hline
        HCOOCH$_3$              & \multirow{6}{*}{30}   & 1165.3 & 8.5817 & 26.94 & 0.1992\\
        HCOOCH$_3$ : CO         &                       & 1161.6 & 8.6084 & 12.01 & 0.0887\\
        HCOOCH$_3$ : H$_2$CO    &                       & 1167.7 & 8.5634 & 22.73 & 0.1653*\\
        HCOOCH$_3$ : CH$_3$OH   &                       & 1165.8 & 8.5782 & 21.59 & 0.1596\\
        HCOOCH$_3$ : H$_2$O     &                       & 1173.0 & 8.5253 & 12.75 & 0.0927\\
        HCOOCH$_3$ : Apolar     &                       & 1169.1 & 8.5534 & 18.23 & 0.1331\\
        \hline
        HCOOCH$_3$              & \multirow{6}{*}{50}   & 1165.3 & 8.5817 & 26.31 & 0.1945\\
        HCOOCH$_3$ : CO         &                       & 1167.4 & 8.5658 & 23.14 & 0.1704\\
        HCOOCH$_3$ : H$_2$CO    &                       & 1167.4 & 8.5658 & 22.50 & 0.1637*\\
        HCOOCH$_3$ : CH$_3$OH   &                       & 1166.0 & 8.5763 & 19.69 & 0.1455\\
        HCOOCH$_3$ : H$_2$O     &                       & 1172.5 & 8.5288 & 12.38 & 0.0901\\
        HCOOCH$_3$ : Apolar     &                       & 1167.9 & 8.5622 & 18.70 & 0.1366\\
        \hline
        HCOOCH$_3$              & \multirow{6}{*}{80}   & 1164.8 & 8.5853 & 25.35 & 0.1875\\
        HCOOCH$_3$ : CO         &                       & 1167.4 & 8.5658 & 23.16 & 0.1706\\
        HCOOCH$_3$ : H$_2$CO    &                       & 1167.0 & 8.5693 & 21.70* & 0.1580*\\
        HCOOCH$_3$ : CH$_3$OH   &                       & 1166.0 & 8.5764 & 17.49 & 0.1293\\
        HCOOCH$_3$ : H$_2$O     &                       & 1172.3 & 8.5305 & 13.23 & 0.0964\\
        HCOOCH$_3$ : Apolar     &                       & 1166.5 & 8.5728 & 18.94 & 0.1385\\
        \hline
        HCOOCH$_3$              & \multirow{6}{*}{100}  & 1165.8 & 8.5782 & \phantom{1}2.854 & 0.0210\\
                                &                       & 1177.1 & 8.4956 & $-$ & $-$ \\
        HCOOCH$_3$ : CO         &                       & 1163.8 & 8.5924 & 24.00 & 0.1776\\
        HCOOCH$_3$ : H$_2$CO    &                       & 1176.8 & 8.4973 & \phantom{1}2.113 & 0.0153\\
        HCOOCH$_3$ : CH$_3$OH   &                       & 1165.5 & 8.5799 & 16.70 & 0.1235\\
        HCOOCH$_3$ : H$_2$O     &                       & 1171.8 & 8.5340 & 14.51 & 0.1060\\
        HCOOCH$_3$ : Apolar     &                       & 1165.5 & 8.5799 & 16.09 & 0.1186\\
        \hline
        HCOOCH$_3$              & \multirow{10}{*}{120} & 1165.8 & 8.5782 & \phantom{1}2.742 & 0.0202\\
                                &                       & 1176.8 & 8.4973 & $-$ & $-$ \\
        HCOOCH$_3$ : CO         &                       & 1171.1 & 8.5393 & \phantom{1}4.508 & 0.0328\\
                                &                       & 1165.8 & 8.5782 & $-$ & $-$ \\
        HCOOCH$_3$ : H$_2$CO    &                       & 1174.2 & 8.5165 & \phantom{1}3.921 & 0.0285\\
                                &                       & 1166.0 & 8.5764 & $-$ & $-$ \\
        HCOOCH$_3$ : CH$_3$OH   &                       & 1165.5 & 8.5799 & $-$** & $-$** \\
        HCOOCH$_3$ : H$_2$O     &                       & 1166.2 & 8.5746 & 14.03 & 0.1027\\
        HCOOCH$_3$ : Apolar     &                       & 1176.8 & 8.4973 & \phantom{1}1.258 & 0.0091\\
                                &                       & 1164.5 & 8.5870 & $-$ & $-$ \\
        \hline
        \noalign{\smallskip}
    \end{tabular}
    \tablefoot{* FWHM result of two or more blended peaks. ** Due to sever blending with the matrix constituent no FWHM determination.}
    \label{tab.MF_O-CH3_rock}
\end{table*}

\section{The C$-$O stretching mode}
\begin{figure*}[ht!]
    \begin{tabular}{cc}
        \vspace{0.5cm}
        \includegraphics[width=0.475\hsize]{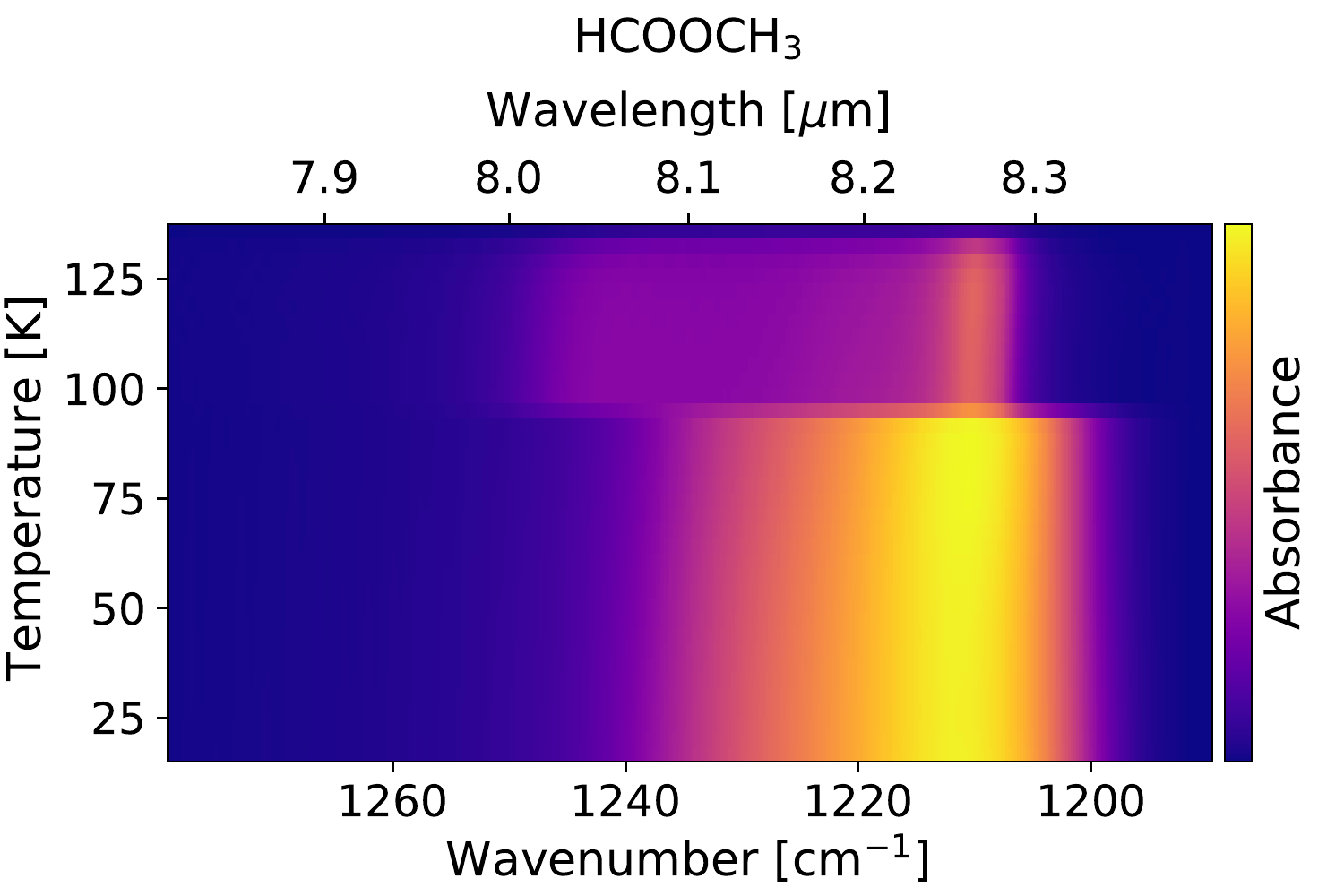} & \includegraphics[width=0.475\hsize]{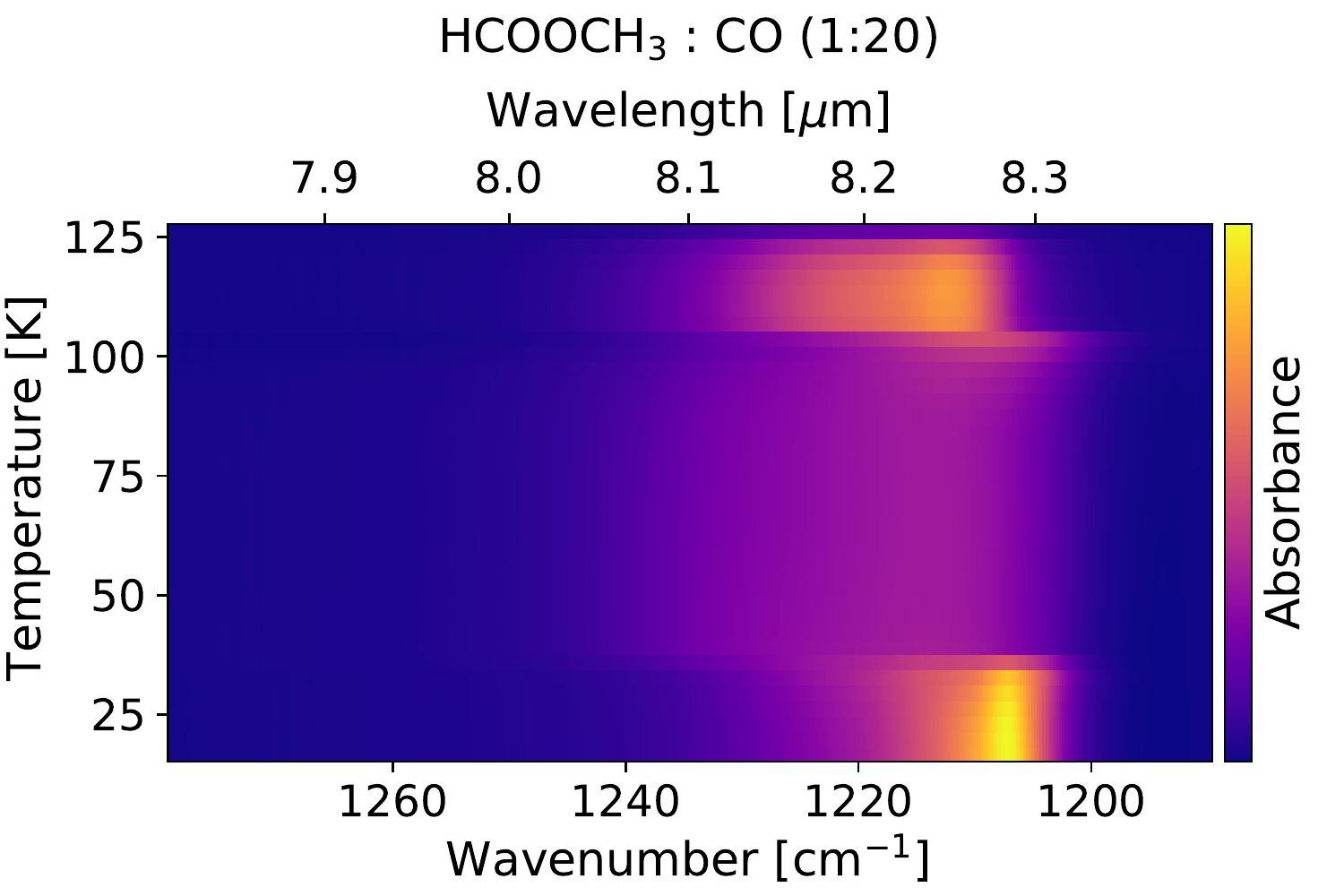} \\
        \vspace{0.5cm}
        \includegraphics[width=0.475\hsize]{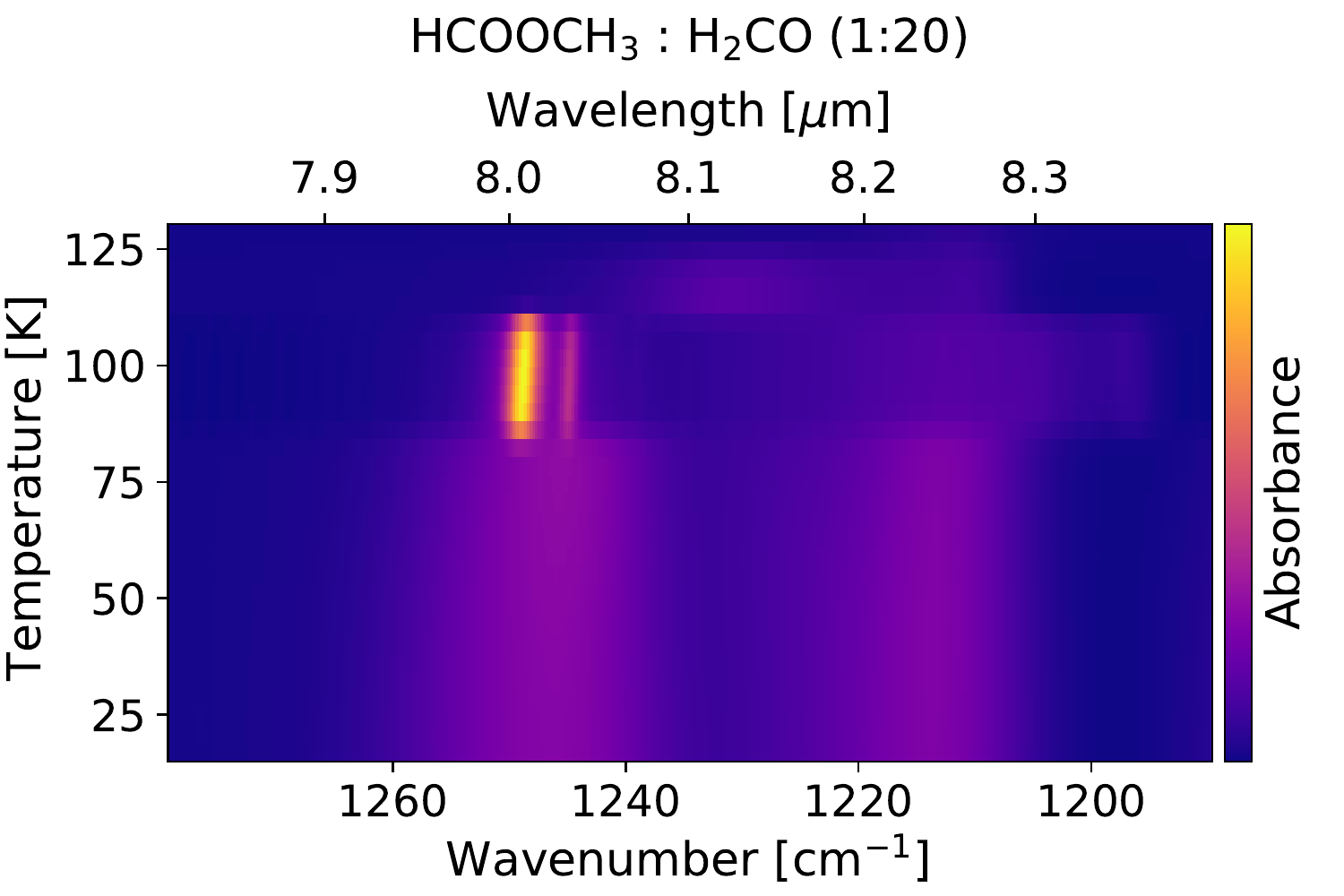} & \includegraphics[width=0.475\hsize]{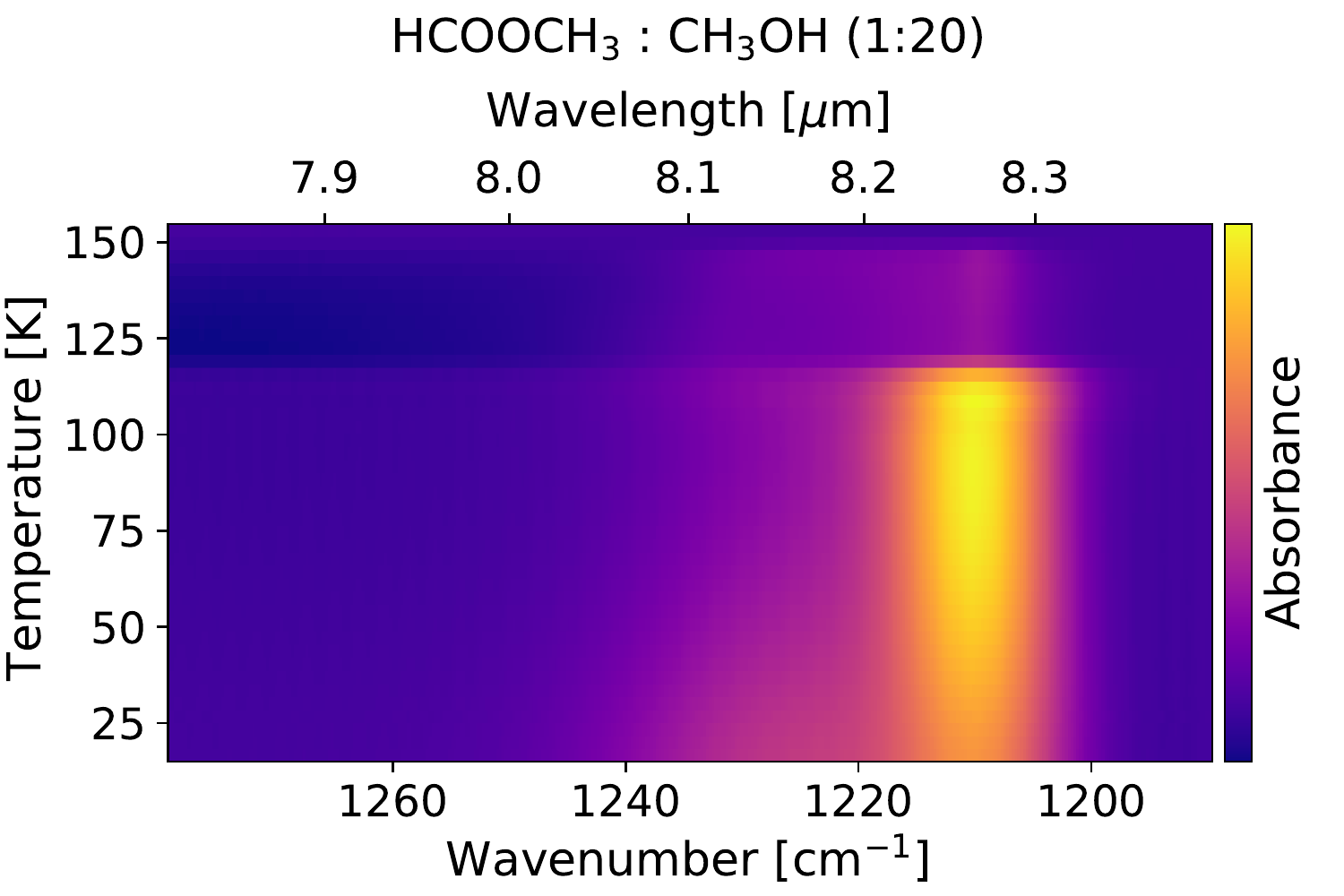} \\
        \includegraphics[width=0.475\hsize]{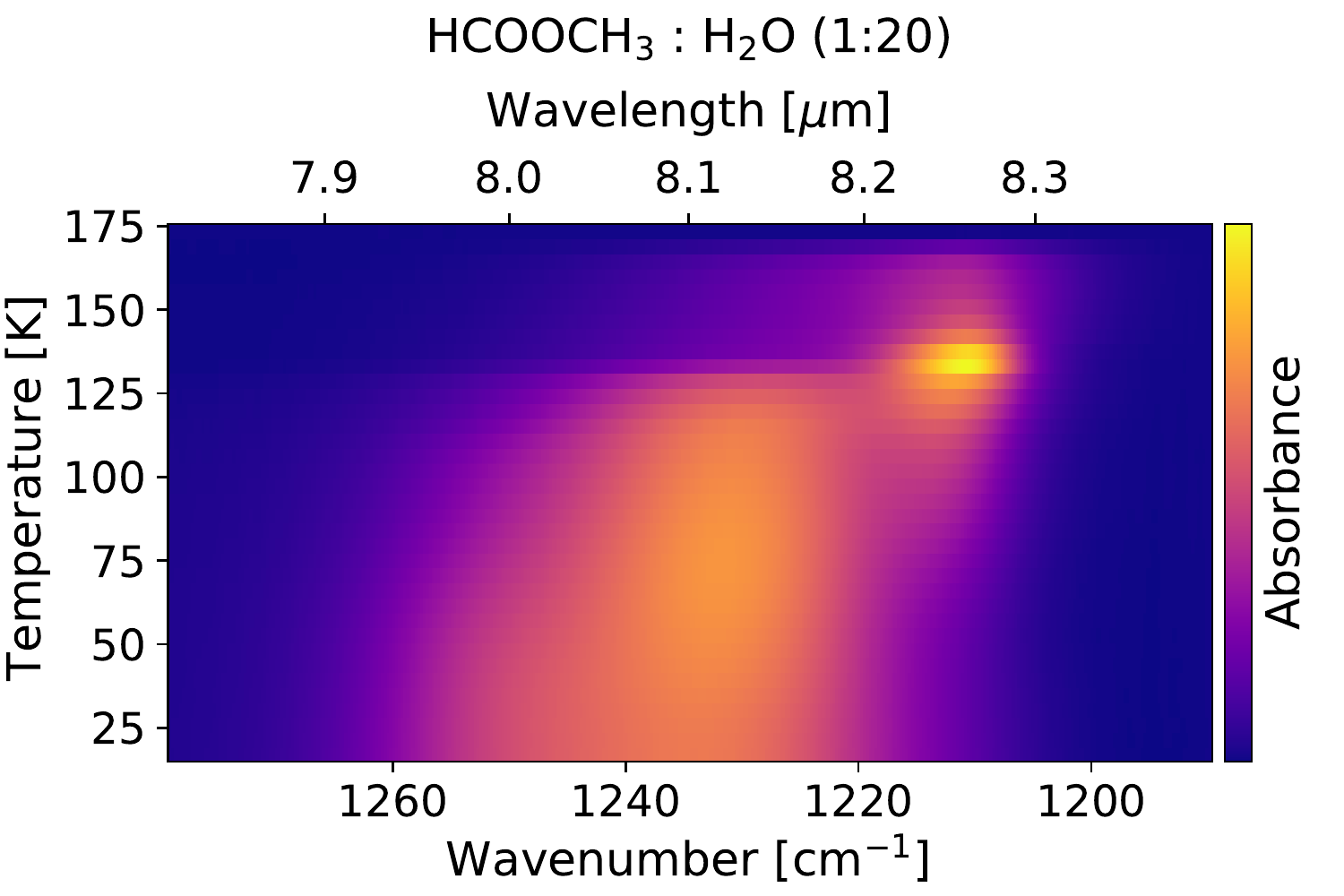} & \includegraphics[width=0.475\hsize]{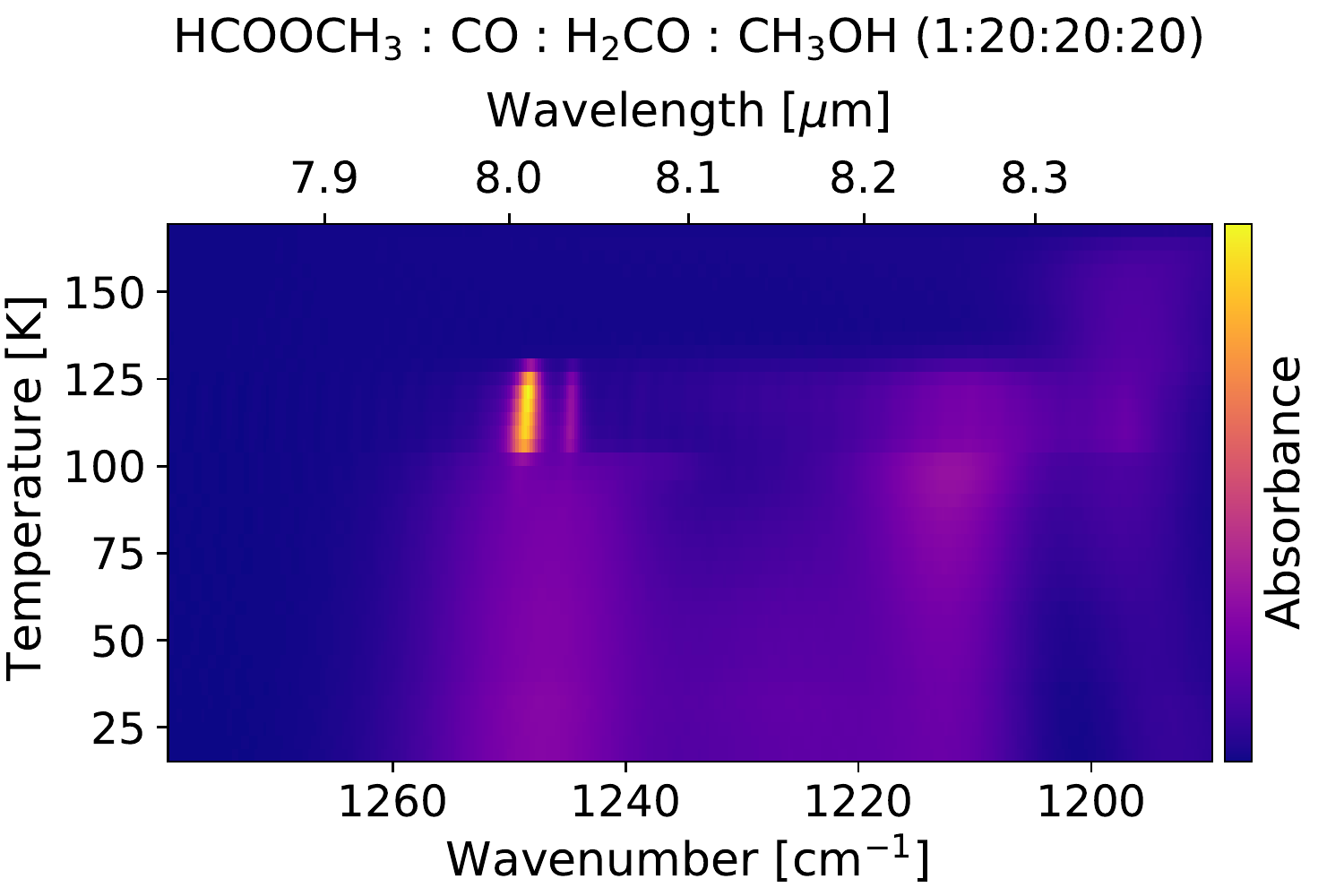} \\
    \end{tabular}
\caption{Temperature versus Spectra (TvS) heatmaps of the C$-$O stretching mode (1211.3 cm$^{-1}$ / 8.256 $\mu$m) of HCOOCH$_3$ in the astronomically relevant ice matrices as a function of temperature.}
\label{fig:MF_1212_heatmaps}
\end{figure*}

\begin{figure*}[ht!]
    \includegraphics[width=0.975\hsize]{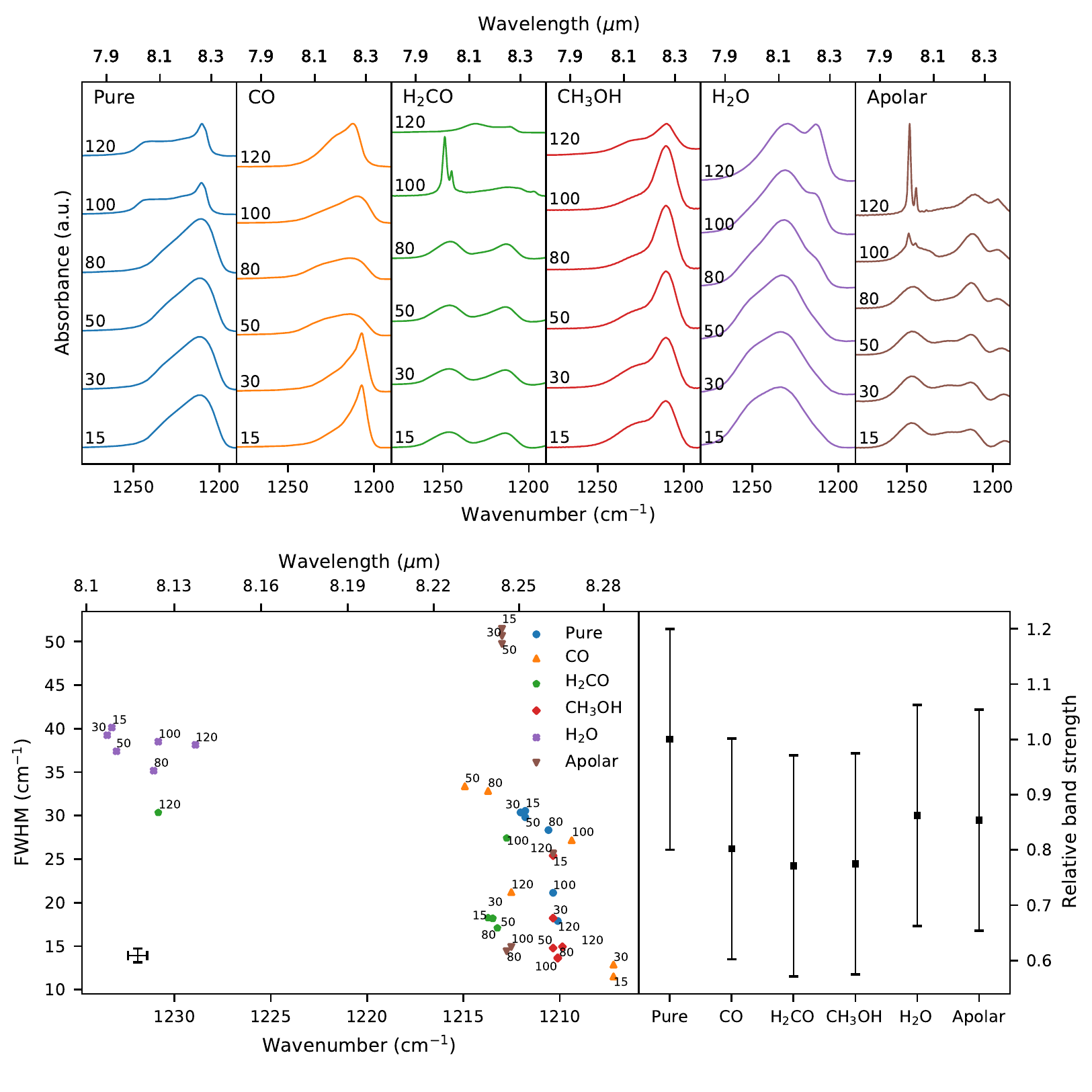}
    \caption{Top panel: Absorption feature of the C$-$O stretching vibrational mode (1211.3 cm$^{-1}$ / 8.256 $\mu$m) for each of the mixtures at selected temperatures. Bottom left panel: FWHM as a function of peak position at the selected temperatures. The error bar of each individual point is given in the bottom left. Bottom right panel: The relative band strengths for methyl formate in the different ice matrices at 15 K with respect to the pure band strength.}
    \label{fig:MF_1212}
\end{figure*}

\begin{table*}
    \centering
    \caption{Peak position and FWHM of the methyl formate C$-$O stretching mode (1211.3 cm$^{-1}$ / 8.256 $\mu$m) in various matrices.}
    \begin{tabular}{l c l l l l}
        \hline
        \hline
        \multirow{2}{*}{Mixture} & Temperature &  \multicolumn{2}{c}{$\lambda_{\rm{peak}}$} &  \multicolumn{2}{c}{FWHM} \\ 
        & (K) & (cm$^{-1}$) & ($\mu$m) & (cm$^{-1}$)  &  ($\mu$m) \\ 
        \hline
        HCOOCH$_3$              & \multirow{6}{*}{15}   & 1211.8 & 8.2522 & 30.55 & 0.2063\\
        HCOOCH$_3$ : CO         &                       & 1207.2 & 8.2835 & 11.53 & 0.0788\\
        HCOOCH$_3$ : H$_2$CO    &                       & 1213.7 & 8.2391 & 18.26 & 0.1234\\
        HCOOCH$_3$ : CH$_3$OH   &                       & 1210.3 & 8.2621 & 25.40 & 0.1718\\
        HCOOCH$_3$ : H$_2$O     &                       & 1233.2 & 8.1087 & 40.13 & 0.2621\\
        HCOOCH$_3$ : Apolar     &                       & 1213.0 & 8.2440 & 51.47* & 0.3389*\\
        \hline
        HCOOCH$_3$              & \multirow{6}{*}{30}   & 1212.0 & 8.2506 & 30.37 & 0.2051\\
        HCOOCH$_3$ : CO         &                       & 1207.2 & 8.2835 & 12.88 & 0.0880\\
        HCOOCH$_3$ : H$_2$CO    &                       & 1213.5 & 8.2408 & 18.20 & 0.1230\\
        HCOOCH$_3$ : CH$_3$OH   &                       & 1210.3 & 8.2621 & 18.22 & 0.1239\\
        HCOOCH$_3$ : H$_2$O     &                       & 1233.5 & 8.1071 & 39.24 & 0.2563\\
        HCOOCH$_3$ : Apolar     &                       & 1213.0 & 8.2440 & 50.66* & 0.3338*\\
        \hline
        HCOOCH$_3$              & \multirow{6}{*}{50}   & 1211.8 & 8.2522 & 29.82 & 0.2015\\
        HCOOCH$_3$ : CO         &                       & 1214.9 & 8.2309 & 33.38 & 0.2242\\
        HCOOCH$_3$ : H$_2$CO    &                       & 1213.5 & 8.2408 & 18.17 & 0.1229\\
        HCOOCH$_3$ : CH$_3$OH   &                       & 1210.3 & 8.2621 & 14.78 & 0.1008\\
        HCOOCH$_3$ : H$_2$O     &                       & 1233.0 & 8.1102 & 37.40 & 0.2445\\
        HCOOCH$_3$ : Apolar     &                       & 1213.0 & 8.2440 & 49.73* & 0.3279*\\
        \hline
        HCOOCH$_3$              & \multirow{6}{*}{80}   & 1210.6 & 8.2604 & 28.34 & 0.1917\\
        HCOOCH$_3$ : CO         &                       & 1213.7 & 8.2391 & 32.85 & 0.2208\\
        HCOOCH$_3$ : H$_2$CO    &                       & 1213.2 & 8.2424 & 17.08 & 0.1156\\
        HCOOCH$_3$ : CH$_3$OH   &                       & 1210.1 & 8.2637 & 13.60 & 0.0928\\
        HCOOCH$_3$ : H$_2$O     &                       & 1231.1 & 8.1230 & 35.17 & 0.2312\\
        HCOOCH$_3$ : Apolar     &                       & 1212.8 & 8.2457 & 14.40 & 0.0978\\
        \hline
        HCOOCH$_3$              & \multirow{6}{*}{100}  & 1210.4 & 8.2621 & 21.13 & 0.1426\\
        HCOOCH$_3$ : CO         &                       & 1209.4 & 8.2687 & 27.18 & 0.1840\\
        HCOOCH$_3$ : H$_2$CO    &                       & 1212.8 & 8.2457 & 27.42 & 0.1857\\
        HCOOCH$_3$ : CH$_3$OH   &                       & 1210.1 & 8.2637 & 13.66 & 0.0932\\
        HCOOCH$_3$ : H$_2$O     &                       & 1230.8 & 8.1246 & 38.50* & 0.2548*\\
        HCOOCH$_3$ : Apolar     &                       & 1212.5 & 8.2473 & 14.90 & 0.1013\\
        \hline
        HCOOCH$_3$              & \multirow{7}{*}{120}  & 1210.1 & 8.2637 & 17.89 & 0.1211\\
        HCOOCH$_3$ : CO         &                       & 1212.5 & 8.2473 & 21.22 & 0.1430\\
        HCOOCH$_3$ : H$_2$CO    &                       & 1230.8 & 8.1246 & 30.35 & 0.2026\\
        HCOOCH$_3$ : CH$_3$OH   &                       & 1209.9 & 8.2654 & 14.93 & 0.1017\\
        HCOOCH$_3$ : H$_2$O     &                       & 1228.9 & 8.1373 & 38.13* & 0.2537*\\
                                &                       & 1213.0 & 8.2440 & $-$ & $-$ \\
        HCOOCH$_3$ : Apolar     &                       & 1210.3 & 8.2621 & 25.64* & 0.1760*\\
        \hline
        \noalign{\smallskip}
    \end{tabular}
    \tablefoot{* FWHM result of two or more blended peaks. ** Due to sever blending with the matrix constituent no FWHM determination.}
    \label{tab.MF_C-O_str}
\end{table*}

\section{The C$=$O stretching mode}
\begin{figure*}[ht!]
    \begin{tabular}{cc}
        \vspace{0.5cm}
        \includegraphics[width=0.475\hsize]{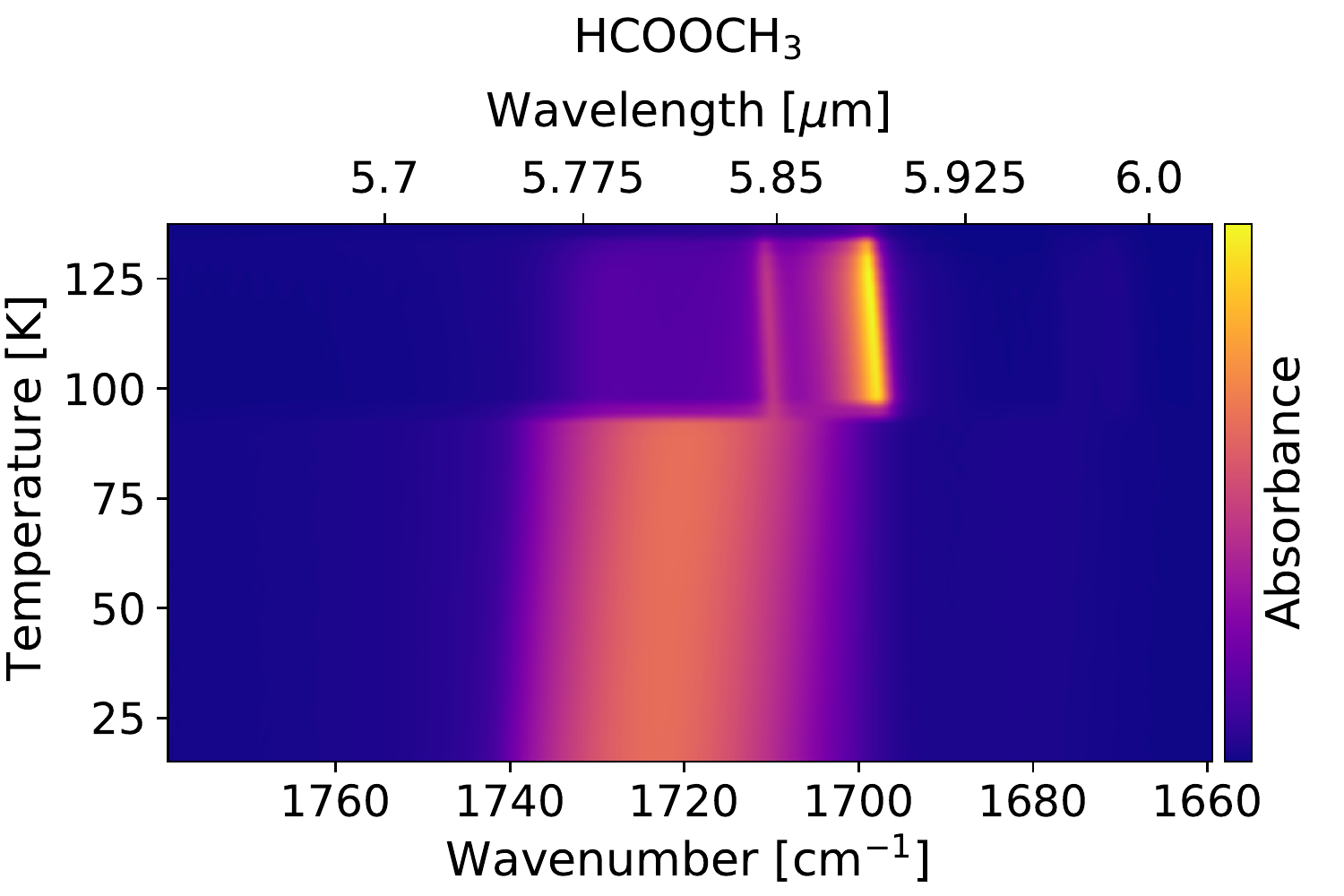} & \includegraphics[width=0.475\hsize]{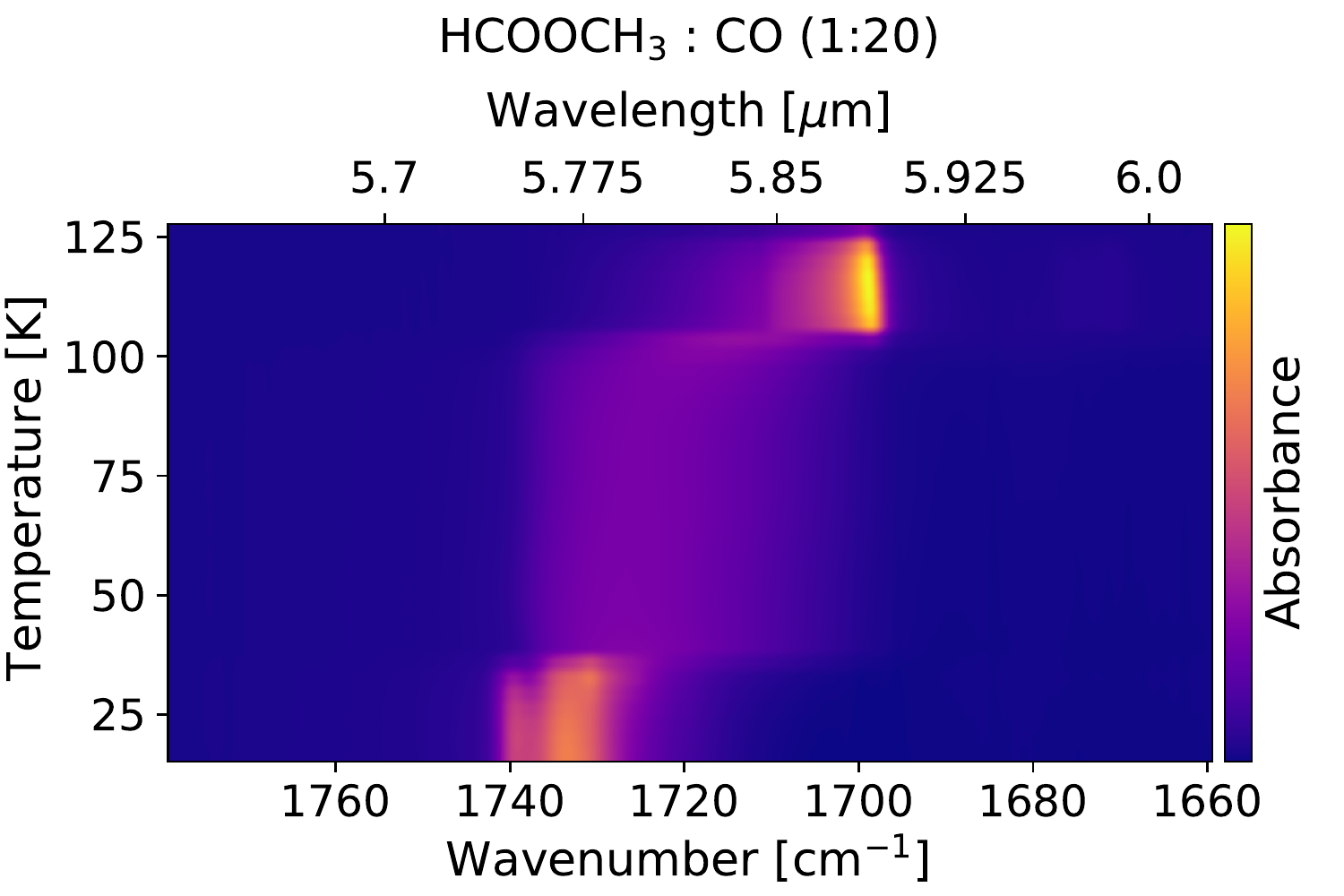} \\
        \includegraphics[width=0.475\hsize]{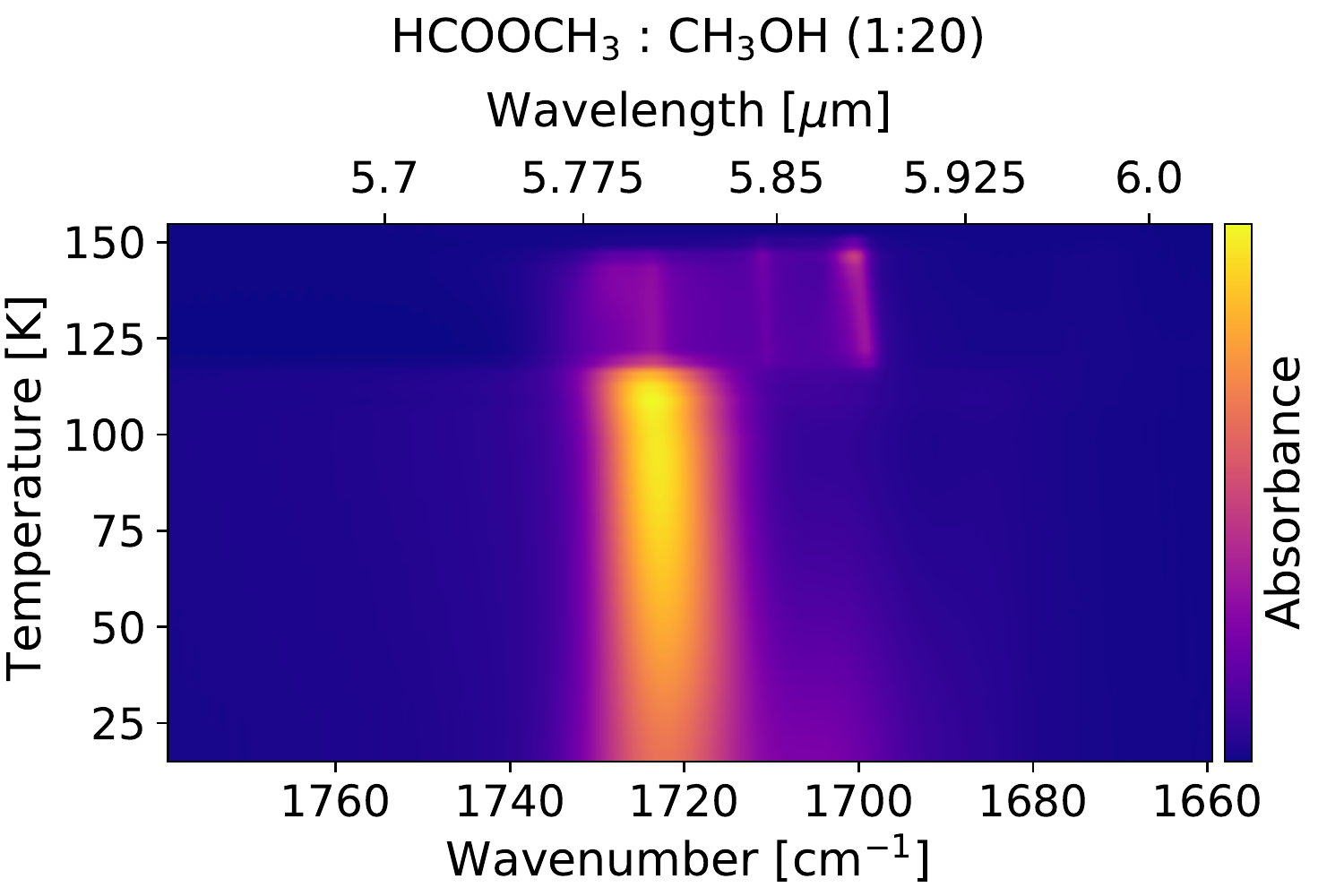} & \includegraphics[width=0.475\hsize]{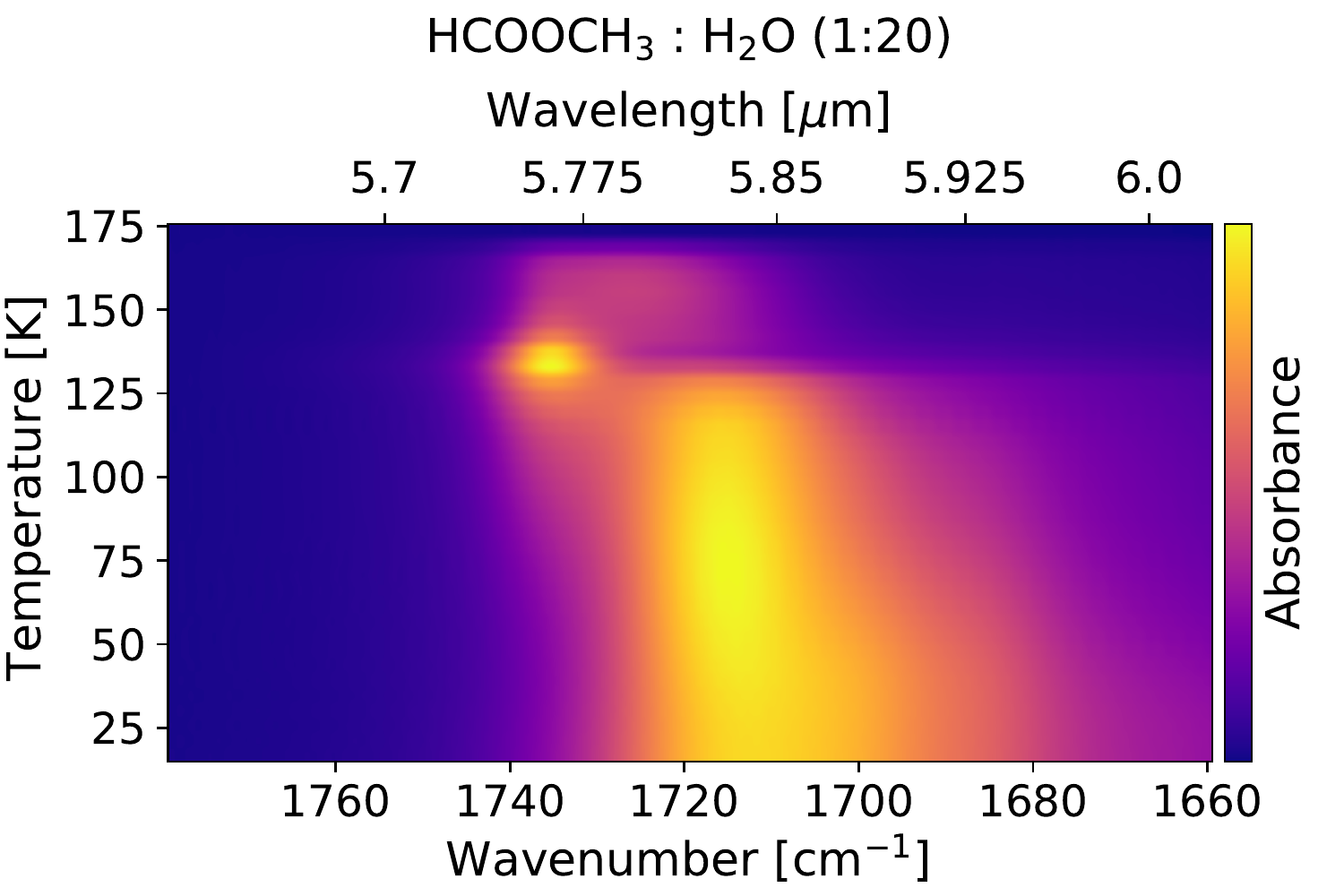} \\
    \end{tabular}
\caption{Temperature versus Spectra (TvS) heatmaps of the C=O stretching mode (1723.1 cm$^{-1}$ / 5.804 $\mu$m) of HCOOCH$_3$ in the astronomically relevant ice matrices as a function of temperature.}
\label{fig:MF_1723_heatmaps}
\end{figure*}

\begin{figure*}[ht!]
    \includegraphics[width=0.975\hsize]{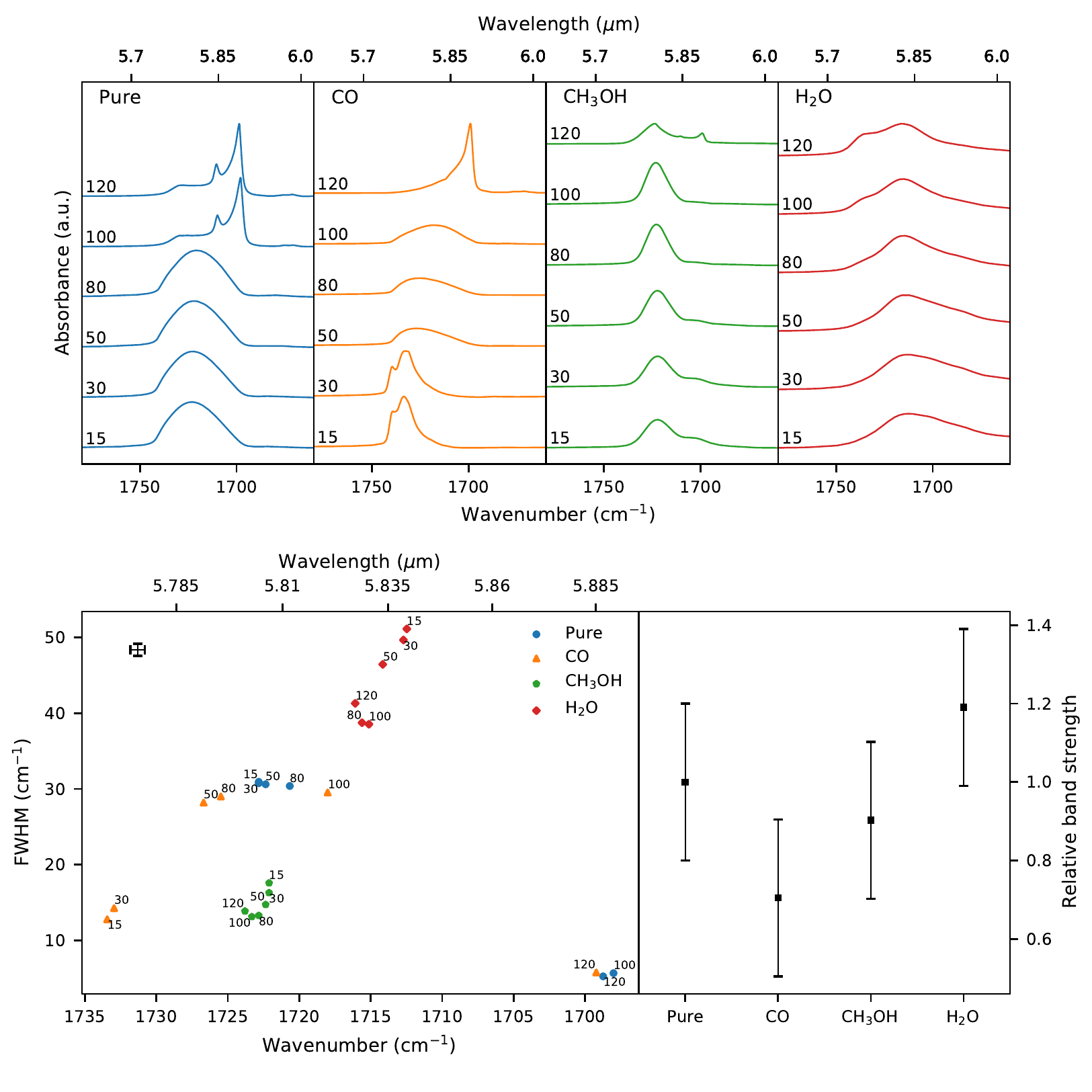}
    \caption{Top panel: Absorption feature of the C=O stretching vibrational mode (1723.1 cm$^{-1}$ / 5.804 $\mu$m) for each of the mixtures at selected temperatures. Bottom left panel: FWHM as a function of peak position at the selected temperatures. The error bar of each individual point is given in the top left. Bottom right panel: The relative band strengths for methyl formate in the different ice matrices at 15 K with respect to the pure band strength.}
    \label{fig:MF_1723}
\end{figure*}

\begin{table*}
    \centering
    \caption{Peak position and FWHM of the methyl formate C=O stretching mode (1723.1 cm$^{-1}$ / 5.804 $\mu$m) in various matrices.}
    \begin{tabular}{l c l l l l}
        \hline
        \hline
        \multirow{2}{*}{Mixture} & Temperature &  \multicolumn{2}{c}{$\lambda_{\rm{peak}}$} &  \multicolumn{2}{c}{FWHM} \\ 
        & (K) & (cm$^{-1}$) & ($\mu$m) & (cm$^{-1}$)  &  ($\mu$m) \\ 
        \hline
        HCOOCH$_3$              & \multirow{4}{*}{15}   & 1722.8 & 5.8044 & 30.94 & 0.1043\\
        HCOOCH$_3$ : CO         &                       & 1733.4 & 5.7689 & 12.80* & 0.0425*\\
        HCOOCH$_3$ : CH$_3$OH   &                       & 1722.1 & 5.8068 & 17.57 & 0.0593\\
        HCOOCH$_3$ : H$_2$O     &                       & 1712.5 & 5.8395 & 51.14* & 0.1762*\\
        \hline
        HCOOCH$_3$              & \multirow{5}{*}{30}   & 1722.8 & 5.8044 & 30.75 & 0.1038\\
        HCOOCH$_3$ : CO         &                       & 1733.0 & 5.7705 & 14.26* & 0.0474*\\
                                &                       & 1739.5 & 5.7489 & $-$ & $-$ \\
        HCOOCH$_3$ : CH$_3$OH   &                       & 1722.1 & 5.8068 & 16.28 & 0.0549\\
        HCOOCH$_3$ : H$_2$O     &                       & 1712.7 & 5.8387 & 49.67* & 0.1711*\\
        \hline
        HCOOCH$_3$              & \multirow{4}{*}{50}   & 1722.4 & 5.8060 & 30.61 & 0.1033\\
        HCOOCH$_3$ : CO         &                       & 1726.7 & 5.7914 & 28.20 & 0.0950\\
        HCOOCH$_3$ : CH$_3$OH   &                       & 1722.4 & 5.8060 & 14.72 & 0.0496\\
        HCOOCH$_3$ : H$_2$O     &                       & 1714.2 & 5.8338 & 46.45* & 0.1597*\\
        \hline
        HCOOCH$_3$              & \multirow{4}{*}{80}   & 1720.7 & 5.8117 & 30.38 & 0.1027\\
        HCOOCH$_3$ : CO         &                       & 1725.5 & 5.7955 & 29.02 & 0.0978\\
        HCOOCH$_3$ : CH$_3$OH   &                       & 1722.8 & 5.8044 & 13.27 & 0.0447\\
        HCOOCH$_3$ : H$_2$O     &                       & 1715.6 & 5.8289 & 38.74* & 0.1325*\\
        \hline
        HCOOCH$_3$              & \multirow{5}{*}{100}  & 1698.0 & 5.8893 & \phantom{1}5.659 & 0.0196\\
                                &                       & 1709.8 & 5.8486 & $-$ & $-$ \\
        HCOOCH$_3$ : CO         &                       & 1718.0 & 5.8207 & 29.54 & 0.1000\\
        HCOOCH$_3$ : CH$_3$OH   &                       & 1723.3 & 5.8028 & 13.11 & 0.0442\\
        HCOOCH$_3$ : H$_2$O     &                       & 1715.1 & 5.8305 & 38.54* & 0.1312*\\
        \hline
        HCOOCH$_3$              & \multirow{6}{*}{120}  & 1698.7 & 5.8868 & \phantom{1}5.239 & 0.0181\\
                                &                       & 1710.5 & 5.8461 & $-$ & $-$ \\
        HCOOCH$_3$ : CO         &                       & 1699.2 & 5.8851 & \phantom{1}5.769 & 0.0200\\
        HCOOCH$_3$ : CH$_3$OH   &                       & 1723.8 & 5.8011 & 13.85 & 0.0466\\
                                &                       & 1699.2 & 5.8851 & $-$ & $-$ \\
        HCOOCH$_3$ : H$_2$O     &                       & 1716.1 & 5.8272 & 41.30* & 0.1398*\\
        \hline
        \noalign{\smallskip}
    \end{tabular}
    \tablefoot{* FWHM result of two or more blended peaks. ** Due to sever blending with the matrix constituent no FWHM determination.}
    \label{tab.MF_C=O_str}
\end{table*}

\end{appendix}
\end{document}